\pdfoutput=1
\documentclass[12pt,a4paper]{article}

\usepackage{jheplike,ifpdf,tocloft,rotating,mathrsfs,mathtools}

\hypersetup{pdftitle={},pdfcreator={},linkcolor=[rgb]{0.15,0.35,0.75},colorlinks=true,citecolor=[rgb]{0.675,0,0.2},urlcolor=[rgb]{0.15,0.35,0.65}}
\let\olditemize\itemize\renewcommand{\itemize}{\vspace{-2pt}\olditemize\setlength{\itemsep}{1pt}\setlength{\parskip}{0pt}\setlength{\parsep}{-0pt}}
\let\oldenumerate\enumerate\renewcommand{\enumerate}{\vspace{-4pt}\oldenumerate\setlength{\itemsep}{1pt}\setlength{\parskip}{0pt}\setlength{\parsep}{0pt}}

\DeclareMathOperator*{\Res}{\mathrm{Res}}
\newcommand{\eq}[1]{\vspace{-0.5pt}\begin{equation}#1\vspace{-0.5pt}\end{equation}}
\newcommand{\fwbox}[2]{\text{\makebox[#1][c]{$\hspace{-150pt}\displaystyle#2\hspace{-150pt}$}}}
\newcommand{\fwboxL}[2]{\text{\makebox[#1][l]{$#2$}}}
\newcommand{\fwboxR}[2]{\text{\makebox[#1][r]{$#2$}}}
\newcommand{\fig}[3]{\raisebox{#1}{\includegraphics[scale=#2]{#3}}}
\newcommand{\mi}{\raisebox{0.75pt}{\scalebox{0.75}{$\hspace{-2pt}\,-\,\hspace{-0.5pt}$}}}
\newcommand{\pl}{\raisebox{0.75pt}{\scalebox{0.75}{$\hspace{-2pt}\,+\,\hspace{-0.5pt}$}}}

\renewcommand{\hat}{\widehat}
\renewcommand{\tilde}{\widetilde}
\newcommand{\ab}[1]{\langle #1\rangle}
\newcommand{\x}[2]{(#1,#2)}
\newcommand{\merge}{\raisebox{-1.5pt}{\scalebox{1.5}{$\otimes$}}}
\newcommand{\newcap}{\mathrm{\raisebox{0.75pt}{{$\,\bigcap\,$}}}}
\newcommand{\tcap}{\scalebox{1}{$\!\newcap\!$}}
\newcommand{\tncap}{\scalebox{0.8}{$\!\newcap\!$}}

\renewcommand{\r}[1]{{\color{hred}(#1)}}
\newcommand{\proj}[1]{\left[#1\right]}
\newcommand{\tme}[1]{[#1]}
\newcommand{\tmec}[1]{\overline{[#1]}}
\newcommand{\tens}{\!\hspace{-1pt}\otimes\!\hspace{-1pt}}
\newcommand{\cop}[2]{\left[#1\right]\tens\left[#2\right]}
\newcommand{\Li}[2]{\hspace{1pt}\mathrm{Li}_{#1}(#2)}
\newcommand{\even}{\hspace{1pt}\text{e}}
\newcommand{\odd}{\hspace{1pt}\text{o}}
\newcommand{\eps}{\delta}
\newcommand{\Zeta}[1]{\hspace{1pt}\zeta_{#1}}
\newcommand{\logsquare}[1]{\log^{\hspace{-0.25pt}2}\hspace{-1pt}(#1)}
\newcommand{\logsquarediv}{\logsquare{\eps}}
\newcommand{\logdiv}{\log(\eps)}
\newcommand{\logkdiv}[1]{\log^{\hspace{-0.25pt}#1}\hspace{-1pt}(\eps)}
\newcommand{\logk}[2]{\log^{\hspace{-0.25pt}#1}\hspace{-1pt}(#2)}
\renewcommand{\u}{u_1}\renewcommand{\v}{u_2}\newcommand{\w}{u_3}
\newcommand{\yu}{y_1}\newcommand{\yv}{y_2}\newcommand{\yw}{y_3}
\newcommand{\cusp}{\gamma_{\text{{\rm cusp}}}}
\newcommand{\td}{d}
\newcommand{\iu}{i}
\newcommand{\ipi}{\pi\iu}
\newcommand{\asyO}[1]{\mathcal{O}\left( #1 \right)}
\newcommand{\Mellin}[1]{\mathcal{M}\{#1\}}
\newcommand{\defas}{\equiv}
\newcommand{\LScoeff}[1]{\hat{#1}}
\newcommand{\pFq}[3]{{}_2F_1\left( { #1 \atop #2} \middle| #3\right)}
\newcommand{\HyperInt}{\href{http://bitbucket.org/PanzerErik/hyperint/}{\texttt{\textup{HyperInt}}}}
\newcommand{\Maple}{\textsf{Maple}}
\newcommand{\Mathematica}{\textsc{Mathematica}}

\newcommand{\impureFunctionSymbol}{\mathcal{Y}}
\newcommand{\divergentFunctionSymbol}{\mathcal{X}}
\newcommand{\hInt}[2]{H_{#1}^{#2}}

\definecolor{mhvBlue}{rgb}{0.3,0.2,0.75}
\definecolor{fRed}{rgb}{0.48,0.02824,0.18824}
\definecolor{cut2}{rgb}{0.18824,0.18824,0.48}
\definecolor{cut1}{rgb}{0.48,0.02824,0.18824}
\definecolor{hblue}{rgb}{0,0,0.575}
\definecolor{hred}{rgb}{0.575,0.0,0.225}
\definecolor{hteal}{rgb}{0.0,0.545,0.7451}
\definecolor{dred}{rgb}{0.575,0.4,0.45}
\definecolor{divblue}{rgb}{0,0,0.575}
\definecolor{wred}{rgb}{0.575,0.0,0.225}

\title{~\\[50pt]{\LARGE \mbox{Manifestly Dual-Conformal Loop Integration}}\\[-24pt]}
\author[1]{\vspace{-10pt}Jacob~L.~Bourjaily}\affiliation[1]{Niels Bohr International Academy and Discovery Center, University of Copenhagen\\The Niels Bohr Institute, Blegdamsvej 17, DK-2100, Copenhagen \O, Denmark}\emailAdd{bourjaily@nbi.ku.dk}
\affiliation[2]{SLAC National Accelerator Laboratory, Stanford University, Stanford, CA 94309, USA}
\author[2]{\!\!\!\!,\,\,\,Falko~Dulat}\emailAdd{dulatf@slac.stanford.edu}
\author[3]{\!\!\!\!,\,\,\,Erik~Panzer}
\affiliation[3]{All Souls College, University of Oxford, OX1~4AL, Oxford, UK}\emailAdd{erik.panzer@all-souls.ox.ac.uk}
\abstract{
Local, manifestly dual-conformally invariant loop integrands are now known for all finite quantities associated with observables in planar, maximally supersymmetric Yang-Mills theory through three loops. These representations, however, are not infrared-finite term by term and therefore require regularization; and even using a regulator consistent with dual-conformal invariance, ordinary methods of loop integration would na\"{i}vely obscure this symmetry. In this work, we show how any planar loop integral through at least two loops can be systematically regulated and evaluated directly in terms of strictly finite, manifestly dual-conformal Feynman-parameter integrals. We apply these methods to the case of the two-loop ratio and remainder functions for six particles, reproducing the known results in terms of individually regulated local loop integrals, and we comment on some of the novelties that arise for this regularization scheme not previously seen at one loop. 
}
\preprint{\begin{flushright}SLAC-PUB-17371\end{flushright}}

\begin{document}
\maketitle\thispagestyle{empty}

\setcounter{page}{1}
\vspace{-12pt}\section{Introduction and Overview}\label{sec:introduction}\vspace{-6pt}

There has been incredible progress in our understanding of perturbative quantum field theory in recent years. This is especially (and increasingly) true for perturbative theories involving massless particles, in four dimensions, with (maximal) supersymmetry, and a planar limit. But many of the insights gained from studying particularly simple quantum field theories such as maximally supersymmetric ($\mathcal{N}\!=\!4$) Yang-Mills theory (SYM) in the planar limit have much wider (potential or realized) applications.

Among the greatest source of progress has been the realization that the difficulties of renormalization, regularization, and loop integration can (and should) be separated from those of constructing loop {\it integrands}. Prior to loop integration, loop integrands are rational functions like tree-amplitudes which can be determined by their residues (or `cuts'); and, provided that certain qualifications about scheme dependence or regulated Feynman rules are taken into account, loop integrands may be meaningfully defined and computed without dealing with ultraviolet or infrared divergences. Indeed, there have been enormous advances in our ability to represent perturbative scattering amplitudes at the loop-integrand-level in a wide variety of quantum field theories. And this progress has led to the discovery of many important insights with broad applications. For example, the discovery of on-shell recursion of tree-amplitudes \cite{ArkaniHamed:2010kv}, dual-conformal symmetry \cite{Drummond:2006rz,Alday:2007hr,Drummond:2007au,Drummond:2008vq}, the infinite-dimensional (Yangian) symmetry for planar SYM \cite{Drummond:2009fd}, and the correspondence with Grassmannian geometry \cite{ArkaniHamed:2009dg,ArkaniHamed:2009dn,ArkaniHamed:2009sx,ArkaniHamed:2012nw,Bourjaily:2010kw,Bourjaily:2012gy,Bourjaily:2016mnp} were all discovered from loop-integrand-level investigations. 

However, even for the simplest ultraviolet-finite theories involving massless particles, one must eventually deal with the fact that scattering amplitudes themselves are not observable: they suffer from infrared divergences that require regularization. And yet, it has long been expected that at least some of the symmetries of loop integrands such as dual-conformal invariance should survive to infrared-finite quantities related to amplitudes such as ratio and remainder functions. It is not unreasonable to hope that there exists some formalism for computing infrared-finite quantities, such as ratio functions, in terms that do not require regularization, but we have little to add to such a hope---except to say that even if such a formalism were to exist, the best developed methods of loop integration---Feynman parameterization, Mellin-Barnes, etc.---are poorly suited to preserve symmetries such as dual-conformal invariance (at least as they have been traditionally described).  

In this work, we describe how dual-conformal invariance of planar loop integrands can preserved through infrared regularization and Feynman parameterization. The existence of an infrared regulator consistent with dual-conformal invariance was shown in \mbox{ref.\ \cite{Bourjaily:2013mma}}. This regulator works by giving (leg-label-dependent) masses to all external legs proportional to a parameter `$\eps$' which is both dimensionless and invariant under dual-conformal transformations. In the limit of $\eps\!\to\!0$, any planar loop integral relevant to SYM is expected to become regularized as a polynomial in $\logdiv$. When this is the case, we show how the coefficients of any $\logkdiv{k}$ divergence can be expressed in terms of finite, Feynman-parametric integrals with denominators that depend exclusively on `parity-even' cross-ratios---those rationally related to Mandelstam invariants. 

We use these ideas to re-compute the two-loop ratio and
remainder~\cite{DelDuca:2009au,DelDuca:2010zg,Golden:2013xva} functions for six particles in planar SYM, using the integrand-level representation of amplitudes given in \mbox{ref.\ \cite{Bourjaily:2015jna}}. We do this in part because these functions are well known and verifiable; but we expect that these methods will prove useful for a wider class of still unexplored cases. Although the dual-conformal integration techniques outlined here seem valuable quite generally, it is clear that applications to higher multiplicity, for example, will still require new insights regarding the `right' kinematic variables and the best way to represent integrals that are not polylogarithmic. We must leave these issues to future work, but we expect that the parametric integral representations obtained using the strategy we describe here will, at the very least, recast the issues involved in going further in a sharper light.

Before moving on, it is worth mentioning that starting from loop integrands may not prove the most efficient way of computing divergence-free quantities related to observables. Indeed, in the case of six (or seven) particle amplitudes in planar SYM---the primary example studied in this work---the symbolic bootstrap approach described in \mbox{refs.\ \cite{Dixon:2011pw,Dixon:2013eka,Dixon:2014voa,Caron-Huot:2016owq,Drummond:2014ffa,Dixon:2016nkn,Li:2016ctv,Almelid:2017qju,Chicherin:2017dob,Caron-Huot:2018dsv}} has proven dramatically more powerful than integrand-to-integral strategies. This strategy eschews reference to loop integrands altogether, determining finite quantities directly from the space of functions expected to arise from loop integration---fixed by some number of globally-defined physical constraints. Indeed, it is hard to envision our computation of the two loop ratio and remainder functions for six particles as being competitive with the symbolic bootstrap which has already been used to determine these functions through seven loops \cite{sixLoops}. Our main purpose here is to improve integration technology for local Feynman integrals, which have much broader applications than is currently understood using symbolic bootstraps. Moreover, the Feynman integral approach provides the most concrete evidence that the bootstrap results are both correct and complete.  ~\\[-6pt]

This paper is organized as follows. In section~\ref{sec:regularization_and_integration} we describe how planar loop integrands (at least through two loops) can in general be regularized and integrated in terms of (Feynman-)parameter integrals that depend exclusively on `parity-even' dual-conformal cross-ratios---those rationally-related to ordinary Mandelstam invariants. This will involve the use of the so-called `dual-conformal regulator' introduced in \mbox{ref.\ \cite{Bourjaily:2013mma}} which we review in section~\ref{subsec:dci_regulator}. We describe the mechanics of Feynman-parameterization in dual-coordinates and the embedding formalism in section~\ref{subsec:feynman_parameterization}, and we demonstrate---by direct construction through two loops---that these Feynman-parametric representations can always be transformed in a way that expresses any planar integral in terms of a basis of integrals that {\it exclusively} (and manifestly) depend on parity-even conformal cross-ratios. For the sake of clarity, we outline how this works at one loop in section~\ref{subsubsec:feynman_parameter_ints_at_one_loop} and at two loops in section~\ref{subsubsec:feynman_parameter_ints_at_two_loop}---with several more illustrative examples discussed in section~\ref{subsubsec:exempli_gratia}. 

Using the dual-conformal regulator, any ultraviolet-finite integral can be expressed as
a polynomial in $\logdiv$;\footnote{In principle, there could also be poles in $\eps$. Such poles would not complicate our analysis, but seem not to arise for integrals in which we are interested.} the coefficients in this expansion are guaranteed to be dual-conformally invariant. In section~\ref{subsec:divergent_coefficients_extraction} we outline a general and systematic strategy to extract each coefficient of $\logkdiv{k}$ in the $\eps\!\to\!0$ limit in terms of finite, dual-conformal, (Feynman-)parametric integrals. In section~\ref{subsec:analytic_integration_methods} we discuss how such parametric integrals may be systematically evaluated (in terms of hyperlogarithms, if possible) using algorithms for the iterative integration of hyperlogarithms~\cite{Brown:2011ik,Anastasiou:2013srw,Panzer:2014gra} implemented in tools such as {\HyperInt}~\cite{Panzer:2014caa}.

In section~\ref{sec:loop_integrands} we review the notation and formalism involved in the representation of local loop integrands for scattering amplitudes (and related functions) in planar SYM. We start with a description of one-loop amplitude integrands in section~\ref{subsec:one_loop_amplitude_integrands}, and two loops in section~\ref{subsec:two_loop_amplitude_integrands}. While mostly review, section~\ref{sec:loop_integrands} should establish all the essential ingredients required in the primary applications discussed in section~\ref{sec:six_point_example}, where we apply these ideas to reproduce the two-loop ratio and remainder functions for six particles in planar SYM, starting from a local integrand expression and directly integrating each term. Beyond (merely) a demonstration of the viability of this approach, this example illustrates several interesting and unanticipated novelties about the dual-conformal regularization scheme not seen at one loop. Examples of the individual terms that contribute to the six-point ratio function at two loops are discussed in section~\ref{subsec:examples}; the structure of the (cancelling) term-wise divergences is described in section~\ref{subsubsec:impurities_and_other_novelties} together with the appearance of lower-weight contributions and `impurities'. In section~\ref{subsec:two_loop_log} we describe the form of the two-loop-logarithm of the six-particle MHV amplitude and discuss its relationship to the remainder function and the cusp anomalous dimension. All the notation and conventions needed for this analysis are described in Appendix~\ref{appendix:conventions_and_definitions}. Details of the individual contributions to the ratio function are discussed in Appendix~\ref{appendix:explicit_integrals}, including technical aspects of how they were obtained in Appendix~\ref{appendix:technical_aspects_of_integration}. Included as ancillary files to this work's submission to the {\tt arXiv} are complete expressions for these contributions---including both the dual-conformally-invariant (Feynman-)parametric integrals and the explicit hyperlogarithms that result from parametric integration. These files are also available under DOI \href{http://doi.org/10.5287/bodleian:BRyawJrRN}{10.5287/bodleian:BRyawJrRN}. The details of how these files are organized is described in Appendix~\ref{appendix:organization_of_ancillary_files}.

We conclude in section~\ref{sec:conclusions} with a brief discussion how these ideas may be generalized, and the new (or newly-sharpened) issues that must be addressed to exploit the full potential of this technology.

\newpage
\vspace{-6pt}\section{Dual-Conformal Regularization and Feynman Integration}\label{sec:regularization_and_integration}\vspace{-6pt}

In this section, we review the dual-conformal regularization scheme introduced in \mbox{ref.\ \cite{Bourjaily:2013mma}}, and show how (at least through two loops) all dual-conformal integrals---including those that require infrared regularization---can be expressed in terms of {\it manifestly} dual-conformal Feynman parameter integrals (possibly also depending on a single, scale-invariant regularization parameter). Moreover, we will show that any one- or two-loop dual-conformal integrand can be expanded into a basis of (not-necessarily pure) Feynman parameter integrals {\it depending exclusively on parity-even, $x$-space cross-ratios} (that is, ordinary Mandelstam invariants), with coefficients built from more general dual-conformal cross-ratios (which may or may not be parity-even). We will prove this by direct construction through two loops; and we expect this fact may be generalized to higher loops.

As dual-conformal symmetry will play an important role in our present work, let us briefly review it here---mostly to introduce some essential notation. Dual-conformal symmetry is ordinary conformal symmetry, but on the space of dual-momentum coordinates defined by associating the momentum $p_a$ for the $a^{\text{th}}$ external particle to the difference $p_a\!=\!(x_{a+1}\mi x_a)$ (with cyclic labeling understood). This is obviously translationally invariant; and provided all the momenta are taken to be incoming, momentum conservation is automatic in $x$-coordinate space.

Inverse propagators may be expressed in terms of dual coordinates as follows:
\eq{\x{a}{b}\!=\!\x{b}{a}\!\equiv\!(x_b\mi x_a)^2\!=\!(p_a\pl\ldots\pl p_{b-1})^2\!\equiv\!s_{a\cdots b-1}\quad\text{and}\quad \x{\ell}{a}\!\equiv\!(x_{\ell}\mi x_a)^2\,.\label{x_brackets_defined}}
Thus, $\x{a}{b}$ is simply an ordinary Mandelstam invariant---often written as `$x_{ab}^2$' in the literature. (We choose not to use that notation here mostly for simplicity.) Notice that for a massless momentum $p_a^2\!=\!\x{a}{a\,\pl1}\!=\!0$.

\vspace{-6pt}\subsection{Dual-Conformal Regularization: Definitions and Review}\label{subsec:dci_regulator}\vspace{-2pt}

We are interested in regulating those infrared divergences associated with the masslessness of external momenta---those that would be regularized by giving each momentum some small mass $m$ and expanding in the limit $m\!\to\!0$. This is close in spirit to the Higgs regularization scheme described in \mbox{ref.\ \cite{Alday:2009zm}}---which makes
the propagators adjacent to external legs massive---and results in a regularized expression involving a polynomial in $\log(m)$. Although obviously well-motivated, the principal problem with such a regularization scheme is that $m$ carries mass dimension---severely breaking any potential (dual-)conformal invariance of the result.  Dimensional regularization similarly introduces a mass scale, obscuring ultimate conformal invariance of infrared-safe functions. 

The key idea behind the `dual-conformal regulator' introduced in \mbox{ref.\ \cite{Bourjaily:2013mma}} is to give external particles masses, but in a way that introduces no new scales into the problem and leaves intact all scaling-weights under dual-conformal transformations. The later requirement forces us to give different particles different masses. Specifically, a dual-conformal regulator may be defined by taking external particles off the mass shell according to:\footnote{For exactly four light-like particles, the momentum dependent pre-factor in (\ref{momentum_definition_of_dci_regulator}) is more subtle.}
\eq{p_a^2\mapsto p_a^2+\eps\frac{(p_{a-1}\pl p_a)^2(p_{a}\pl p_{a+1})^2}{(p_{a-1}\pl p_a\pl p_{a+1})^2}\,,\label{momentum_definition_of_dci_regulator}}
where `$\eps$'\footnote{When the dual-conformal regulator was introduced in \mbox{ref.\ \cite{Bourjaily:2013mma}}, the parameter we call `$\eps$' was denoted `$\epsilon$'. We have changed this convention to avoid confusion with other regularization schemes.} is a manifestly dimensionless, multiplicative number---taken to be small.

This regularization scheme was initially described in dual-momentum space, where each $x_a$ was shifted in the direction of its cyclic neighbor $x_{a+1}$ by
\eq{x_a\mapsto x_{\hat{a}}\equiv x_a+\eps(x_{a+1}\mi x_a)\frac{\x{a\mi2}{a}}{\x{a\mi2}{a\,\pl1}}\,.\label{x_space_dci_shift}}
This rule matches that given in equation (\ref{momentum_definition_of_dci_regulator}) up to terms of $\mathcal{O}(\eps^2)$ and so the two rules may be considered effectively equivalent. Notice that in terms of \mbox{$\x{a}{b}\!\equiv\!(x_b\mi x_a)^2$} defined in (\ref{x_brackets_defined}), $p_a^2\!\equiv\!\x{a}{a\,\pl1}$, and so (\ref{momentum_definition_of_dci_regulator}) translates to
\eq{\x{{\color{hred}a}}{{\color{hred}a\,\pl1}}\mapsto\x{{\color{hred}\hat{a}}}{{\color{hred}\hat{a\,\pl1}}}=\x{{\color{hred}a}}{{\color{hred}a\,\pl1}}+\eps\frac{\x{{\color{hblue}a\mi1}}{{\color{hred}a\,\pl1}}\x{{\color{hred}a}}{{\color{hblue}a\,\pl2}}}{\x{{\color{hblue}a\mi1}}{{\color{hblue}a\,\pl2}}}\,.\label{two_bracket_regularization}}
This makes it manifest that $\x{\hat{a}}{\hat{a\,\pl1}}$ carries the same conformal weights as $\x{a}{a\,\pl1}$ (would have had, had it not vanished). What this means is that after the shift (\ref{x_space_dci_shift}), $\x{\hat{a}}{\hat{b}}\!\neq\!0$ for all $a,b$ (without imposing additional constraints); and cross-ratios involving `$\x{\hat{a}}{\hat{a\,\pl1}}$' can be expanded in powers of $\eps$ times un-shifted cross-ratios that do not vanish even when all external momenta are massless. Because only consecutive two-brackets $\x{a}{a\,\pl1}$ effectively require regularization, we will drop the hats from such brackets whenever the meaning is sufficiently clear.

(It is worth mentioning that the dual-conformal regulator defined above is only one choice among many potential alternatives. For example, different external particles could be assigned distinct parameters $\delta_a$ in (\ref{momentum_definition_of_dci_regulator})---and these factors could even be modified by arbitrary combinations of conformally-invariant cross-ratios. The choice we make here (following that of \mbox{ref.\ \cite{Bourjaily:2013mma}}) seems the simplest, but it may be worth exploring alternative regularization schemes.)

Given a loop integrand expressed in dual-momentum coordinates, the replacement (\ref{x_space_dci_shift}) will regulate all regions of infrared-divergence. In contrast to an infrared cutoff or dimensional regularization, the dual-conformal regulator leaves intact the original Feynman integration contour---it merely changes the integration measure multiplicatively, by a conformally-invariant factor:
\eq{I=\!\!\!\int\!\prod_{i=1}^Ld^4\ell_i\,\, \mathcal{I}\,\,\mapsto\,\, I^\eps\equiv\!\!\!\int\!\prod_{i=1}^L\left[d^4\ell_i\left(\prod_{a}\frac{\x{\ell_i}{a}}{\x{\ell_i}{\hat{a}}}\right)\right]\mathcal{I}\,.\label{dci_regulator_at_integrand_level}}

One advantage of considering the regulator defined in this way is that it makes it clear that not all propagators in a Feynman integral really need to be regulated. The places in loop-momentum space associated with infrared divergences are the soft-collinear regions for which \mbox{$\ell\!\to\!\epsilon x_a\pl (1\mi\epsilon)x_{a+1}$}, where the multiplicative factor in (\ref{dci_regulator_at_integrand_level}) becomes $\mathcal{O}(\eps)$; everywhere else, it is $1\pl\mathcal{O}(\eps)$. Specifically, this means that for any finite integral, $I_{\text{fin}}^\eps\!=\!I_{\text{fin}}\pl\mathcal{O}(\eps)$. Thus, unless a propagator $\x{\ell}{a}$ is specifically associated with a region of infrared-divergence, the integrand involving $\x{\ell}{\hat{a}}$ and $\x{\ell}{a}$ will differ at most by terms of $\mathcal{O}(\eps)$.

The punchline of the preceding discussion is that we may, without any loss of generality, consider any vanishing brackets of the form (\ref{two_bracket_regularization}) as being replaced by $\eps$ times ratios of non-vanishing brackets. Moreover, only those regions associated with infrared divergences need to be regulated---as the difference will always contribute terms proportional to $\eps$ in the integral (which vanish in the limit $\eps\!\to\!0$).

\vspace{-2pt}\subsubsection{Illustration: Regularization of One-Loop Amplitudes}\label{subsubsec:dci_regulator_at_one_loop}\vspace{-2pt}

The most important property of the dual-conformal regulator is that $\eps$ is the only new parameter; like a mass regulator, infrared-divergent integrals become polynomial in $\logdiv$ in the $\eps\!\to\!0$ limit---but with all coefficients in this expansion being conformally invariant. It may be worthwhile to describe a few examples at one loop to illustrate the kinds of regulated expressions that arise. 

Consider the case of the scalar box integrals at one loop. These integrals combined with parity-odd pentagons form a complete basis of integrands relevant for SYM. Every {\it integrand} in this basis (once canonically normalized) is manifestly dual-conformally invariant---but many are infrared-divergent upon integration.

A scalar box is infrared-finite if and only if none of the momenta flowing into its corners are massless. Such an integral is often called a `four-mass' box. Normalized in the standard way\footnote{Careful readers should note that the expression we give here is {\it minus} the standard one. We are also rescaling the standard measure $d^4\!\ell\!\to\!d^4\!\ell/\pi^2$ to simplify our expressions.} to have unit-magnitude maximal-co-dimension residues, the four-mass box integral corresponds to:
\eq{I^{4\text{m}}(u,v)\equiv I_{a,b,c,d}\equiv\!\int\!\!d^4\ell\,\frac{\x{a}{c}\x{b}{d}\Delta[u,v]}{\x{\ell}{a}\x{\ell}{b}\x{\ell}{c}\x{\ell}{d}}\,,\label{four_mass_box_loop_integral}\vspace{-2.5pt}}
where the dual-conformally invariant cross-ratios $u,v$ and the normalization factor $\Delta[u,v]$ are defined by:
\eq{u\equiv\frac{\x{a}{b}\x{c}{d}}{\x{a}{c}\x{b}{d}},\quad v\equiv\frac{\x{a}{d}\x{b}{c}}{\x{a}{c}\x{b}{d}},\quad\Delta[u,v]\equiv\sqrt{(1\mi u\mi v)^2\mi 4 u\,v}\,.\label{uv_delta_defined}}
We will show in the next subsection how the integral (\ref{four_mass_box_loop_integral}) can be performed using Feynman parameters {\it directly} in terms of the two dual-conformal cross-ratios $u,v$. But for now, let us merely quote the result (see e.g.~\cite{Bourjaily:2013mma}):
\eq{I^{4m}(u,v)=\Li{2}{\tilde{u}}+\Li{2}{\tilde{v}}+\frac{1}{2}\log(u)\log(v)-\log(\tilde{u})\log(\tilde{v})-\Zeta{2}\,,\label{four_mass_box_integrated}\vspace{-2.5pt}}
with
\eq{\tilde{u}\equiv\frac{1}{2}\big(1- u+ v+\Delta[u,v]\big)\quad\text{and}\quad\tilde{v}\equiv\frac{1}{2}\big(1+ u - v+ \Delta[u,v]\big)\,.}

Because the dual-conformal regulator amounts to adding masses to all corners (and taking the limit as $\eps\!\to\!0$), the regulated expressions for all infrared-divergent scalar boxes can be obtained from \eqref{four_mass_box_integrated} by simply Taylor-expanding in small $\eps$. For example, consider the case where $b\!=\!a\,\pl1$, so that $\x{a}{b}\!\to\!0$. As this involves exactly one `massless' corner, this case is called a `three-mass' box. When $b\!=\!a\pl1$, the cross-ratio $u$ defined in \eqref{uv_delta_defined} requires regularization. (The cross-ratio $v$ remains unchanged to $\mathcal{O}(\eps)$.) The regularization of $u$ is achieved exactly according to the rule \eqref{two_bracket_regularization}, resulting in:
\eq{u=\frac{{\color{hred}\x{a}{a\,\pl1}}\x{c}{d}}{\x{a}{c}\x{a\,\pl1}{d}}\mapsto\eps\frac{{\color{hred}\x{a\mi1}{a\,\pl1}\x{a}{a\,\pl2}}\x{c}{d}}{{\color{hred}\x{a\mi1}{a\,\pl2}}\x{a}{c}\x{a\,\pl1}{d}}\equiv \delta\,u'\,.}
Series-expanding (\ref{four_mass_box_integrated}) in small $\eps$, we obtain the regulated expression
\eq{\hspace{-10pt}I^{3\text{m}}(u',v)\equiv I^{\eps}_{a,a+1,c,d}=\Li{2}{1\mi v}+\frac{1}{2}\log(u')\log(v)+{\color{divblue}\frac{1}{2}\log(v)\logdiv}\label{three_mass_box_first_time}}
for the three-mass box, which coincides with $I^{4\text{m}}(\eps\,u',v)$ up to terms that vanish in the limit $\eps\rightarrow 0$. Regulated expressions for all other one-loop scalar boxes can be obtained by regulating $u,v$ as above and series-expanding (\ref{four_mass_box_integrated}) in the same way. (See \mbox{ref.\ \cite{Bourjaily:2013mma}} for regulated expressions for all one-loop boxes.)

Although the canonically normalized one-loop scalar box integrands are manifestly dual-conformally invariant, they are not directly expressed in terms of dual-conformal cross-ratios. In the next section, we describe how, after Feynman parameterization (and some Feynman-parameter integrations), this can always be made manifest. This is important, because na\"{i}ve Feynman-parameterization would {\it dramatically} spoil dual-conformal-invariance, as Feynman-parameterized denominators would involve {\it sums} of terms with {\it different} conformal weights. The fact that this problem seems universally avoidable seems important---beyond merely simplifying our work below. We will prove this by direct construction for all integrands required through two loops, and we expect it to hold more generally.

\newpage
\vspace{-6pt}\subsection{Feynman Parameterization and Conformal Rescaling}\label{subsec:feynman_parameterization}\vspace{-6pt}

As mentioned above, we find it possible to express any dual-conformal integrand at one or two loops directly as Feynman parameter integrals which depend exclusively on dual-conformal cross-ratios (and the regularization parameter $\eps$, if necessary). Moreover, all such integrals can be expanded in terms of those which depend exclusively on parity-even, $x$-space cross-ratios (those built rationally from Mandelstam invariants), with coefficients involving more general cross-ratios. Roughly speaking, this works because the Feynman parameters can always be rescaled by factors of $\x{a}{b}$ in a way that eliminates any conformal scaling from the denominator. To understand why and how this works, it will be good to consider some examples. 

In the next subsection, we review how Feynman parameter integration works in dual-momentum coordinates within the embedding formalism (see e.g.\ \mbox{ref.\ \cite{Weinberg:2010fx}}). Given a few basic operations, we will prove the claim above by explicit construction through two loops---with illustrative examples and applications along the way.

\vspace{-6pt}\subsubsection{Feynman Parameter Integrals in Dual Coordinates}\label{subsubsec:feynman_parameter_integral_basics}\vspace{-6pt}

In this section, we will mostly review well-known (if still not widely familiar) facts about how Feynman parameter integration works in the embedding formalism using dual-momentum coordinates. For a more thorough discussion, we refer the reader to \mbox{refs.\ \cite{Hodges:2010kq,Mason:2010pg}}; the most important role of this subsection is to establish some essential notation and a few key facts that will be important for us later on. 

Before getting started, however, we should perhaps clarify our terminology. What we will call `Feynman parameters' are sometimes more specifically referred to as `Schwinger parameters' or even `$\alpha$ parameters' \cite{Smirnov:2004ym,Smirnov:2006ry}---distinguished by how these integrals are de-projectivized. For the most part, de-projectivization is a trivial distinction; but to be clear, when we have reason to de-projectivize any integral explicitly, we will always use $\delta(\alpha_i\mi1)$ instead of Feynman's choice of $\delta((\sum_i\alpha_i)\mi1)$. With this understood, we will denote the projective integration measure over $(n\,\pl1)$ Feynman parameters $\{\alpha_0,\ldots,\alpha_n\}$ by `$\proj{d^n\vec{\alpha}}$':
\vspace{-4pt}\eq{\proj{d^n\vec{\alpha}}\equiv\frac{d\alpha_0\cdots d\alpha_n}{\mathrm{vol}(GL(1))}\equiv d\alpha_0\cdots d\alpha_n\,\delta(\alpha_i\mi1)\quad\text{(for any $i$)}\,.\label{projective_measure_defined}\vspace{-4pt}}

We will always consider inverse propagators to be defined within the embedding formalism in which they may (without any loss of generality) be represented linearly in a vector space of inverse propagators. No actual basis (nor its rank) will play any role for us here---merely the fact that inverse propagators may be added linearly. (Inverse propagators represented in momentum-twistor space automatically have such linearity; we have chosen to avoid that formalism here for the sake of familiarity to most readers---and because nothing about twistor space will be important for the propagator structure being considered below.)

For us, the embedding formalism is merely used to simplify how inverse propagators get combined through Feynman parameterization. So long as inverse propagators may be added linearly, then in the standard Feynman parameterization,
\eq{\frac{1}{\x{\ell}{a_1}\cdots\x{\ell}{a_n}}=\Gamma(n)\int\limits_0^\infty\!\!\proj{d^{n-1}\vec{\alpha}}\frac{1}{\big(\alpha_0\x{\ell}{a_1}+\ldots+\alpha_{n-1}\x{\ell}{a_n}\big)^n}\,,}
the sum of inverse propagators in the denominator on the right hand side may be expressed more simply as,
\eq{\hspace{-12pt}\alpha_0\x{\ell}{a_1}\pl\ldots\pl\alpha_{n-1}\x{\ell}{a_n}\!\equiv\!\x{\ell}{Y(\vec{\alpha})},\!\hspace{5pt}\text{where}\hspace{5pt}
    Y(\vec{\alpha})\!\equiv\!\alpha_0(a_1)\pl\ldots\pl\alpha_{n-1}(a_n).\hspace{-10pt}}
That is, we may simply add dual-coordinates linearly when Feynman-parameterizing propagators in loop integrands.

As always, Feynman parameterization trivializes the integral over $d^4\ell$---trading an integration over loop momenta for another one over Feynman parameters. It turns out that everything we need to know about the map from loop-integrals to Feynman-parameter integrals follows from the simple fact \cite{Hodges:2010kq}:
\eq{\Gamma(4)\!\!\int\!\!d^4\ell\,\frac{1}{\x{\ell}{Y}^4}=2\,\frac{1}{\x{Y}{Y}^2}\,,\label{fundamental_loop_integral_in_feynman_parameters}}
which, for our purposes, we may consider an identity. (More interested readers should consult \mbox{refs.\ \cite{Hodges:2010kq,Mason:2010pg}} for more thorough discussions and derivations in \mbox{momentum-}twistor space.) Iterated differentiation of each side of (\ref{fundamental_loop_integral_in_feynman_parameters}) with respect to $Y$ and contracting with arbitrary reference dual-points `$N$'---that is, acting with $\x{N}{\frac{d}{dY}}$---allows us to derive analogs of (\ref{fundamental_loop_integral_in_feynman_parameters}) for integrals with more internal propagators:
\eq{\begin{split}\Gamma(5)\!\!\int\!\!d^4\ell\,\frac{\x{\ell}{N}}{\x{\ell}{Y}^5}&=4\frac{\x{Y}{N}}{\x{Y}{Y}^3}\,;\\
\Gamma(6)\!\!\int\!\!d^4\ell\,\frac{\x{\ell}{N_1}\x{\ell}{N_2}}{\x{\ell}{Y}^6}&=4\left(3\frac{\x{Y}{N_1}\x{Y}{N_2}}{\x{Y}{Y}^4}-\frac{\x{N_1}{N_2}}{\x{Y}{Y}^3}\right);\end{split}\label{pentagon_and_hexagon_loop_integrals}}
etc. We may refer to the right-hand sides of (\ref{fundamental_loop_integral_in_feynman_parameters}) and (\ref{pentagon_and_hexagon_loop_integrals}) as `box', `pentagon', and `hexagon' Feynman parameter integrands, respectively---reflecting the number of loop-dependent propagators in the initial integrands (and not the number of particles). It turns out that only these three cases will be required in the representation of amplitudes in planar SYM through two loops (for arbitrary multiplicity). 

In the following subsections, we illustrate how Feynman parameterization in the embedding formalism leads to manifestly dual-conformal expressions for loop integrals---represented as Feynman-parameter integrals. This is achieved by strategically rescaling the Feynman parameters so that, once rescaled, every factor in the denominator depends {\it only} on dual-conformal cross-ratios. At one loop, this is nearly too trivial to warrant a detailed discussion; but we include it here as an illustrative example of the much less trivial examples we will consider at two loops.

\vspace{-2pt}\subsubsection{Manifestly Conformal Feynman Integrals at One Loop}\label{subsubsec:feynman_parameter_ints_at_one_loop}\vspace{-2pt}

Let us now briefly review how one-loop integrals can be integrated via Feynman parameterization, and how the Feynman parameters involved may be rescaled in such a way that the resulting representations are manifestly dual-conformally invariant. This will be admittedly somewhat trivial, but it will allow us to illustrate some of the elegance of the Feynman parameter representations (in the embedding formalism) and highlight what we desire at two loops (and beyond). Moreover, as we have discussed in \mbox{section \ref{subsec:dci_regulator}}, understanding how this works in the four-mass case will directly lead to dual-conformally-regulated expressions of all other scalar boxes required to represent amplitudes in maximally supersymmetric theories at one loop. 

Recall that the four-mass box integral is defined in loop-momentum space by
\eq{I^{4\text{m}}(u,v)\equiv\!\int\!\!d^4\ell\,\frac{\x{a}{c}\x{b}{d}\Delta[u,v]}{\x{\ell}{a}\x{\ell}{b}\x{\ell}{c}\x{\ell}{d}}\,,\label{four_mass_box_loop_integral_second_time}}
where $\Delta[u,v]$ and the conformal cross-ratios $u,v$ were defined in (\ref{uv_delta_defined}). While manifestly conformally invariant, the integral representation (\ref{four_mass_box_loop_integral_second_time}) is unsatisfactory for at least two important reasons: first, it is a four-fold rational integral representation of a function (\ref{four_mass_box_integrated}) with transcendental-weight two; and secondly, while it is manifestly dual-conformally invariant, it is not {\it directly} expressed in terms of dual-conformally invariant cross-ratios. The first complaint may be addressed through Feynman parameterization (and integration), but at the cost---however temporarily---of (dramatically) enhancing the second problem. Let us see how each of these issues may be resolved in turn by explicit analysis. 

Using Feynman parameterization to represent the propagators in (\ref{four_mass_box_loop_integral_second_time}) and performing the $\ell$-integration (using (\ref{fundamental_loop_integral_in_feynman_parameters})), we have:
\eq{I^{4\text{m}}(u,v)=\Gamma(4)\int\limits_0^{\infty}\!\!\proj{d^3\vec{\alpha}}d^4\ell\,\frac{\x{a}{c}\x{b}{d}\Delta[u,v]}{\x{\ell}{Y}^4}=\int\limits_0^{\infty}\!\!\proj{d^3\vec{\alpha}}\frac{2\x{a}{c}\x{b}{d}\Delta[u,v]}{\x{Y}{Y}^2}\,,\label{four_mass_feynman_param_details}}
where we have defined
\eq{\begin{split}Y&\equiv{\color{hred}\alpha_0}(a)+{\color{hblue}\alpha_1(b)+\alpha_2(c)+\alpha_3(d)}\,\\&\equiv{\color{hred}\alpha_0}(a)+{\color{hblue}(Q)}\,\end{split}\label{y_for_four_mass}}
so that the linearity of $\x{Y}{Y}$ in ${\color{hred}\alpha_0}$ is made manifest: 
\eq{\x{Y}{Y}=2\,{\color{hred}\alpha_0}\x{a}{{\color{hblue}Q}}+\x{{\color{hblue}Q}}{{\color{hblue}Q}}\,.}
Because of this linearity in ${\color{hred}\alpha_0}$ of $\x{Y}{Y}$, integration over ${\color{hred}\alpha_0}$ is trivially recognized as a total derivative, resulting in
\eq{\begin{split}I^{4\text{m}}(u,v)&=\int\limits_0^\infty\!\!\proj{d^2\vec{\alpha}}\!\!\int\limits_0^\infty\!\!d{\color{hred}\alpha_0}\,\frac{2\x{a}{c}\x{b}{d}\Delta[u,v]}{(2\,{\color{hred}\alpha_0}\x{a}{{\color{hblue}Q}}+\x{{\color{hblue}Q}}{{\color{hblue}Q}})^2}\,\\
&=\int\limits_0^\infty\!\!\proj{d^2\vec{\alpha}}\frac{\x{a}{c}\x{b}{d}\Delta[u,v]}{\x{a}{{\color{hblue}Q}}\x{{\color{hblue}Q}}{{\color{hblue}Q}}}\,.\end{split}\label{four_mass_box_feynman_part1}}
Notice that by introducing Feynman parameters and recognizing one integration as a total derivative, we have addressed our first complaint above: equation (\ref{four_mass_box_feynman_part1}) is now a two-fold integral representation of a function known to have transcendental-weight two (see equation (\ref{four_mass_box_integrated})). {\it However}, in the process of introducing Feynman parameters, we have dramatically obscured the dual-conformal invariance of the result. To see why, notice that the integrand (\ref{four_mass_box_feynman_part1}) involves the factors 
\eq{\begin{split}
\x{a}{{\color{hblue}Q}}&=\alpha_1\x{a}{b}+\alpha_2\x{a}{c}+\alpha_3\x{a}{d}\,,\\
\x{{\color{hblue}Q}}{{\color{hblue}Q}}&=2\,\big(\alpha_1\alpha_2\x{b}{c}+\alpha_1\alpha_3\x{b}{d}+\alpha_2\alpha_3\x{c}{d}\big)\,,\end{split}\label{denominator_factors_of_box_before_rescaling}}
which are {\it sums} of terms with {\it different} conformal-weights under rescalings! Thus, however compact, (\ref{four_mass_box_feynman_part1}) is still quite far from being {\it manifestly} conformally invariant. Nevertheless, it turns out to be easy to rescale the Feynman parameters $\alpha_i$ in a way that completely cures this non-manifest invariance. This would be achieved provided that, after rescaling each Feynman parameter by some kinematic-dependent factor, \mbox{$\alpha_i\!\mapsto\!s^i(\vec{x})\alpha_i$}, each factor in the denominator, (\ref{denominator_factors_of_box_before_rescaling}), transforms into a function with uniform conformal weights. One choice of rescalings (among many) that achieves this is the following:
\eq{\alpha_1\mapsto\alpha_1\x{c}{d},\quad\alpha_2\mapsto\alpha_2\x{b}{d},\quad\alpha_3\mapsto\alpha_3\x{b}{c}\,.\label{four_mass_box_rescaling}}
Under this rescaling (and allowing for the slight abuse of notation),
\eq{\begin{split}
\x{a}{{\color{hblue}Q}}&\mapsto\x{a}{c}\x{b}{d}{\color{hred}(\alpha_1u+\alpha_2+\alpha_3 v)}\,,\\
\x{{\color{hblue}Q}}{{\color{hblue}Q}}&\mapsto2\,\x{b}{c}\x{b}{d}\x{c}{d}{\color{hred}(\alpha_1\alpha_2+\alpha_1\alpha_3+\alpha_2\alpha_3)}\,,\end{split}\label{denominator_factors_of_box_with_rescaling}}
with $u,v$ defined as in (\ref{uv_delta_defined}). Including the appropriate Jacobian associated with the rescaling (\ref{four_mass_box_rescaling}) (which in this case is a factor of $\x{b}{c}\x{b}{d}\x{c}{d}$ in the numerator), and factoring out the two-bracket-dependent pre-factors in (\ref{denominator_factors_of_box_with_rescaling}), we find:
\eq{I^{4\text{m}}(u,v)=\frac{1}{2}\int\limits_0^{\infty}\!\!\proj{d^2\vec{\alpha}}\frac{\Delta[u,v]}{(\alpha_1\,u+\alpha_2+\alpha_3\, v)(\alpha_1\alpha_2+\alpha_1\alpha_3+\alpha_2\alpha_3)}\,.\label{four_mass_feynman_parameter}}

As (\ref{four_mass_feynman_parameter}) is a two-fold integral representation of the weight-two function (\ref{four_mass_box_integrated}) {\it directly} expressed in terms of dual-conformally-invariant cross-ratios, we have addressed both of the undesirable features of the initial integral (\ref{four_mass_box_loop_integral_second_time}) discussed above. 

This example illustrates how Feynman parameterization has the potential to (deeply) obscure dual-conformal invariance, but that this problem can sometimes (always?) be addressed by rescaling Feynman parameters appropriately. This could have been addressed earlier in our analysis by ensuring that $Y(\vec{\alpha})$ in (\ref{y_for_four_mass}) had been assigned uniform conformal weights---which would indeed have been possible. 

The enhanced flexibility in finding suitable rescalings obtained after $\alpha_0$ had been integrated-out was somewhat superficial in this example. However, it becomes much more important at higher loops. To see why, consider if point $(a)$ had been (the dual point associated with) another loop momentum, $(\ell_2)$, say. In such a case, $\x{a}{Q}$ would take the form of another propagator, while $\x{Q}{Q}$ would become an $\ell_2$-independent factor in the integrand; it would be a bad idea (from the perspective of subsequent $\ell_2$ integral) to render these equal in weight through various $\ell_2$-dependent rescalings of Feynman parameters.

\vspace{-2pt}\subsubsection{Manifestly Conformal Feynman Integrals at Two Loops}\label{subsubsec:feynman_parameter_ints_at_two_loop}\vspace{-2pt}

In this section, we show that introducing Feynman parameters (loop-by-loop) and performing suitable rescalings can make manifest the dual-conformal invariance of all (possibly regulated) integrals required in the representation of amplitudes or ratio functions at two loops. We will start with the most general case required at two loops: the general double-pentagon integral,
\eq{\fwbox{125pt}{\fig{-24.62pt}{1}{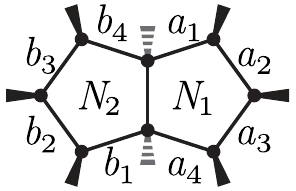}}\label{double_pentagon_figure}}

As described in \mbox{ref.\ \cite{Bourjaily:2015jna}}, all four-dimensional two-loop integrands with SYM power-counting (or better) can be represented in terms of double-pentagons---possibly as contact-terms. To be clear, this includes all (possibly elliptic) double-box and penta-box integrals. Although this truly general case will prove the applicability of our approach, such generality will prevent many simplifications that arise for actual integrals in which one may be interested---for example, the `chiral' double-pentagons of ref.~ \cite{ArkaniHamed:2010gh} (which are actually used in the representation of two-loop integrands in \mbox{ref.\ \cite{Bourjaily:2015jna}}) admit much further simplification. Thus, to highlight the additional simplifications that may arise in concrete cases, we will end this subsection with a thorough analysis of some specific integrals relevant to amplitudes involving six particles. 

Let us consider the general double-pentagon integral (\ref{double_pentagon_figure}) of the form:
\eq{\begin{split}\hspace{-105pt}I\!&=\!\int\!\!d^4\ell_1d^4\ell_2\,\,\frac{\x{\ell_1}{N_1}\x{\ell_2}{N_2}}{\x{\ell_1}{a_1}\x{\ell_1}{a_2}\x{\ell_1}{a_3}\x{\ell_1}{a_4}\x{\ell_1}{\ell_2}\x{\ell_2}{b_1}\x{\ell_2}{b_2}\x{\ell_2}{b_3}\x{\ell_2}{b_4}}\,\hspace{-25pt}\\\hspace{-20pt}&\equiv\!\int\!\!d^4\ell_1d^4\ell_2\,\,\mathcal{I}\,.\end{split}\label{general_double_pentagon_integrand}}
The numerators in (\ref{general_double_pentagon_integrand}) are required for the integrand to behave asymptotically like a scalar box (for each loop) in the ultraviolet, but our analysis will not require any particular form for the numerators: the integrand need not have `unit' leading singularities, nor must it even be dual-conformally invariant. Indeed, our analysis will make little use of the numerators at all---focusing instead on the denominator-structure. In general, we will find a representation of (\ref{general_double_pentagon_integrand}) as an expansion in terms of manifestly dual-conformally invariant, parity-even Feynman parameter integrands; the coefficients of (\ref{general_double_pentagon_integrand}) in this expansion will be dual-conformal if (\ref{general_double_pentagon_integrand}) were, but they need not be. 

In a moment, it will be important for us that we may assume that none of $\x{a_i}{a_j},\x{a_i}{b_j},\x{b_i}{b_j}$ vanish. This requires no loss of generality because we may always consider the integral as dual-conformally regularized if necessary. (However, if the integral had had massless corners that did not require regularization, several key simplifications would be possible that we are not currently permitting ourselves.) 

Let us now begin to do the space-time integrals in (\ref{general_double_pentagon_integrand}) one loop at a time by introducing Feynman parameters. Let us begin with the $\ell_1$ integration, introducing
\eq{\begin{split}Y_1&\equiv{\color{hblue}\alpha_1(a_1)+\alpha_2(a_2)+\alpha_3(a_3)+\alpha_4(a_4)}+{\color{hred}\gamma_1}(\ell_2)\,\\&\equiv{\color{hblue}(Q_1)}+{\color{hred}\gamma_1}(\ell_2)\,\end{split}}
so that
\eq{\x{Y_1}{Y_1}=\x{{\color{hblue}Q_1}}{{\color{hblue}Q_1}}+2\,{\color{hred}\gamma_1}\x{\ell_2}{{\color{hblue}Q_1}}\,.}
Now, using (\ref{pentagon_and_hexagon_loop_integrals}) to do the space-time integral over $\ell_1$, and then recognizing the ${\color{hred}\gamma_1}$ integral as a total derivative, we see that (\ref{general_double_pentagon_integrand}) becomes:
\eq{\begin{split}&\!\!\int\!\!d^4\ell_1\,\mathcal{I}=\!\!\int\limits_{0}^{\infty}\!\!\proj{d^3\!\vec{\alpha}}\!\!\int\limits_0^{\infty}\!\!d{\color{hred}\gamma_1}\frac{4\big(\x{{\color{hblue}Q_1}}{N_1}+2\,{\color{hred}\gamma_1}\x{\ell_2}{N_1}\big))\x{\ell_2}{N_2}}{\big(\x{{\color{hblue}Q_1}}{{\color{hblue}Q_1}}+2\,{\color{hred}\gamma_1}\x{\ell_2}{{\color{hblue}Q_1}}\big)^3\x{\ell_2}{b_1}\x{\ell_2}{b_2}\x{\ell_2}{b_3}\x{\ell_2}{b_4}}\,\\
&\!\!\!\!=\!\!\int\limits_{0}^{\infty}\!\!\proj{d^3\!\vec{\alpha}}\!\left(\!\frac{\x{{\color{hblue}Q_1}}{N_1}}{\x{{\color{hblue}Q_1}}{{\color{hblue}Q_1}}^2\x{\ell_2}{{\color{hblue}Q_1}}}\pl\frac{\x{\ell_2}{N_1}}{2\x{{\color{hblue}Q_1}}{{\color{hblue}Q_1}}\x{\ell_2}{{\color{hblue}Q_1}}^2}\!\right)\!\!\frac{\x{\ell_2}{N_2}}{\x{\ell_2}{b_1}\x{\ell_2}{b_2}\x{\ell_2}{b_3}\x{\ell_2}{b_4}}\,.
\end{split}\label{double_pentagon_post_ell1_integration}}

Both terms in (\ref{double_pentagon_post_ell1_integration}) have the same set of $\ell_2$ propagators---merely with different powers of $\x{\ell_2}{{\color{hblue}Q_1}}$. Thus, introducing Feynman parameters for the $\ell_2$ integral will be structurally similar in both cases; provided we de-projectivize these parameters by setting the coefficient of ${\color{hblue}(Q_1)}$ to one, these two cases will be even more uniform in their treatment. Thus, let us define the Feynman parameter factor for the $\ell_2$ integral in (\ref{double_pentagon_post_ell1_integration}) according to
\eq{\begin{split}Y_2&\equiv{\color{hblue}(Q_1)+\beta_1(b_1)+\beta_2(b_2)+\beta_3(b_3)}+{\color{hred}\gamma_2}(b_4)\,\\
&\equiv{\color{hblue}(Q_2)}+{\color{hred}\gamma_2}(b_4)\,\end{split}}
so that
\eq{\x{Y_2}{Y_2}=\x{{\color{hblue}Q_2}}{{\color{hblue}Q_2}}+2\,{\color{hred}\gamma_2}\x{{\color{hblue}Q_2}}{b_4}\,.}
Gathering terms and performing the ${\color{hred}\gamma_2}$ integral (a total derivative) results in:
\eq{\begin{split}\hspace{-20pt}I&=\!\!\int\limits_{0}^{\infty}\!\!\proj{d^3\vec{\alpha}}\!d^3\!\vec{\beta}\!\!\int\limits_0^{\infty}\!\!d{\color{hred}\gamma_2}\frac{2}{\x{{\color{hblue}Q_1}}{{\color{hblue}Q_1}}}\left(2\frac{\x{{\color{hblue}Q_1}}{N_1}\x{Y_2}{N_2}}{\x{{\color{hblue}Q_1}}{{\color{hblue}Q_1}}\x{Y_2}{Y_2}^3}\pl3\frac{\x{Y_2}{N_1}\x{Y_2}{N_2}}{\x{Y_2}{Y_2}^4}\mi\frac{\x{N_1}{N_2}}{\x{Y_2}{Y_2}^3}\right)\,,\hspace{-24pt}\\
&\equiv\!\!\int\limits_{0}^{\infty}\!\!\proj{d^3\vec{\alpha}}\!d^3\!\vec{\beta}\frac{1}{f_1\,f_2\,f_3}\!\left(\frac{n_{12}}{f_1\,f_2}+\frac{n_{13}}{f_1\,f_3}+\frac{n_{22}}{f_2^2}+\frac{n_{23}}{f_2\,f_3}+\frac{n_{33}}{f_3^2}-\frac{n_2}{f_2}\right)\,,
\end{split}\label{schematic_double_pentagon_penult}}
where
\eq{f_1\equiv\frac{1}{2}\x{{\color{hblue}Q_1}}{{\color{hblue}Q_1}},\quad f_2\equiv\frac{1}{2}\x{{\color{hblue}Q_2}}{{\color{hblue}Q_2}},\quad f_3\equiv\x{{\color{hblue}Q_2}}{b_4}\,,\label{general_double_pentagon_denominators}}
and where we have identified the numerators in (\ref{schematic_double_pentagon_penult}) as corresponding to:
\eq{\begin{array}{@{}r@{$\,\,$}l@{$\;\;\;\hspace{20pt}$}r@{$\,\,$}l@{}}n_{12}&\equiv\x{{\color{hblue}Q_1}}{N_1}\x{{\color{hblue}Q_2}}{N_2}\,,&n_{23}&\equiv\frac{1}{2}\Big(\x{{\color{hblue}Q_2}}{N_1}\x{b_4}{N_2}+\x{b_4}{N_1}\x{{\color{hblue}Q_2}}{N_2}\Big)\,,\\
n_{13}&\equiv\x{{\color{hblue}Q_1}}{N_1}\x{b_4}{N_2}\,,&n_{33}&\equiv\x{b_4}{N_1}\x{b_4}{N_2}\,,\\
n_{22}&\equiv\x{{\color{hblue}Q_2}}{N_1}\x{{\color{hblue}Q_2}}{N_2}\,,&n_{2\phantom{2}}&\equiv\x{N_1}{N_2}\,.\\
\end{array}}

Equation (\ref{schematic_double_pentagon_penult}) is a six-fold Feynman-parametric integral representation of the general double-pentagon (\ref{general_double_pentagon_integrand}), where we have made absolutely no assumptions about the form of the tensor numerators involving $N_1,N_2$. The representation (\ref{schematic_double_pentagon_penult}) is still not {\it manifestly} dual-conformally invariant because the pieces, $f_i$, $n_i$, $n_{ij}$ are not. Let us now show that the Feynman parameters may be rescaled in a way that renders the denominators manifestly dual-conformal. So long as no two-brackets vanish, we may rescale the Feynman parameters as follows:
\eq{\alpha_1\mapsto\alpha_i/\x{a_i}{b_4},\qquad \beta_i\mapsto\beta_i/\x{b_i}{b_4}\,.\label{general_double_pentagon_parameter_rescaling}}
Under this rescaling
\eq{f_3\mapsto(\alpha_1\pl\alpha_2\pl\alpha_3\pl\alpha_4\pl\beta_1\pl\beta_2\pl\beta_3),}
and $f_1,f_2$ become uniform in dual-conformal scaling weight---for example, we could chose to factor them according to
\eq{f_{1,2}\mapsto\frac{\x{a_1}{a_2}}{\x{a_1}{b_4}\x{a_2}{b_4}}\big(\alpha_1\,\alpha_2+\ldots\big)\equiv \frac{\x{a_1}{a_2}}{\x{a_1}{b_4}\x{a_2}{b_4}}f_{1,2}'\,,}
where every term in the parentheses has a coefficient directly expressible as products of parity-even, dual-conformally invariant cross-ratios. After absorbing the non-conformal pre-factors of the rescaled denominators---together with the global Jacobian factor from the rescaling (\ref{general_double_pentagon_parameter_rescaling})---into new (also rescaled) numerators, the result will be manifestly dual-conformal provided the original integral were. Because the numerators are simply polynomials in the Feynman parameters, we may consider the final integrals as sums of terms which depend exclusively on parity-even dual-conformal cross-ratios with some kinematic-coefficients---being what it was desired to prove.

There are many places for simplifications to arise in the above analysis that we were unable to exploit in the interest of generality. To better illustrate these potential simplifications, and to gain more intuition about how the analysis described above works in detail, it will be helpful to consider a few more concrete cases.

\vspace{-2pt}\subsubsection{{\it Exempli Gratia}{\rm:} The Double-Pentagon Integrals $\Omega^{(2)}$ and $\tilde{\Omega}^{(2)}$}\label{subsubsec:exempli_gratia}\vspace{-2pt}

A double-pentagon integral (that is not a contact-term) must have at least six external legs; there are exactly two (cyclic classes of) infrared-finite double-pentagon `hexagon functions'---those associated with Feynman diagrams involving six particles. As hexagon functions have played an important role in recent years (see \mbox{refs.\ \cite{Dixon:2011pw,Dixon:2013eka,Dixon:2014iba,Dixon:2014voa,Caron-Huot:2016owq,Dixon:2016nkn,sixLoops}}), they will comprise our primary concrete examples in the following sections. Let us therefore start by considering the simplest of these at two loops. 

The two distinct, two-loop hexagon functions associated with Feynman diagrams are conventionally called $\Omega^{(2)}(\u,\v,\w)$ and $\tilde{\Omega}^{(2)}(\u,\v,\w)$ (see e.g.\ \cite{Caron-Huot:2018dsv} for a more recent discussion), where $\{\u,\v,\w\}$ are cyclically-related dual-conformal cross-ratios defined for six particles: 
\vspace{-2pt}\eq{\u\equiv\frac{\x{1}{3}\x{4}{6}}{\x{1}{4}\x{3}{6}},\quad \v\equiv\frac{\x{2}{4}\x{1}{5}}{\x{2}{5}\x{1}{4}},\quad \w\equiv\frac{\x{3}{5}\x{2}{6}}{\x{3}{6}\x{2}{5}}\,.\label{definition_of_uvw_body}\vspace{-2pt}}
As cyclic rotations merely permute their cross-ratio arguments,\footnote{Actually, $\tilde{\Omega}^{(2)}$ includes a part that changes sign under a three-fold cyclic rotation, so it should more specifically be described as being a function of $\{\u,\v,\w\}$ and a disambiguating sign $\pm1$.} we will often more simply write $\Omega^{(2)}$ and $\tilde{\Omega}^{(2)}$ when the order of their arguments is the canonical one. 

As integrands, these two cases are nearly identical---with exactly the same propagators, and differing only in their tensor numerators. In loop-momentum space, $\Omega^{(2)}$ corresponds to the integral:
\vspace{-6pt}\eq{\fwboxR{0pt}{\Omega^{(2)}(\u,\v,\w)\Leftrightarrow}\fwbox{105pt}{\fig{-29.8pt}{1}{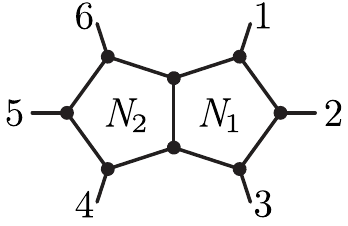}}\,.\label{omega2_figure}\vspace{-6pt}}
$\tilde{\Omega}^{(2)}$ would be the same, but with $N_2$ replaced by its (parity) conjugate $\overline{N_2}$. Concretely expressed in dual-momentum coordinates, they correspond to
\eq{\begin{split}\Omega^{(2)}&\equiv\!\!\int\!\!d^4\ell_1d^4\ell_2\frac{\x{\ell_1}{N_1}\x{\ell_2}{N_2}}{\x{\ell_1}{1}\x{\ell_1}{2}\x{\ell_1}{3}\x{\ell_1}{4}\x{\ell_1}{\ell_2}\x{\ell_2}{4}\x{\ell_2}{5}\x{\ell_2}{6}\x{\ell_2}{1}}\,,\\
\tilde{\Omega}^{(2)}&\equiv\!\!\int\!\!d^4\ell_1d^4\ell_2\frac{\x{\ell_1}{N_1}\x{\ell_2}{\overline{N_2}}}{\x{\ell_1}{1}\x{\ell_1}{2}\x{\ell_1}{3}\x{\ell_1}{4}\x{\ell_1}{\ell_2}\x{\ell_2}{4}\x{\ell_2}{5}\x{\ell_2}{6}\x{\ell_2}{1}}\,,
\end{split}\label{omega2_and_omega2tilde_integrands}}
where the tensor-numerators are most naturally defined using momentum-twistor variables (that we briefly review in \mbox{Appendix \ref{appendix:conventions_and_definitions}}); but for the sake of concreteness, they can be expressed in terms of (the dual-momentum-coordinate points associated with) the bi-twistors:
\vspace{-3pt}\eq{\hspace{-20pt}N_1\equiv(612)\tncap(234),\;\; N_2\equiv(345)\tncap(561)\ab{13\,46},\;\;\overline{N_2}\equiv(46)\ab{13\,(345)\tncap(561)}\,.\hspace{-20pt}\label{omega_numerators_defined}\vspace{-3pt}}
Less explicitly, but directly described in dual coordinates, these numerators $N_i,\overline{N_i}$ may be defined as the (complex-conjugate) pair of points in $x$-space that are light-like separated from the four propagators involving $\ell_i$. These are not rationally related to the six dual-coordinates $x_a$ (despite being rationally related to the momentum-twistors associated with these points---as is clear from (\ref{omega_numerators_defined})).

As the actual form of the numerators will play only a very small role in the Feynman-parameterization and rescaling, our analysis for both integrals will be quite similar. In both cases, the integral is infrared-finite, and thus no regularization is required. This will prevent us from choosing the arguably simple Feynman parameter rescaling used above in (\ref{general_double_pentagon_parameter_rescaling}), but it will also mean that greater simplifications are possible along the way. For example, allowing some two-brackets to vanish will allow us to recognize more Feynman-parameter integrals as total derivatives, resulting in better (meaning, lower-dimensional) parametric representations of the final integrals.

The $\ell_1$-dependent parts of both integrals in (\ref{omega2_and_omega2tilde_integrands}) are identical, and so let us start there. Following the analysis above, we introduce Feynman parameters for the $\ell_1$ propagators according to:
\eq{\begin{split}Y_1&\equiv{\color{hblue}\alpha_{1}(1)+\alpha_{2}(4)+\beta_1(2)+\beta_2(3)}+{\color{hred}\gamma_0}(\ell_2)\,,\\
&\equiv{\color{hblue}(Q_1)}+{\color{hred}\gamma_0}(\ell_2)\,.\end{split}}
The attentive reader will notice that we have chosen to label our Feynman parameters in a somewhat unusual way---the reason for which we hope to become clearer when we generalize this analysis to higher loops.

Relative to what we saw for the general case, we may immediately recognize one great simplification: 
\eq{\x{Y_1}{N_1}={\color{hred}\gamma_0}\x{\ell_2}{N_1}\,,}
because $\x{{\color{hblue}Q_1}}{N_1}\!=\!0$. (As mentioned above, this is {\it precisely} the defining property of the dual point $N_1$ (up to complex conjugation).) Thus, recognizing that the ${\color{hred}\gamma_0}$ integral is a total derivative, 
\eq{4\int\limits_0^\infty\!\!d{\color{hred}\gamma_0}\,\frac{\x{Y_1}{N_1}}{\x{Y_1}{Y_1}^3}=4\int\limits_0^\infty\!\!d{\color{hred}\gamma_0}\,\frac{{\color{hred}\gamma_0}\x{\ell_2}{N_1}}{(\x{{\color{hblue}Q_1}}{{\color{hblue}Q_1}}+2\,{\color{hred}\gamma_0}\x{\ell_2}{{\color{hblue}Q_1}})^3}=\frac{1}{2}\frac{\x{\ell_2}{N_1}}{\x{{\color{hblue}Q_1}}{{\color{hblue}Q_1}}\x{\ell_2}{{\color{hblue}Q_1}}^2}\,,}
we see that (\ref{omega2_and_omega2tilde_integrands}) become:
\eq{\begin{split}\Omega^{(2)}&=\frac{1}{2}\int\limits_0^{\infty}\!\!d^2\vec{\alpha}\proj{d^1\!\vec{\beta}}\frac{1}{\x{{\color{hblue}Q_1}}{{\color{hblue}Q_1}}}\!\!\int\!\!d^4\ell_2\frac{\x{\ell_2}{N_1}\x{\ell_2}{N_2}}{\x{\ell_2}{{\color{hblue}Q_1}}^2\x{\ell_2}{1}\x{\ell_2}{4}\x{\ell_2}{5}\x{\ell_2}{6}}\,;\\
\tilde{\Omega}^{(2)}&=\frac{1}{2}\int\limits_0^{\infty}\!\!d^2\vec{\alpha}\proj{d^1\!\vec{\beta}}\frac{1}{\x{{\color{hblue}Q_1}}{{\color{hblue}Q_1}}}\!\!\int\!\!d^4\ell_2\frac{\x{\ell_2}{N_1}\x{\ell_2}{\overline{N_2}}}{\x{\ell_2}{{\color{hblue}Q_1}}^2\x{\ell_2}{1}\x{\ell_2}{4}\x{\ell_2}{5}\x{\ell_2}{6}}\,.
\end{split}\label{omega2_post_ell1_integration}}

We can now recognize the $\ell_2$ integral (\ref{omega2_post_ell1_integration}) as an instance of a `hexagon' integral from (\ref{pentagon_and_hexagon_loop_integrals}) after Feynman parameterization. Let us label the Feynman parameters for the $\ell_2$ integrals according to
\eq{\begin{split}Y_2&\equiv{\color{hblue}({\color{hblue}Q_1})+\alpha_{3}(1)+\alpha_{4}(4)}+{\color{hred}\gamma_1}(5)+{\color{hred}\gamma_2}(6)\,\\
&\equiv{\color{hblue}(Q_2)}+{\color{hred}\gamma_1}(5)+{\color{hred}\gamma_2}(6)\,.\end{split}\label{omega2_y2}}
As before, we have de-projectivized the Feynman parameter integral by setting the coefficient of $({\color{hblue}Q_1})$ in (\ref{omega2_y2}) to 1 for $Y_2$. However, in contrast to the analysis of the general case, we have singled-out {\it two} Feynman parameters, ${\color{hred}\gamma_1},{\color{hred}\gamma_2}$---which will be integrated-out momentarily. This is related to another concrete illustration of the simplifications that are afforded to us in a concrete case: exploiting the fact that $\x{5}{6}\!=\!0$ (without any regularization required), we have that 
\eq{\x{Y_2}{Y_2}=\x{{\color{hblue}Q_2}}{{\color{hblue}Q_2}}+2\,{\color{hred}\gamma_1}\x{{\color{hblue}Q_2}}{5}+2\,{\color{hred}\gamma_2}\x{{\color{hblue}Q_2}}{6}\,.}
This is helpful because there is no term in $\x{Y_2}{Y_2}$ proportional to ${\color{hred}\gamma_1\gamma_2}$, making it possible to recognize both integrations as total derivatives. To see this, notice that using $Y_2$ as defined in (\ref{omega2_y2}), the Feynman parameter integration of $\ell_2$ in (\ref{omega2_post_ell1_integration}) results in:
\eq{\begin{split}
\Omega^{(2)}&=\int\limits_0^{\infty}\!\!d^4\!\vec{\alpha}\!\proj{d^1\!\vec{\beta}}d^2{\color{hred}\vec{{\color{hred}\gamma}}}\,\frac{1}{\x{{\color{hblue}Q_1}}{{\color{hblue}Q_1}}}\left(6\frac{\x{Y_2}{N_1}\x{Y_2}{N_2}}{\x{Y_2}{Y_2}^4}-2\frac{\x{N_1}{N_2}}{\x{Y_2}{Y_2}^3}\right)\!;\\
\widetilde{\Omega}^{(2)}&=\int\limits_0^{\infty}\!\!d^4\!\vec{\alpha}\!\proj{d^1\!\vec{\beta}}d^2{\color{hred}\vec{{\color{hred}\gamma}}}\,\frac{1}{\x{{\color{hblue}Q_1}}{{\color{hblue}Q_1}}}\left(6\frac{\x{Y_2}{N_1}\x{Y_2}{\overline{N_2}}}{\x{Y_2}{Y_2}^4}-2\frac{\x{N_1}{\overline{N_2}}}{\x{Y_2}{Y_2}^3}\right)\!.\hspace{-20pt}\end{split}\label{omega2_post_ell2_integration}}
Using the fact that the numerators in the first terms of (\ref{omega2_post_ell2_integration}) simplify considerably, 
\eq{\begin{split}
\x{Y_2}{N_1}\x{Y_2}{N_2}&=\Big({\color{hred}\gamma_1}\x{5}{N_1}+{\color{hred}\gamma_2}\x{6}{N_1}\Big)\Big(\beta_1\x{2}{N_2}+\beta_2\x{3}{N_2}\Big)\,,\\\x{Y_2}{N_1}\x{Y_2}{\overline{N_2}}&=\Big({\color{hred}\gamma_1}\x{5}{N_1}+{\color{hred}\gamma_2}\x{6}{N_1}\Big)\Big(\beta_1\x{2}{\overline{N_2}}+\beta_2\x{3}{\overline{N_2}}\Big)\,,\end{split}}
(which is part of the inherent magic of the (`chiral') numerators $N_1,\overline{N_2}$ defined in (\ref{omega_numerators_defined}),) we recognize that the ${\color{hred}\vec{{\color{hred}\gamma}}}$ part of (\ref{omega2_post_ell2_integration}) is schematically of the form
\eq{\int\limits_0^{\infty}\!\!d^2{\color{hred}\vec{{\color{hred}\gamma}}}\left(\!6\frac{n_1\,{\color{hred}\gamma_1}\pl n_2\,{\color{hred}\gamma_2}}{(g_0\pl g_1\,{\color{hred}\gamma_1}\pl g_2\,{\color{hred}\gamma_2})^4}\mi2\frac{n_0}{(g_0\pl g_1\,{\color{hred}\gamma_1}\pl g_2\,{\color{hred}\gamma_2})^3}\right)=\frac{1}{g_0\,g_1\, g_2}\Big(\frac{n_1}{g_1}+\frac{n_2}{g_2}-n_0\Big).\label{schematic_gamma_integral}}
Comparing (\ref{schematic_gamma_integral}) with (\ref{omega2_post_ell2_integration}) (and being careful about factors of 2), we have
\eq{\begin{split}\Omega^{(2)}(\u,\v,\w)&\equiv\int\limits_0^{\infty}\!\!d^4\!\vec{\alpha}\!\proj{d^1\!\vec{\beta}}\frac{1}{f_1\,f_2\,g_1\,g_2}\left(\frac{n_1}{g_1}+\frac{n_2}{g_2}-n_0\right)\!,\\\widetilde{\Omega}^{(2)}(\u,\v,\w)&\equiv\int\limits_0^{\infty}\!\!d^4\!\vec{\alpha}\!\proj{d^1\!\vec{\beta}}\frac{1}{f_1\,f_2\,g_1\,g_2}\left(\frac{\tilde{n}_1}{g_1}+\frac{\tilde{n}_2}{g_2}-\tilde{n}_0\right)\!,\end{split}\label{omega2_pre_rescaling}}
in which we may recognize,
\eq{\begin{array}{l@{$\;\;\;\;$}l@{$\;\;\;\;$}l}f_i\equiv\frac{1}{2}\x{{\color{hblue}Q_i}}{{\color{hblue}Q_i}}\,,& g_1\equiv\x{{\color{hblue}Q_2}}{5},& g_2\equiv\x{{\color{hblue}Q_2}}{6}\,,\\
n_0\equiv\x{N_1}{N_2}\,,&n_1\equiv\x{5}{N_1}\x{{\color{hblue}Q_1}}{N_2}\,,&n_2\equiv\x{6}{N_1}\x{{\color{hblue}Q_1}}{N_2}\,,\\
\tilde{n}_0\equiv\x{N_1}{\overline{N_2}}\,,&\tilde{n}_1\equiv\x{5}{N_1}\x{{\color{hblue}Q_1}}{\overline{N_2}}\,,&\tilde{n}_2\equiv\x{6}{N_1}\x{{\color{hblue}Q_1}}{\overline{N_2}}\,.\end{array}\label{omega2_fgns_prescaling}}

The parametric integrals in (\ref{omega2_pre_rescaling}) are simple enough, but they are not yet manifestly dual-conformally invariant. If we wished to regulate all the massless legs in these integrals, we could use the same Feynman-parameter rescaling as described in our analysis of the general double-pentagon integral (\ref{general_double_pentagon_parameter_rescaling})---which would indeed render the representations (\ref{omega2_pre_rescaling}) manifestly dual-conformally invariant. However, because we know that both $\Omega^{(2)}$ and $\tilde{\Omega}^{(2)}$ are in fact infrared-finite, we should instead tailor how we rescale Feynman parameters a bit differently. 

As always, the problem is to rescale all the Feynman parameters in the representation such that every factor in the denominator is uniform in conformal weights. There are (as always) many solutions to this problem; in our present case, the following rescalings will suffice:
\eq{\begin{array}{l@{$\;\;\;\;$}l@{$\;\;\;\;$}l}\displaystyle\alpha_1\mapsto\alpha_1\frac{1}{\x{{\color{hred}1}}{{\color{hred}4}}}\,,&\displaystyle\alpha_2\mapsto\alpha_2\frac{\x{1}{5}\x{2}{6}}{\x{1}{{\color{hred}4}}\x{2}{5}\x{{\color{hred}4}}{6}}\,,&\displaystyle\beta_1\mapsto\beta_1\frac{\x{1}{5}}{\x{1}{{\color{hred}4}}\x{{\color{hred}2}}{5}}\,,\\[12pt]\displaystyle\alpha_3\mapsto\alpha_3\frac{1}{\x{{\color{hred}1}}{{\color{hred}4}}}\,,&\displaystyle\alpha_4\mapsto\alpha_4\frac{\x{1}{5}\x{2}{6}}{\x{1}{{\color{hred}4}}\x{2}{5}\x{{\color{hred}4}}{6}}\,,&\displaystyle\beta_2\mapsto\beta_2\frac{\x{1}{5}\x{2}{6}}{\x{1}{{\color{hred}4}}\x{2}{5}\x{{\color{hred}3}}{6}}\,.
\end{array}\label{omega2tilde_rescalings}}
Upon rescaling the Feynman parameters, redefining $f_i,g_i$ to be their rescaled versions modulo uniform pre-factors (absorbed into new numerator factors), we find the same essential form as above (motivating the abuse of notation in the following):
\eq{\begin{split}\Omega^{(2)}(\u,\v,\w)&\equiv\int\limits_0^{\infty}\!\!d^4\!\vec{\alpha}\!\proj{d^1\!\vec{\beta}}\frac{1}{f_1\,f_2\,g_1\,g_2}\left(\frac{n_1}{g_1}+\frac{n_2}{g_2}-n_0\right)\!,\\\widetilde{\Omega}^{(2)}(\u,\v,\w)&\equiv\int\limits_0^{\infty}\!\!d^4\!\vec{\alpha}\!\proj{d^1\!\vec{\beta}}\frac{1}{f_1\,f_2\,g_1\,g_2}\left(\frac{\tilde{n}_1}{g_1}+\frac{\tilde{n}_2}{g_2}-\tilde{n}_0\right)\!,\end{split}\label{omega2_post_rescaling}}
where
\eq{\begin{array}{l@{$\;\;\;\;\;\;$}l}f_1\equiv\alpha_1\alpha_2\pl\u\,\alpha_1\beta_2\pl\v\,\alpha_2\beta_1\,,&g_1\equiv\alpha_1\pl\alpha_3\pl\beta_1\pl\w\,\beta_2\,,\\
f_2\equiv f_1\pl \alpha_1\alpha_4\pl\alpha_2\alpha_3\pl\alpha_3\alpha_4\pl\u\,\alpha_3\beta_2\pl\v\,\alpha_4\beta_1\,,
&g_2\equiv\alpha_2\pl\alpha_4\pl\beta_1\pl\beta_2\,,\end{array}\label{omega2_fs_and_gs}}
and
\eq{\begin{array}{l}\fwboxL{280pt}{n_0\equiv\w\,,}\\
n_1\equiv(1\mi\v)\w\,\beta_1\pl\frac{1}{2}\w(1\mi\u\mi\v\pl\w\pl\Delta_6)\,\beta_2\,,\\
n_2\equiv\frac{1}{2}(1\mi\u\mi\v\pl\w\mi\Delta_6)\,\beta_1\pl(1\mi\u)\w\,\beta_2\,,\end{array}\label{omega2_nums_post_rescaling}}
while,
\eq{\begin{array}{l}\fwboxL{280pt}{\tilde{n}_0\equiv(1\mi\u\mi\v\pl\u\v)\,,}\\
\tilde{n}_1\equiv\frac{1}{2}(1\mi\v)(1\mi\u\mi\v\pl\w\mi\Delta_6)\beta_1\pl(1\mi\u)(1\mi\v)\w\beta_2\,,\\
\tilde{n}_2\equiv(1\mi\u)(1\mi\v)\beta_1\pl\frac{1}{2}(1\mi\u)(1\mi\u\mi\v\pl\w\pl\Delta_6)\beta_2\,.\end{array}\label{omega2tilde_nums_post_rescaling}}
Here, the numerators are manifestly dual-conformally invariant functions of the cross-ratios, including the `parity-odd' factor of $\Delta_6$ defined by
\eq{\Delta_6\equiv\sqrt{(1\mi\u\mi\v\mi\w)^2\mi4\,\u\v\w}\,.\label{delta6_defined}}
Importantly, $\Delta_6$ should be understood to change sign under a rotation of the dual points by three---and this sign reflects the fact that it is odd under parity. See \mbox{Appendix \ref{appendix:conventions_and_definitions}} for variables in which this sign may be disambiguated.  

In general, we may define the `even/odd' parts of any six-point function by,
\eq{I\equiv I^{\even}+I^{\odd},\quad\text{where}\quad I^{\even}\equiv\frac{1}{2}(1\pl r^3)I,\quad I^{\odd}\equiv\frac{1}{2}(1\mi r^3)I\,,\label{odd_even_defined_first_time}}
where $r\!:\!x_a\!\mapsto\!x_{a+1}$ defines a rotation of all external dual-coordinate points. As all the cross-ratios $u_a$ map into themselves under a three-fold rotation, while $\Delta_6$ changes sign, it is easy to extract the even/odd parts of each integral. For example, 
\eq{\frac{1}{2}(1\mi r^3)\big[\widetilde{\Omega}^{(2)}\big]=\frac{\Delta_6}{2}\int\limits_0^{\infty}\!\!d^4\!\vec{\alpha}\!\proj{d^1\!\vec{\beta}}\frac{1}{f_1\,f_2\,g_1\,g_2}\left(\frac{(\v\mi1)\beta_1}{g_1}+\frac{(1\mi\u)\beta_2}{g_2}\right)\!.
\label{odd_part_of_omega2tilde}}
We could write an analogous expression for $\Omega^{(2)}$, but we may observe that the {\it initial} (loop-momentum-space) definition of $\Omega^{(2)}$ in equation (\ref{omega2_and_omega2tilde_integrands}) is in fact symmetric under a rotation by three. Thus, the `odd' part of $\Omega^{(2)}$ {\it must} vanish:
\eq{\frac{1}{2}(1\mi r^3)\big[\Omega^{(2)}\big]=\frac{\Delta_6}{2}\int\limits_0^{\infty}\!\!d^4\!\vec{\alpha}\!\proj{d^1\!\vec{\beta}}\frac{1}{f_1\,f_2\,g_1\,g_2}\left(\frac{\w\beta_2}{g_1}-\frac{\beta_1}{g_2}\right)=0\,.
\label{odd_part_of_omega2}}
This implies a Feynman-parameter integral-level identity among the terms in the representations of (\ref{omega2_post_rescaling}), allowing us to write the considerably more compact expression:
\eq{\hspace{-20pt}\Omega^{(2)}\!\equiv\!\int\limits_0^{\infty}\!\!d^4\!\vec{\alpha}\!\proj{d^1\!\vec{\beta}}\!\frac{1}{f_1\,f_2\,g_1\,g_2}\!\left(\!\frac{(1\mi\v)\w\,\beta_1}{g_1}\pl\frac{(1\mi\u\mi\v\pl\w)\beta_1\pl(1\mi\u)\w\,\beta_2}{g_2}\mi\w\!\right)\!.\hspace{-20pt}\label{omega2_simplified}}

While these Feynman-parameter integral representations of $\Omega^{(2)}$ are and $\tilde{\Omega}^{(2)}$ do not yet make their transcendental weight manifest (both being five-fold integral representations of weight-four functions), they are manifestly functions of dual-conformally invariant cross-ratios. Moreover, standard integration approaches can now be used to convert these representations into, e.g., hyperlogarithms (see, e.g., \mbox{ref.\ \cite{Bourjaily:2018aeq}}).

\phantomsection
\markright{}\addcontentsline{toc}{paragraph}{\hspace{82pt} All-Loop `Ladder' Generalizations: $\!\Omega^{(L)}\!\!,\,\tilde{\Omega}^{(L)}$}
\vspace{-2pt}\subsubsection*{{\it Exempli Gratia}{\rm:} All-Loop (`Ladder') Generalizations: $\Omega^{(L)}$}\label{subsubsec:exempli_gratia_2}\vspace{-2pt}

Interestingly, the analysis above generalizes to all higher loop `ladder' integrals, denoted $\Omega^{(L)}$ in \mbox{ref.\ \cite{Caron-Huot:2018dsv}}, in a very simple way. These integrals involve the same double-pentagon denominators (and tensor-numerators) as $\Omega^{(2)}$ and $\tilde{\Omega}^{(2)}$, but with an arbitrary ladder of boxes between them. To be clear, $\Omega^{(L)}$ is defined in loop-momentum space as the integral, 
\eq{\hspace{-20pt}\Omega^{(L)}\equiv\!\!\int\!\!d^{4L}\vec{\ell}\frac{\x{\ell_1}{N_1}\x{1}{4}^{L-2}\x{\ell_L}{N_2}}{\x{\ell_1}{2}\x{\ell_1}{3}\big(\prod_{i=1}^L\x{\ell_i}{1}\x{\ell_i}{4}\big)\big(\x{\ell_1}{\ell_2}\cdots\x{\ell_{L-1}}{\ell_L}\big)\x{\ell_L}{5}\x{\ell_L}{6}}\!.\hspace{-20pt}\label{omegaL_loop_space_definition}}
Graphically, this corresponds to the Feynman integral,
\vspace{-6pt}\eq{\fwboxR{0pt}{\Omega^{(L)}(\u,\v,\w)\Leftrightarrow}\fwbox{165pt}{\fig{-29.8pt}{1}{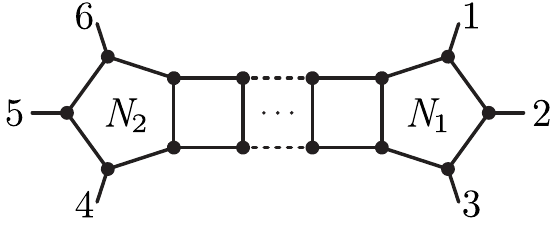}}\,.\label{omegaL_figure}\vspace{-6pt}}
The integral $\tilde{\Omega}^{(L)}$ would be defined analogously---with $N_2\!\leftrightarrow\!\overline{N_2}$. Following the same sequence of Feynman parameterizations---and recognizing every intermediate box integral along the ladder as having one Feynman-parameter integration which is a total derivative (resulting in a two-fold representation of each), and rescaling the Feynman parameters in the same way we did before, we obtain the following $(2L\pl1)$-fold Feynman-parameter representation of $\Omega^{(L)}$:
\eq{\hspace{-20pt}\Omega^{(L)}\!\!\equiv\!\!\int\limits_0^{\infty}\!\!d^{2L}\!\vec{\alpha}\,d\beta\,\frac{1}{f_1\!\cdots\! f_L\,g_1\,g_2}\left(\!\frac{(1\mi \v)u_3\beta}{g_1}\pl\frac{(1\mi \u\mi \v\pl \w)\beta\pl(1\mi\u)u_3}{g_2}\mi u_3\right)\!,\hspace{-20pt}\label{omegaL_feynman_rep}}
where
\vspace{-12pt}\eq{\begin{split}f_k&\equiv\u(\alpha^1_1\pl\ldots\pl\alpha^k_1)\pl\v(\alpha_2^1\pl\ldots\pl\alpha_2^k)\beta\pl\sum_{i,j=1}^k\alpha^i_1\alpha^j_2\,,\\[-12pt]
g_1&\equiv\w\pl\beta\pl(\alpha^1_1\pl\ldots\pl\alpha^L_1)\,,\\
g_2&\equiv1\pl \beta\pl(\alpha^1_2\pl\ldots\pl\alpha^L_2)\,.\end{split}\label{all_loop_omega2_denominators}}
Here, relative to the analysis for $\Omega^{(2)}$, we have de-projectivized the $\beta_i$ integral by setting $\beta_1\!\to\!\beta,\beta_2\!\to\!1$, and we have relabelled $\alpha_{2k-1}\!\to\!\alpha^{k}_1,\alpha_{2k}\!\to\!\alpha^{k}_2$. Notice that the numerator structure is completely $L$-independent. As such, the $L$-loop integral of $\tilde{\Omega}^{(L)}$ is directly analogous to \eqref{omegaL_feynman_rep}, with identical denominator factors as those defined in \eqref{all_loop_omega2_denominators}, but where the numerator factors $n_i$ would be given by $\tilde{n}_i$ defined in equation~\eqref{omega2tilde_nums_post_rescaling}.

Direct integration of the representation \eqref{omegaL_feynman_rep} using the package {\HyperInt} \cite{Panzer:2014caa} has been completed through four loops analytically, matching the known results obtained via the hexagon bootstrap program and differential equations \cite{Caron-Huot:2018dsv}. (Curiously, although our analysis seemed to require $L\!\geq\!2$, setting $L\!=\!1$ in \eqref{omegaL_feynman_rep} results in a representation of the unique (up to rotation) finite hexagon Feynman integral at one loop---also referred to as $\Omega^{(1)}$ in \cite{Caron-Huot:2018dsv}.) We expect that differential equations for \eqref{omegaL_feynman_rep} could be found directly in this representation.

\phantomsection
\markright{}\addcontentsline{toc}{paragraph}{\hspace{82pt} An `Elliptic' Regularized Hexagon Function}
\vspace{-2pt}\subsubsection*{{\it Exempli Gratia}{\rm:} An `Elliptic' Regularized Two-Loop Hexagon Function}\label{subsubsec:exempli_gratia_3}\vspace{-2pt}

Let us conclude our discussion of concrete examples with one that explicitly requires regularization of infrared divergences. Specifically, consider the double-box integral, 
\vspace{-6pt}\eq{\fwbox{65pt}{\fig{-24.62pt}{1}{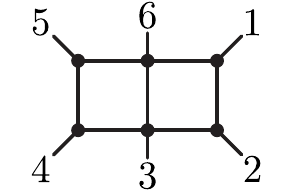}}\Leftrightarrow I\!\equiv\!\int\!\!d^4\ell_1d^4\ell_2\frac{\x{1}{3}\x{2}{5}\x{4}{6}}{\x{\ell_1}{1}\x{\ell_1}{2}\x{\ell_1}{3}\x{\ell_1}{\ell_2}\x{\ell_2}{4}\x{\ell_2}{5}\x{\ell_2}{6}}\,.\label{double_box_example}}
This integral is $\logkdiv{4}$-divergent. (In dimensional regularization, it would have a term proportional to $1/\epsilon^4$.)

The collinear regions associated with divergences are precisely the massless legs $\{p_1,p_2,p_4,p_5\}$---the momenta $\{p_3,p_6\}$ are not associated with infrared-divergent regions in loop-momentum space, as they do not flow into three-point vertices of the graph. Thus, only four of the six massless momenta require regularization. Following the discussion of the dual-conformal regulator in \mbox{section~\ref{subsec:dci_regulator}}, we see that among the six vanishing two-brackets $\x{a}{a\,\pl1}$ the following replacements should be made:
\eq{\x{{\color{hred}1}}{{\color{hred}2}}\!\mapsto\!\eps\frac{\x{{\color{hred}1}}{{\color{hblue}3}}\x{{\color{hred}2}}{{\color{hblue}6}}}{\x{{\color{hblue}3}}{{\color{hblue}6}}},\;\x{{\color{hred}2}}{{\color{hred}3}}\!\mapsto\!\eps\frac{\x{{\color{hred}2}}{{\color{hblue}4}}\x{{\color{hblue}1}}{{\color{hred}3}}}{\x{{\color{hblue}1}}{{\color{hblue}4}}},\;\x{{\color{hred}4}}{{\color{hred}5}}\!\mapsto\!\eps\frac{\x{{\color{hred}4}}{{\color{hblue}6}}\x{{\color{hblue}3}}{{\color{hred}5}}}{\x{{\color{hblue}3}}{{\color{hblue}6}}},\;\x{{\color{hred}5}}{{\color{hred}6}}\!\mapsto\!\eps\frac{\x{{\color{hblue}1}}{{\color{hred}5}}\x{{\color{hblue}4}}{{\color{hred}6}}}{\x{{\color{hblue}1}}{{\color{hblue}4}}}.\nonumber}
With these replacements understood, we may proceed with the Feynman parameterization in the ordinary way. 

For the $\ell_1$ integral we introduce
\eq{\begin{split}Y_1&\equiv(2)+\alpha_{1}(1)+\alpha_2(3)+\gamma_1(\ell_2)\,\\
&\equiv(Q_1)+\gamma_1(\ell_2)\,;\end{split}\label{y1_elliptic}}
and for the $\ell_2$ integral we define
\eq{\begin{split}Y_2&\equiv(Q_1)+\alpha_{3}(4)+\alpha_4(6)+\gamma_2(5)\,\\
&\equiv(Q_2)+\gamma_2(5)\,.\end{split}\label{y2_elliptic}}
Following the now familiar sequence of steps, we obtain the following four-fold Feynman-parameter representation for the integral \eqref{double_box_example}:
\begin{equation*}
	I=4\!\int\!\!\td^4\vec{\alpha}
	\frac{
		\x{1}{3}\x{2}{5}\x{4}{6}
	}{
		\x{Q_1}{Q_1}\x{Q_2}{Q_2}\x{Q_2}{5}
	}\,.
\end{equation*}
Rescaling the Feynman parameters according to,
\begin{equation*}
	\alpha_1\!\mapsto\!\alpha_1\frac{\x{2}{4}}{\x{1}{4}},\quad
	\alpha_2\!\mapsto\!\alpha_2\frac{\x{2}{6}}{\x{3}{6}},\quad
	\alpha_3\!\mapsto\!\alpha_3\frac{\x{2}{6}}{\x{4}{6}},\quad
	\alpha_4\!\mapsto\!\alpha_4\frac{\x{2}{4}}{\x{4}{6}},
\end{equation*}
results in the following, manifestly dual-conformally invariant (regulated) representation of the original integral \eqref{double_box_example}:
\begin{equation}
	I=
	\int \frac{\td^4\vec{\alpha}}{f_1\,f_2\,f_3}\,,
	\quad\text{where}\qquad
	\begin{aligned}
		f_1&\equiv\alpha_1\alpha_2+{\color{divblue}\eps(\alpha_1+\alpha_2)},\\
		f_2&\equiv\u\,f_1+\alpha_1\alpha_3+\alpha_2\alpha_4+\alpha_3+\alpha_4+\alpha_3\alpha_4,\\
		f_3&\equiv1+\alpha_1\v+\alpha_2\w+{\color{divblue}\eps(\alpha_3\w+\alpha_4\v)}.
	\end{aligned}
	\label{final_elliptic_example}%
\end{equation}

What is especially interesting about this example is that when $\eps\!\neq\!0$, the integral (\ref{final_elliptic_example}) is in fact non-polylogarithmic: it has {\it no} residues with maximal co-dimension, and a co-dimension three residue of (\ref{final_elliptic_example}) taken about $\vec{f}\!=\!0$ results in an elliptic integral over the remaining variable \cite{CaronHuot:2012ab}. This could have in fact been noticed immediately from the way that the dual-conformal regulator regulates an integral such as (\ref{double_box_example}): by adding masses to each of the corners of the double box, the regulated expression is essentially a fully-massive double-box (which is well known to be non-polylogarithmic \cite{CaronHuot:2012ab}).

Nevertheless, in the limit of $\eps\!\to\!0$ the integral (\ref{final_elliptic_example}) becomes a simple polynomial in $\logdiv$. This may or may not be surprising at first glance, but it turns out to be easy to understand and in fact prove in broad generality. 

In the next section, we will prove that any dual-conformally regulated expression becomes a polynomial in $\logdiv$ in the limit of $\eps\!\to\!0$. Moreover, we will describe how the coefficient of each $\logkdiv{k}$ in this expansion can be extracted analytically in terms of finite (and dual-conformally invariant) Feynman-parameter integrals.

\newpage
\vspace{-2pt}\subsection{Analytic Extraction of Divergent and Finite Coefficients}\label{subsec:divergent_coefficients_extraction}\vspace{-2pt}

With the dual-conformal regulator $\eps$, infrared-divergent integrals admit an expansion
\begin{equation}
	I(\eps) 
	= \sum_{k=0}^N I_k\cdot (\log\eps)^k
	\ +\ \text{terms vanishing with $\eps\rightarrow 0$,}
	\label{eq:I-log-expansion}%
\end{equation}
where the coefficients $I_k$ of the polynomial in $\logdiv$ denote $\eps$-independent, dual-conformally invariant functions. For the concrete calculations described in the previous section, we found it most efficient to first compute $I(\eps$) with its full dependence on $\eps$, and then expand this result for small $\eps$. However, the functions $I(\eps)$ tend to be much more complicated than the coefficients $I_k$ that we are after, wherefore this approach is clearly not optimal.

In this section, we sketch an alternative strategy, which circumvents the computation of $I(\eps)$ itself and instead permits direct access to the coefficients $I_k$. As we will demonstrate below, this method transfers the added complexity due to the dependence on $\eps$ (an additional, mass-like parameter) to an analytic regulator. The latter seems easier to mesh with current parametric integration tools, and we therefore hope that the following will be useful in applications.

Consider a dual-conformally regulated Feynman-parametric integral of the form\footnote{This assumption constitutes no restriction, because any integral involving a product of several factors (each linear in $\eps$), like \eqref{final_elliptic_example}, may be brought into the form \eqref{skeleton_regulated_feynman_parameter_int} by introducing more Feynman parameters to combine the $\eps$-dependent denominators into a single linear form.}
\begin{equation}
	I(\eps)
	= \int\!\!\!
	\frac{\td\vec{\alpha}}{(g(\vec{\alpha})\,\eps+f(\vec{\alpha}))^\lambda}\,.
	\label{skeleton_regulated_feynman_parameter_int}%
\end{equation}
It is a fundamental property of the Mellin transform $\Mellin{I}(z)$ that its poles correspond directly to terms in the asymptotic expansion of $I(\eps)$ at $\eps=0$. Let us briefly recall the basics, while we refer to \cite{FlajoletGourdonDumas:MellinAsyHarmonic} for a detailed account. The Mellin transform\footnote{In the fundamental strip, the integrals over $\eps$ and $\vec{\alpha}$ commute by dominated convergence.}
\eq{
	\Mellin{I}(z)
	\defas \int_0^{\infty}\!\!\!\!\td\eps\,\, \eps^{z-1} I(\eps)
	= \frac{\Gamma(z)\Gamma(\lambda-z)}{\Gamma(\lambda)}
	\int\!\!\! \frac{\td \vec{\alpha}}{g(\vec{\alpha})^{z} f(\vec{\alpha})^{\lambda-z}}
}
converges in some strip $\mu\!<\!\Re(z)\!<\!\nu$ and extends to a meromorphic function of $z$ with poles at rational numbers. This fundamental strip and the loci of these poles of such analytically regularized parametric integrals can be calculated from the Newton polytopes of $g$ and $f$. The Mellin inversion is a contour integral along a parallel to the imaginary axis, within the fundamental strip:
\eq{
	I(\eps)
	=
	\!\!\!\int\limits_{-\iu\infty+\sigma}^{\iu\infty+\sigma}\!\!\!\!\frac{\td z}{2\ipi}\,
	\eps^{-z}
	\Mellin{I}(z)
	\quad\text{where}\quad
	\mu<\sigma<\nu.
}
To obtain the asymptotic expansion at $\eps\!=\!0$, we close the contour to the left and pick up all residues at poles to the left of the line $\Re(z)\!=\!\mu$:
\begin{equation}
	I(\eps)
	=
	\sum_{\Re(z^*) \leq \mu}
	\Res_{z=z^*}\Bigg(
		\eps^{-z}\,
		\Mellin{I}(z)
	\Bigg).
	\label{eq:small-eps-from-residues}%
\end{equation}
In our applications, this expansion is always dominated by a pole at $z^{\ast}\!=\!\mu\!=\!0$, which immediately implies the claimed expansion of the form \eqref{eq:I-log-expansion}. The coefficients $I_k$ of $\log^k(\eps)$ therein are determined, via \eqref{eq:small-eps-from-residues}, by the polar part of the Laurent series
\eq{
	\Mellin{I}(z)
	=\sum_{k\geq -N} \LScoeff{I}_k \cdot z^{k-1},
	\quad\text{namely through}\quad
	I_k = \frac{(-1)^k}{k!} \LScoeff{I}_{-k}.
	\label{eq:laurent_series}%
}
As discussed in \cite[section~5.8]{Smirnov:2012gma}, the coefficients $\LScoeff{I}_k$ can be obtained with standard techniques such as sector decomposition~\cite{Binoth:2000ps,Binoth:2003ak,Borowka:2012yc,Smirnov:2008py,Bogner:2007cr}; but a decomposition can be avoided with analytic regularization~\cite{Panzer:2014gra}, and Mellin-Barnes techniques~\cite{Anastasiou:2013srw} have been proposed as well.
All of these result in convergent parameteric integral representations for the coefficients $\LScoeff{I}_k$. Given that our input integrand is a rational function of integration parameters and only dual-conformally invariant cross-ratios, also the derived integrands for $\LScoeff{I}_k$ will be dual-conformally invariant.

If $I(\eps)$ is finite at $\eps\!=\!0$, we note that $\Mellin{I}(\eps)\!=\!I(0)/z\pl\asyO{z^0}$ has a first order pole with residue $I_0\!=\!\LScoeff{I}_0\!=\!I(0)$ as expected. However, if $\Mellin{I}(z)$ acquires higher order poles at $z\!=\!0$, these will induce terms proportional to powers of $\logdiv$.

To illustrate this procedure, let us consider the dual-conformally regulated, so-called `two-mass hard' integral: 
\eq{
	I^{\text{2mh}}(u,v)
	\defas
	\frac{1}{2}
	\int\limits_0^{\infty}\!\!\!\proj{\td^2\vec{\alpha}}\,
	\frac{1}{
		({\color{divblue}(u\alpha_1 + v\alpha_2)\delta}+\alpha_3)
		(\alpha_1\alpha_2 + \alpha_1\alpha_3 + \alpha_2\alpha_3)
	}
	.
	\label{eq:regulated_2mh_integral}%
}
Its Mellin transform $\Mellin{I^{\text{2mh}}}(z)$ is $\Gamma(z)\Gamma(1\mi z)\!=\!z^{-1}\pl z \Zeta{2}\pl\asyO{z^3}$ times
\eq{\frac{1}{2}\int\limits_0^{\infty}\!\!\proj{\td^2\vec{\alpha}}\,\frac{\alpha_3^{z-1}(u\alpha_1 \pl v\alpha_2)^{-z}}{\alpha_1\alpha_2 \pl \alpha_1\alpha_3 \pl \alpha_2\alpha_3}= \frac{1}{z^2}- \frac{\log(u\, v)}{2z}+ \frac{\log(u)\log(v)}{2}+ \asyO{z},}
which can for example be computed directly with {\HyperInt}\footnote{In this simple case, this expansion can also be computed with tools like \cite{HuberMaitre:HypExp}, after recognizing the integral as the hypergeometric function $\pFq{z,z}{2z}{1\mi\tfrac{v}{u}}$ times $\Gamma^2(z)/\left(\Gamma(2z) u^z \right)$.} or by using standard algorithms for the evaluation of Mellin-Barnes integrals~\cite{Smirnov:2004ym,Smirnov:2006ry,Smirnov:2012gma}.

In this case, the fundamental strip of $\Mellin{I^{\text{2mh}}}(z)$ is $(\mu,\nu)=(0,1)$ with a third order pole at $z=0$. Using \eqref{eq:laurent_series}, we conclude therefore that the asymptotic expansion is
\eq{
	I^{\text{2mh}}(u,v)	
	=\frac{1}{2}\log(u)\log(v)
	+\Zeta{2}
	+{\color{divblue}\frac{1}{2}\log(u\,v)\logdiv}
	+{\color{divblue}\frac{1}{2}\logsquarediv}
	+ \asyO{\eps^{>0}}.
	\label{eq:regulated-2mh-expansion}%
}

\vspace{-2pt}\subsection{Feynman Parameter Integration}\label{subsec:analytic_integration_methods}\vspace{-6pt}

The Feynman parameter integrals encountered above can be calculated using hyperlogarithms \cite{Brown:2008um,Anastasiou:2013srw,Panzer:2014gra} as implemented for example in {\HyperInt} \cite{Panzer:2014caa}---provided, that is, that a linearly-reducible integration order exists. This condition requires that all iterated residues encountered through integration (by iterated expansion into partial fractions) have denominators that factor linearly in the next integration variable. Note that when this criterion is not fulfilled initially, it can sometimes be achieved through a transformation of the integration variables (see e.g.\ \mbox{refs.\ \cite{Panzer:2014gra,Bourjaily:2018aeq,Besier:2018jen}}).

Most Feynman integrals are not expressible in terms of multiple polylogarithms at all \cite{Bourjaily:2018yfy}, and therefore cannot admit any linearly reducible parametrization. As a first step beyond polylogs, some Feynman integrals have recently been evaluated in terms of elliptic polylogarithms; e.g.\ \cite{brown2011multiple,Adams:2017tga,Adams:2017ejb,Broedel:2014vla,Broedel:2015hia,Broedel:2017jdo,Bourjaily:2017bsb}.
We expect that the double-pentagon described in section~\ref{subsubsec:feynman_parameter_ints_at_two_loop} also admits such a representation. In fact, already the regulated double-box integral \eqref{final_elliptic_example} described at the end of section~\ref{subsubsec:exempli_gratia} is elliptic for non-zero $\eps$ (see e.g.\ \mbox{refs.\ \cite{CaronHuot:2012ab,Bourjaily:2017bsb,Bourjaily:2018ycu}}). 

However, our prime interest here are the two-loop hexagon integrals defined in \eqref{individual_integrands_defined}, and in fact all of them turn out to be linearly reducible out of the box (without changes of the integration variables). Moreover, this in fact holds for the regulated integrals as functions of $\eps$. We therefore did not apply the procedure illustrated in the previous section and, in the first step, kept the full dependence on $\eps$.

The integrals are then obtained as linear combinations of multiple polylogarithms~\\[-8pt]
\vspace{-4pt}\begin{equation*}
	\Li{n_1,\ldots,n_d}{f_1,\ldots,f_d} = \sum_{0<k_1<\cdots<k_d} \frac{f_1^{k_1} \cdots f_d^{k_d}}{k_1^{n_1} \cdots k_d^{n_d}}\vspace{-3pt}
\end{equation*}
whose arguments $f_i\!=\!f_i(u_1,u_2,u_3,\eps)$ are algebraic functions of cross-ratios and $\eps$. If we expand at $\eps\!=\!0$, such polylogarithms become polynomials in $\logdiv$ with coefficients that are multiple polylogarithms whose arguments are algebraic functions of the cross-ratios alone, plus terms that vanish at $\eps\!=\!0$. For example,
\vspace{-4pt}\begin{align*}
	\Li{1}{\sqrt{1\mi2\eps \Delta_6\pl\eps^3 \v}} 
	&= \mi\log\!\big( 1\mi \sqrt{1\mi2\eps \Delta_6\pl\eps^3 \v}\big)
	= \mi\log\left(\eps \Delta_6 + \mathcal{O}(\eps^2) \right) \\
	&= \mi \logdiv \mi \log\!\big(\sqrt{(1\mi\u\mi\v\mi\w)^2\mi4\u\v\w}\big) \pl \mathcal{O}(\eps).\\[-24pt]~
\end{align*}
In some cases, there exist parametrizations of the cross-ratios such that these arguments become rational functions of the parameters. This is convenient because polylogarithms with such arguments can be expanded into a basis of hyperlogarithms~\\[-8pt]
\eq{\int_{0<t_1<\cdots<t_n<z} \frac{\td t_1}{t_1-f_1} \cdots \frac{\td t_n}{t_n-f_n}.\label{sum_rep_of_hyperlogs}}

In the case of six particles, several such parametrizations are known--for example, the $y$ variables \cite{Dixon:2011pw}, hedgehog variables \cite{Parker:2015cia} or related cluster variables \cite{Golden:2013xva}. In any of these variables, $\Delta_6$ defined in \eqref{delta6_defined} is a perfect-square. In our calculation, we used coordinates on the moduli space $\mathfrak{M}_{0,6}$ as defined in \eqref{txy_definition}. For a recent study of rationalizing parametrizations beyond six particles, we refer to \cite{Bourjaily:2018aeq}.

\vspace{-2pt}\section[{Local Integrands for Loop Amplitudes, Ratio Functions, etc.}]{Review: Local Integrands for Loop Amplitudes, etc.}\label{sec:loop_integrands}\vspace{-6pt}

In this section, we review the ingredients required to represent loop amplitudes and ratio functions at the {integrand}-level according to generalized (or prescriptive) unitarity. While most of this section will consist of a rapid discussion of well-established ideas, the key notation and ingredients needed for various representations of one and two-loop amplitudes will be important to us when we discuss concrete examples and applications; we hope this review will establish all the necessary formalism required for the present work.

Powerful methods now exist to construct local, integrand-level representations of perturbative scattering amplitudes in a wide class of quantum field theories. Among the most universally applicable of these is generalized unitarity \cite{Bern:1994zx,Bern:1994cg,Britto:2004nc,Anastasiou:2006jv,Bern:2007ct,Cachazo:2008vp,Berger:2008sj,Abreu:2017xsl}, which follows from the observation that loop {\it integrands} are rational functions and therefore can be \mbox{(re-)}constructed from their residues (or `cuts')---singularities where internal propagators go on-shell. Indeed, for any perturbative amplitude in {\it any} quantum field theory, loop integrands can be expanded into an arbitrary (complete) basis of ordinary, reference Feynman integrands with coefficients determined by cuts. Although the size and complexity of the basis required depends strongly on the details of the quantum field theory in question and the dimension in which it is defined, there is a straightforward method by which the coefficients of integrands in the basis may be found for any amplitude (see e.g.\ \mbox{refs.\ \cite{Bourjaily:2015jna,Bourjaily:2017wjl,Bourjaily:2016evz,Bourjaily:2018omh}} for more recent work).

A {\it prescriptive} representation is one for which the coefficient of every integrand in the basis is a specific field-theory cut---one for which every other integrand in the basis vanishes. While the details of prescriptive unitarity will not be important to us here, our primary examples at one and two loops will come from this framework. Nevertheless, it is worth emphasizing that nothing about our results depends on the details (nor even the existence) of a prescriptive representation for loop integrands.

Of course, there are many other attractive frameworks to represent and compute loop amplitude integrands (especially in the case of planar maximally-supersymmetric, $\mathcal{N}\!=\!4$, Yang-Mills theory (SYM)) including loop-level recursion relations \cite{ArkaniHamed:2010kv,Benincasa:2015zna}, $Q$-cuts \cite{Baadsgaard:2015twa}, etc. However, here we will focus our attention on {\it local} integrand representations obtained using unitarity-based methods because the integration (and regularization) of integrands involving non-local (or linear) propagators remains an important open problem (but see, e.g.,\ \cite{Lipstein:2012vs,Lipstein:2013xra}).

Before starting our review, we should clarify one potentially confusing point about loop integrands for amplitudes. When we speak of `the' loop integrand for a particular scattering amplitude, we mean a rational function that could be obtained as the sum of Feynman diagrams. (For non-planar quantum field theories or those with ultraviolet divergences, this requires some additional clarification, as `the sum of Feynman diagrams' would itself need clarification---scheme dependence, the routing of loop momenta, etc. But for such cases, by `the' loop integrand, we mean any particular representative obtainable from Feynman diagrams.) What we {\it do not mean} is an integrand-level representation that is merely guaranteed to integrate to the same result. That is, we will not make use of integration-by-parts (IBP) or other post-integration identities to express loop amplitudes in terms of what are sometimes called `master integrals'. Starting from the loop integrands equal to those obtainable from the Feynman expansion, it would be reasonable (and probably worthwhile) to make use of integral-level identities such as IBPs to reduce the number of integrations required. We merely emphasize this distinction in order to clarify that these simplifications are not {\it necessary} for dual-conformal regularization or any of the examples we describe below.

\vspace{-2pt}\subsection{Local Integrands for One-Loop Amplitudes and Ratio Functions}\label{subsec:one_loop_amplitude_integrands}\vspace{-6pt}

Local integrand/integral-level representations for general one-loop amplitudes have been known for a long time. Indeed, a complete basis of one-loop integrands for any particular quantum field theory (in any number of dimensions) is not hard to construct. In terms of such a basis, amplitudes would be represented with loop-independent coefficients determined from tree-level data (see e.g.\ \mbox{refs.\ \cite{Bern:1994cg,Bern:2007dw,Bourjaily:2013mma,Bourjaily:2015jna,Bourjaily:2017wjl}}). Like the amplitudes being represented, many of the integrands in such a basis will be infrared or ultraviolet divergent upon integration and therefore must be regulated. Once a particular scheme is chosen, all the integrands in the basis can readily be integrated and the results tabulated---re-usable for a broad class of scattering processes. Such tabulated expressions---for a variety of different regularization schemes---can be found in many places in the literature (see e.g.\ \cite{Bern:1992em,ArkaniHamed:2008gz,Hodges:2010kq,Mason:2010pg,Bourjaily:2013mma}).

It is well known that maximally supersymmetric theories\footnote{This statement is completely independent of planarity: all one-loop diagrams are planar with respect to some ordering of the external legs.} in four dimensions obey the `no triangle hypothesis' at one loop \cite{Bern:2005bb,BjerrumBohr:2006yw}. What this means is that the space of one-loop integrands with the same (or better) ultraviolet behavior as a scalar box integral will suffice to represent all amplitudes in such a theory. This `scalar box' power-counting means that we need only consider the space of integrands involving $p\!\geq\!4$ (loop-dependent) propagators, with at most $(p\mi4)$ products of inverse propagators in the numerator. A classic result from Passarino and Veltman \cite{Passarino:1978jh} (see also \cite{OPP}) is that all such integrands can be expanded into a basis of {\it scalar} `boxes'---those with 4 propagators and loop-independent numerators---and (some independent subset of) parity-odd `pentagons' which each integrate to zero (because the Feynman loop-integral contour is parity-even). Let $\{\mathcal{I}_{a,b,c,d,e},\mathcal{I}_{a,b,c,d}\}$ a complete basis of parity-odd pentagon and scalar box loop integrands.

We will denote the $L$-loop, $n$-particle N$^{k}$MHV scattering amplitude integrand in planar SYM by $\mathcal{A}_n^{(k),L}$. Non-calligraphic letters will be used for integrated expressions. Then the discussion above implies that any one-loop integrand in planar SYM can be represented by
\eq{\mathcal{A}_n^{(k),1}=\sum_{a,b,c,d,e}f_{a,b,c,d,e}\mathcal{I}_{a,b,c,d,e}+\sum_{a,b,c,d}f_{a,b,c,d}\mathcal{I}_{a,b,c,d}\,.\label{box_and_pentagon_representation}}
The coefficients in this expansion are loop-independent combinations of maximal co-dimension residues (`leading singularities' \cite{Cachazo:2008vp})---possibly normalized by the residues of the integrands in the basis. It is common to rescale the integrands in the basis to have unit-magnitude maximal co-dimension residues; if the integrands are so-normalized, then each turns out to be {\it dual-conformally invariant}---a symmetry that was in fact discovered in this context \cite{Drummond:2006rz,Alday:2007hr,Drummond:2007aua,Drummond:2008vq}.

Scalar box integrands in dual space can be expressed as
\eq{\mathcal{I}_{a,b,c,d}\propto\frac{\x{a}{c}\x{b}{d}}{\x{\ell}{a}\x{\ell}{b}\x{\ell}{c}\x{\ell}{d}}\,,\label{schematic_one_loop_scalar_box}}
where the loop-independent (and dual-conformally invariant) `constant' of proportionality is determined by the condition all the co-dimension four residues of (\ref{schematic_one_loop_scalar_box}) are unit in magnitude. This normalization factor will of course depend on the external momenta; but its precise form will not be important to us now. 

Given that every integrand in the basis is manifestly dual-conformally invariant, it is natural to wonder if this symmetry has any meaning for amplitudes. The first problem is that amplitudes for theories with massless particles are not meaningfully defined without regularization, and the most familiar schemes---e.g. dimensional \cite{tHooft:1972tcz} or Higgs \cite{Alday:2009zm} regularization---severely break dual-conformal symmetry. Nevertheless, it was long-ago suspected that {\it finite} observables in planar SYM, such as ratio or remainder functions, would be dual-conformally invariant \cite{Drummond:2006rz,Drummond:2008vq,Drummond:2011ic}.

Because the infrared divergences of amplitudes involving massless particles are universal (see e.g.\ \cite{Bern:2005iz})---in particular, helicity independent---the ratio of any two amplitudes (with fixed multiplicity and external momenta, but different helicities) will be infrared-finite. The {\it ratio function} is defined as the ratio of an amplitude to the `(N$^{k\!=\!0}$)MHV' amplitude, represented perturbatively according to
\eq{\mathcal{R}_n^{(k)}\equiv\frac{\mathcal{A}_n^{(k)}}{\mathcal{A}_n^{(0)}}=\frac{\mathcal{A}_n^{(k),0}\!\!+a\mathcal{A}_n^{(k),1}\!\!+a^2\mathcal{A}_n^{(k),2}\!\!+\ldots}{\mathcal{A}_n^{(0),0}\!\!+a\mathcal{A}_n^{(0),1}\!\!+a^2\mathcal{A}_n^{(0),2}\!\!+\ldots}\equiv\mathcal{R}_n^{(k),0}\!+a\mathcal{R}_n^{(k),1}\!+a^2\mathcal{R}_n^{(k),2}+\ldots\,.\label{perturbative_ratio_function_definition}}
Here we are using the coupling constant $a\!=\!g^2N_c/(8\pi^2)$.
It is conventional to define all amplitudes relative to the MHV tree-amplitude (which is in fact the identity in momentum-twistor variables \cite{Mason:2009qx,ArkaniHamed:2009vw}); thus, we will use $\mathcal{R}_n^{(k),0}$ and $\mathcal{A}_n^{(k),0}$ interchangeably (with $\mathcal{A}_{n}^{(0),0}\!\equiv\!1$ being understood). From (\ref{perturbative_ratio_function_definition}), the one-loop ratio function is simply given by
\eq{\mathcal{R}_n^{(k),1}\equiv\mathcal{A}_n^{(k),1}-\mathcal{A}_n^{(k),0}\mathcal{A}_n^{(0),1}\,.\label{one_loop_ratio_function}}

Clearly, if the one-loop amplitudes appearing in (\ref{one_loop_ratio_function}) are represented according to (\ref{box_and_pentagon_representation}) and integrated using dimensional or mass regularization, all the terms in (\ref{one_loop_ratio_function}) will deeply obscure any potential conformal invariance of the combination. (See \mbox{ref.\ \cite{Elvang:2009ya}} for early work on making this fact manifest in conventional regularization schemes.) However, using our regulator in equation (\ref{momentum_definition_of_dci_regulator}) we can keep the dual conformal invariance manifest, while slightly obscuring finiteness. The fact that infrared divergences cancel in the ratio function follows then from the (quite non-trivial) observation that in the dual-conformally regulated expression for a one-loop amplitude, all $\logkdiv{k}$ terms are proportional to the tree amplitude. If we let
\eq{A_{n}^{(k),1}\equiv\!\int\!\!d^4\ell\,\mathcal{A}_n^{(k),1}\equiv A^{(k),1}_{n,\text{fin}}+{\color{divblue}A^{(k),1}_{n,\text{div}^1}\logdiv}+{\color{divblue}A^{(k),1}_{n,\text{div}^2}\logsquarediv}\,,}
then we find that \cite{Bourjaily:2013mma}
\eq{{\color{divblue}A^{(k),1}_{n,\text{div}^1}}=-2 \mathcal{A}^{(k),0}_{n}\log\!\left(\!\prod_{a}\frac{\x{a}{a\,\pl2}}{\x{a}{a\,\pl3}}\!\right)\quad\text{and}\quad {\color{divblue}A^{(k),1}_{n,\text{div}^2}}=-n\,\mathcal{A}^{(k),0}_{n}\,.\label{one_loop_divergences_dci_regulator}}

In \mbox{ref.\ \cite{Bourjaily:2013mma}} another representation of one-loop integrands was described which renders the finiteness of the ratio function manifest at the cost of obscuring its dual-conformal invariance. While less familiar, we briefly review that representation here because it will play an important role in the concrete examples discussed later.

It is easy to prove that the space of loop integrands bounded by some degree of ultraviolet behavior is always a (strict) subspace of integrands with worse ultraviolet behavior. In particular, all one-loop integrands with scalar box power-counting can be expanded in terms of those with `scalar triangle' power-counting. Moreover, the triangle power-counting basis requires no pentagon integrals (see e.g.\ \cite{Bourjaily:2017wjl}): all the pentagons in \eqref{box_and_pentagon_representation} are expressible in terms of box integrands with loop {\it dependent} numerators and scalar triangles.

The basis of one-loop integrands with scalar triangle power-counting would consist of terms of the form
\eq{\mathcal{I}^i_{a,b,c,d}\sim\frac{\x{\ell}{N^i}}{\x{\ell}{a}{\x{\ell}{b}\x{\ell}{c}\x{\ell}{d}}}\quad\mathrm{and}\quad\mathcal{I}_{a,b,c}\sim\frac{N}{\x{\ell}{a}\x{\ell}{b}\x{\ell}{c}}\,.\label{scalar_triangle_basis}}
Here, the index `$i$' on the box integrals is used to distinguish the (in this case 2) linearly-independent functions of $\ell$ that can appear in the numerator. In contrast to the scalar box basis where an independent subset of pentagon integrands must be chosen, the scalar triangle basis is not over-complete---thus, integrands in (\ref{scalar_triangle_basis}) have unique coefficients. 

The conformal breaking of the integrands in \eqref{scalar_triangle_basis} can be mitigated slightly by making manifest the fact that these integrals have residues supported at infinity in loop-momentum space. Of course, infinity is just another point in dual-coordinates, which we may denote by $X\!\equiv\!x_{\infty}$. Making the residues supported on \mbox{$x_{\ell}\!\to\!x_{\infty}$} manifest, we may write our basis \eqref{scalar_triangle_basis} in terms of the more conformal-looking functions
\eq{\mathcal{I}^i_{a,b,c,d}(X)\equiv\frac{\x{\ell}{N^i}\x{Y^i}{X}}{\x{\ell}{a}\x{\ell}{b}\x{\ell}{c}\x{\ell}{d}\x{\ell}{X}},\quad\mathcal{I}_{a,b,c}(X)\equiv\frac{\x{X}{N}}{\x{\ell}{a}\x{\ell}{b}\x{\ell}{c}\x{\ell}{X}}\,.\label{x_dependent_basis}}
As before, the numerators appearing in this basis are fixed so that these integrands have unit-magnitude residues on all cuts with maximal co-dimension\footnote{Actually, the requirement of unit leading singularities is not strong enough to fix the numerators completely. The remaining freedom can be eliminated by additionally imposing that no integral have parity-even support on cuts involving $\x{\ell}{X}\!=\!0$. See \mbox{ref.\ \cite{Bourjaily:2013mma}} for more details.}, which determines the  $N^i$ and $Y^i$ to be conjugate solutions of the maximal cut. Explicit expressions for the numerators in (\ref{x_dependent_basis}) will be given when they are needed in \mbox{section \ref{sec:six_point_example}}.

Including the additional $X$-dependent propagators in (\ref{x_dependent_basis}) relative to (\ref{scalar_triangle_basis}) renders them `dual-conformal' with respect to $\ell$ and every other dual-momentum coordinate $x_a$---but now also including $(x_{\infty}\!\equiv\!)X$; this explicit $X$-dependence is the (now only remaining) indication of dual-conformal symmetry breaking.

Despite the fact that this basis spoils dual-conformal invariance, there are at least two reasons why this representation of amplitudes is interesting. The first is that this representation is {\it prescriptive}: every integrand has precisely one cut not shared by any other integrand in the basis, which therefore determines its coefficient. The `scalar box' integrand representation in (\ref{box_and_pentagon_representation}) does not meet this criterion because the pentagon integrals have many cuts supported by the boxes. (However, as the coefficients of pentagons are irrelevant upon integration, this point can reasonably be viewed as merely an academic distinction.)

Because the triangle power-counting basis is prescriptive, any one-loop integrand can be represented uniquely as
\eq{\mathcal{A}_n^{(k),1}=\sum_{a,b,c,d}\Big(f^1_{a,b,c,d}\mathcal{I}^1_{a,b,c,d}+f^2_{a,b,c,d}\mathcal{I}^2_{a,b,c,d}\Big)+\sum_{a,b,c} f_{a,b,c}\mathcal{I}_{a,b,c}\,,\label{chiral_integrand_representation}}
where every coefficient $f^i_{a,b,c,d}$ and $f_{a,b,c}$ is a specific co-dimension four residue of the amplitude. Moreover, it turns out that the only co-dimension four residues supported on three propagators are those `composite' residues corresponding to soft-collinear divergences---for which the residue must always be the tree amplitude by general considerations of infrared structure. That is, the only non-vanishing scalar triangle integrand coefficients needed are those of the form $f_{a-1,a,a+1}$ (for each $a$); and all of these are equal to the tree amplitude: $f_{a-1,a,a+1}\!=\!\mathcal{A}_n^{(k),0}$. Thus,
\eq{\sum_{a,b,c} f_{a,b,c}\mathcal{I}_{a,b,c}\equiv\mathcal{A}_{n}^{(k),0}\!\!\times\!\mathcal{I}_{\text{div}}\quad\mathrm{where}\quad \mathcal{I}_{\text{div}}\equiv\sum_{a=1}^n\frac{\x{X}{a}\x{a\mi1}{a\,\pl1}}{\x{\ell}{a\mi1}\x{\ell}{a}\x{\ell}{a\,\pl1}\x{\ell}{X}}\,.\label{triangle_divergence_identificaiton}}
Here we have introduced the subscript `div' because it turns out that these are the {\it only} infrared-divergent integrands that appear in (\ref{chiral_integrand_representation})! Thus, this representation naturally divides any amplitude according to:
\eq{\mathcal{A}_n^{(k),1}\equiv\mathcal{A}_{n,\text{fin}}^{(k),1}+\mathcal{A}_{n,\text{div}}^{(k),1}\,,\label{one_loop_finite_and_div_split}\vspace{-4pt}}
where
\eq{\mathcal{A}_{n,\text{fin}}^{(k),1}\equiv\sum_{a,b,c,d}\Big(f^1_{a,b,c,d}\mathcal{I}^1_{a,b,c,d}+f^2_{a,b,c,d}\mathcal{I}^2_{a,b,c,d}\Big)\quad\mathrm{and}\quad\mathcal{A}_{n,\text{div}}^{(k),1}\equiv\mathcal{A}_{n}^{(k),0}\!\!\times\!\mathcal{I}_{\text{div}}\,.\label{one_loop_chiral_terms}}

Notice that because in this representation the infrared-divergent parts of any amplitude are {\it manifestly} proportional to the tree, these divergences {\it manifestly} cancel in the one-loop ratio function, (\ref{one_loop_ratio_function}):
\eq{\begin{split}\mathcal{R}_n^{(k),1}&=\mathcal{A}_{n,\text{fin}}^{(k),1}+\mathcal{A}_{n}^{(k),0}\mathcal{I}_{\text{div}}-\mathcal{A}_{n}^{(k),0}\Big(\mathcal{A}_{n,\text{fin}}^{(k),1}+\mathcal{I}_{\text{div}}\Big) \,\\
&=\mathcal{A}_{n,\text{fin}}^{(k),1}-\mathcal{A}_n^{(k),0}\mathcal{A}_{n,\mathrm{fin}}^{(0),1}\,.\end{split}}
(Recall that $\mathcal{A}_{n}^{(0),0}$ is taken to be the identity.) Thus, this representation makes manifest the finiteness of the one-loop ratio function, but at the cost of obscuring its dual-conformal invariance. (For those readers interested in more details, explicit ($X$-dependent) analytic expressions for all the integrals appearing in (\ref{one_loop_chiral_terms}) can be found in \mbox{ref.\ \cite{Bourjaily:2013mma}}.)

(There does exist a complete basis of one-loop, manifestly dual-conformal integrands with scalar box power-counting such that no (non-vanishing) combination of individually infrared-divergent integrands is infrared-finite: the so-called `octagon' basis described in \mbox{ref.\ \cite{ArkaniHamed:2010gh}}. In this basis, the finiteness {\it and} conformal invariance of any ratio function would be made manifest; however, we choose not to review this here for two reasons: first, the octagon basis is (very) far from prescriptive, so that the coefficients of most amplitudes would require great efforts of linear algebra to discover; and secondly, because it is not known whether or not such a basis exists beyond one loop.)

It turns out that finiteness, conformal invariance, and prescriptivity are in less opposition at two loops. As we review in the next section, there exists a prescriptive, manifestly dual-conformally invariant basis of two-loop integrands which renders the exponentiation of infrared divergences manifest at the integrand-level. In this basis, two-loop ratio functions are represented in terms of integrals that are either individually finite, or constructed from pairs of one-loop convergent (integrable) integrands. The problem will be that not all two-loop integrands built from finite one-loop integrands will be finite. To understand why and how this happens, let us now review the form of two-loop integrands described in \mbox{ref.\ \cite{Bourjaily:2015jna}}.

\vspace{-2pt}\subsection{Integrands for Two-Loop Amplitudes and Ratio Functions}\label{subsec:two_loop_amplitude_integrands}\vspace{-6pt}

As mentioned above, it turns out to be much easier to find prescriptive, manifestly dual-conformal invariant representations of two-loop amplitude integrands in planar SYM than at one loop. Moreover, it is possible to preserve manifest dual-conformal invariance of all integrands while simultaneously making manifest the exponentiation (and universality) of infrared divergences at the integrand-level. Unfortunately, it turns out that this is not quite strong enough to imply the finiteness of two-loop ratio functions. Let us briefly review the representation of two-loop amplitude integrands described in \mbox{ref.\ \cite{Bourjaily:2015jna}}, and how this can be used to obtain expressions for ratio functions that are less infrared-divergent than the amplitudes involved.

In ref.~\cite{Bourjaily:2015jna}, two-loop amplitude integrands were constructed in terms of finite and divergent parts according to:
\eq{\mathcal{A}_{n}^{(k),2}\equiv\mathcal{A}_{n,\text{fin}}^{(k),2}\pl\mathcal{A}_{n,\text{div}}^{(k),2}\,,\quad\text{with}\quad\mathcal{A}_{n,\text{div}}^{(k),2}\equiv\mathcal{A}_{n,\text{fin}}^{(k),1}\merge\mathcal{I}_{\text{div}}\pl\mathcal{A}_{n}^{(k),0}\frac{1}{2}\Big(\mathcal{I}_{\text{div}}\merge\mathcal{I}_{\text{div}}\Big)\,.\label{two_loop_exponential_form}}
Here, the `merge'-product `$\merge$' is an integrand-level operation that acts on the $X$-dependent integrands appearing in the one-loop amplitudes represented according to (\ref{one_loop_chiral_terms}). Specifically, the merger of two $X$-dependent one-loop integrands is defined according to:
\eq{\begin{split}\mathcal{I}_L(\ell_1,X)\merge\mathcal{I}_{R}(\ell_2,X)&\equiv\left(\mathcal{I}'_{L}(\ell_1)\frac{\x{N_L}{X}}{\x{\ell_1}{X}}\right)\!\merge\!\left(\frac{\x{X}{N_R}}{\x{X}{\ell_2}}\mathcal{I}'_{R}(\ell_2)\right)\\&\equiv\mathcal{I}'_L(\ell_1)\frac{\x{N_L}{N_R}}{\x{\ell_1}{\ell_2}}\mathcal{I}'_{R}(\ell_2)\,.\vspace{0pt}\end{split}\label{merge_operation_detail}}
Notice that operation is symmetric---provided that symmetrization with respect to loop-momentum labels is understood. We refer the reader to \mbox{ref.\ \cite{Bourjaily:2015jna}} for more details about why this operation is useful in the construction of two-loop integrands; but the general form quoted above will suffice for our purposes. Notice that once a pair of $X$-dependent (and hence non-dual-conformally invariant) integrands are merged, the result is always strictly dual-conformally invariant (and $X$-independent).

Importantly, it can be easily proven (see \cite{Bourjaily:2015jna}) that the product of any pair of one-loop amplitude integrands is equal to their merger. That is,
\eq{\mathcal{A}_{n}^{(k_1),1}\!\mathcal{A}_{n}^{(k_2),1}=\mathcal{A}_{n}^{(k_1),1}\merge\mathcal{A}_{n}^{(k_2),1}\,.\label{product_equals_merger}}
Thus, for any two-loop ratio function (obtained from \eqref{perturbative_ratio_function_definition}), we may use \eqref{product_equals_merger} and the representation of two-loop amplitudes according to \eqref{two_loop_exponential_form} and cancel terms to see that
\vspace{-4pt}\eq{\begin{split}\hspace{-50pt}\mathcal{R}_n^{(k),2}\hspace{-0pt}&=\phantom{\mi}{\color{hred}\Big[}\mathcal{A}_{n}^{(k),2}{\color{hred}\Big]}\mi{\color{hblue}\Big[}\mathcal{A}_{n}^{(k),1}\merge\mathcal{A}_{n}^{(0),1}{\color{hblue}\Big]}\mi\mathcal{A}_{n}^{(k),0}{\color{hteal}\Big[}\mathcal{A}_{n}^{(0),2}\mi\mathcal{A}_{n}^{(0),1}\merge\mathcal{A}_{n}^{(0),1}{\color{hteal}\Big]}\\
&=\phantom{\mi}{\color{hred}\Big[}\mathcal{A}_{n,\text{fin}}^{(k),2}\hspace{0pt}\pl\mathcal{A}_{n,\text{fin}}^{(k),1}\merge\mathcal{I}_{\text{div}}\hspace{0pt}\pl\frac{1}{2}\mathcal{A}_{n}^{(k),0}\mathcal{I}_{\text{div}}\merge\mathcal{I}_{\text{div}}{\color{hred}\Big]}\\
&\phantom{=}\hspace{0.0pt}\,\,\,\mi{\color{hblue}\Big[}\Big(\mathcal{A}_{n,\text{fin}}^{(k),1}\pl\mathcal{A}_{n}^{(k),0}\mathcal{I}_{\text{div}}\Big)\merge\Big(\mathcal{A}_{n,\text{fin}}^{(0),1}\pl\mathcal{I}_{\text{div}}\Big){\color{hblue}\Big]}\\
&\phantom{=}\hspace{0.0pt}\,\,\,\mi\mathcal{A}_{n}^{(k),0}{\color{hteal}\Big[}\mathcal{A}_{n,\text{fin}}^{(0),2}\hspace{0pt}\pl\mathcal{A}_{n,\text{fin}}^{(0),1}\merge\mathcal{I}_{\text{div}}\hspace{0pt}\pl\frac{1}{2}\mathcal{I}_{\text{div}}\merge\mathcal{I}_{\text{div}}\mi\Big(\mathcal{A}_{n,\text{fin}}^{(0),1}\pl\mathcal{I}_{\text{div}}\Big)\merge\Big(\mathcal{A}_{n,\text{fin}}^{(0),1}\pl\mathcal{I}_{\text{div}}\Big){\color{hteal}\Big]}\hspace{-50pt}\\
&=\phantom{\mi}\mathcal{A}_{n,\mathrm{fin}}^{(k),2}\mi\mathcal{A}_{n,\mathrm{fin}}^{(k),1}\merge\mathcal{A}_{n,\mathrm{fin}}^{(0),1}\mi\mathcal{A}_{n}^{(k),0}\Big(\mathcal{A}_{n,\mathrm{fin}}^{(0),2}\mi\mathcal{A}_{n,\mathrm{fin}}^{(0),1}\merge\mathcal{A}_{n,\mathrm{fin}}^{(0),1}\Big)\,.\\[-22pt]~\vspace{-0pt}\end{split}\label{two_loop_ratio_function_with_mergers}\vspace{-10pt}}
This is suggestively close to implying the finiteness of all two-loop ratio functions while preserving manifest dual-conformal invariance. However, it turns out that the merger of two infrared-finite integrals need not be infrared-finite.

Perhaps the most direct way of seeing that \eqref{two_loop_ratio_function_with_mergers} cannot be term-wise infrared-finite is from the fact that the last term in \eqref{two_loop_ratio_function_with_mergers} is very close to the two-loop order of the logarithm of the MHV amplitude \cite{ArkaniHamed:2010gh},\footnote{The factor of $4$ in (\ref{log_of_mhv}) is required to match our present conventions regarding the coupling constant `$a$' (see the discussion around equation \eqref{perturbative_ratio_function_definition}).}
\vspace{-4pt}\eq{\begin{split}4\log\!\Big(\!\mathcal{A}_n^{(0)}\!\Big)\Big.^{L=2\text{}}\hspace{-0pt}&={\color{hred}\Big[}\mathcal{A}_{n}^{(0),2}{\color{hred}\Big]}\mi\frac{1}{2}{\color{hblue}\Big[}\mathcal{A}_{n}^{(0),1}\merge\mathcal{A}_{n}^{(0),1}{\color{hblue}\Big]}\\
&={\color{hred}\Big[}\mathcal{A}_{n,\text{fin}}^{(0),2}\hspace{0pt}\pl\Big(\mathcal{I}_{\text{div}}\merge\mathcal{A}_{n,\text{fin}}^{(0),1}\Big)\hspace{0pt}\pl\frac{1}{2}\Big(\mathcal{I}_{\text{div}}\merge\mathcal{I}_{\text{div}}\Big){\color{hred}\Big]}\\
&\phantom{=}\hspace{26.5pt}\,\,\,\mi\frac{1}{2}{\color{hblue}\Big[}\Big(\mathcal{A}_{n,\text{fin}}^{(0),1}\pl\mathcal{I}_{\text{div}}\Big)\merge\Big(\mathcal{A}_{n,\text{fin}}^{(0),1}\pl\mathcal{I}_{\text{div}}\Big){\color{hblue}\Big]}\\
&=\mathcal{A}_{n,\mathrm{fin}}^{(0),2}-\frac{1}{2}\mathcal{A}_{n,\mathrm{fin}}^{(0),1}\merge\mathcal{A}_{n,\mathrm{fin}}^{(0),1}\,.\\[-22pt]~\vspace{-0pt}\end{split}\label{log_of_mhv}\vspace{-10pt}}
which is $\log^2$-divergent for all $n$ (and also at all loop-orders). The last term in \eqref{two_loop_ratio_function_with_mergers} differs from \eqref{log_of_mhv} by only finite terms: letting
\eq{\mathcal{L}_n\equiv\mathcal{A}_{n,\mathrm{fin}}^{(0),2}-\mathcal{A}_{n,\mathrm{fin}}^{(0),1}\merge\mathcal{A}_{n,\mathrm{fin}}^{(0),1}\,,\vspace{-5pt}}
we have
\eq{\mathcal{L}_n=8\log\!\big(\!\mathcal{A}_n^{(0)}\!\big)-\mathcal{A}_{n,\mathrm{fin}}^{(0),2}\,.}
Thus it absolutely must be the case that the individual terms in \eqref{two_loop_ratio_function_with_mergers} include infrared divergences---which obviously must cancel in the ratio function \eqref{two_loop_ratio_function_with_mergers}.

Despite the non-manifest finiteness of \eqref{two_loop_ratio_function_with_mergers}, it is less divergent term-by-term than the individual amplitudes appearing in the first line of \eqref{two_loop_ratio_function_with_mergers}. Thus, it is a better starting point for concrete analysis. However, we should emphasize that the dual-conformal regulator described in the next section {\it could be} used to regulate the na\"{i}ve representation of the two-loop ratio function (the first line of \eqref{two_loop_ratio_function_with_mergers})---at the cost of dealing with individual terms that involve higher-order divergences.

\newpage
\vspace{-6pt}\section{Primary Illustration: Six-Point NMHV Ratio Function}\label{sec:six_point_example}\vspace{-6pt}

In this section we apply the methods and ideas discussed above to the case of the six-particle NMHV ratio function at two loops. Specifically, we will start with the local (loop-momentum) integrand-level representation for the ratio function described in equation (\ref{two_loop_ratio_function_with_mergers}), regulate all infrared-divergent contributions using dual-conformal regularization, recast each term as a manifestly dual-conformal Feynman parameter integral, and obtain analytic expressions for each contribution after (Feynman-parameter) integration in terms of weight-four hyperlogarithms. Full details of each contribution are included among the ancillary files attached to this work's submission to the {\tt arXiv}, which are thoroughly described in \mbox{Appendix \ref{appendix:explicit_integrals}}. 

Beyond merely a concrete application of the ideas described above to a specific case (which can be easily compared to existing work), this exercise will highlight some novelties about the dual-conformal regulator's use beyond one loop. For example, we will find that this regulator (or at least the particular scheme used here), neither preserves uniform transcendentality nor unit leading singularities. What this means is that even for loop integrals (defined in loop-momentum space) expected to be `pure' and have maximal transcendental-weight, these properties need not be (and in fact are not) preserved by the dual-conformal regulator. 

We will start with a detailed review of the integrand-level ingredients that contribute to the six-particle NMHV ratio function at one and two loops. After defining and enumerating the contributions required, we discuss the novel aspects of how these terms are manifested after regularization and explicit integration. In \mbox{section \ref{subsec:two_loop_log}} we describe how the remainder function is related to the logarithm of the MHV amplitude when regulated in this way. 

\vspace{-2pt}\subsection{Local Integrand Representation of the Six-Point Ratio Function}\label{subsec:six_point_specialization}\vspace{-6pt}

As reviewed in \mbox{section \ref{sec:loop_integrands}}, the two-loop six-point NMHV ratio function can be represented at the (loop-)integrand-level by,
\eq{\begin{split}\mathcal{R}_6^{(1),2}&=\mathcal{A}_{6,\mathrm{fin}}^{(1),2}-\mathcal{A}_{6,\mathrm{fin}}^{(1),1}\merge\mathcal{A}_{6,\mathrm{fin}}^{(0),1}-\mathcal{A}_{6}^{(k),0}\Big(\mathcal{A}_{6,\mathrm{fin}}^{(0),2}-\mathcal{A}_{6,\mathrm{fin}}^{(0),1}\merge\mathcal{A}_{6,\mathrm{fin}}^{(0),1}\Big)\,\\
&\equiv\mathcal{A}_{6,\mathrm{fin}}^{(1),2}-\mathcal{A}_{6,\mathrm{fin}}^{(1),1}\merge\mathcal{A}_{6,\mathrm{fin}}^{(0),1}-\mathcal{A}_{6}^{(k),0}\mathcal{L}_6\,.\end{split}\label{six_point_two_loop_ratio_v0}}
In a moment, we will define the above expression in terms of explicit (`merged') two-loop integrands constructed out of one-loop chiral ($X$-dependent) boxes. (Recall the definition of the merge operator, represented by `$\merge$', given in (\ref{merge_operation_detail}).) But for the sake of clarity, it may be helpful to first review the explicit form of the ingredients that appear in (\ref{six_point_two_loop_ratio_v0}).

Let us start with the contributions in (\ref{six_point_two_loop_ratio_v0}) related to (N$^{k=0})$MHV amplitudes. As discussed in \mbox{ref.\ \cite{Bourjaily:2013mma}}, the only non-vanishing four-propagator residues for one-loop MHV amplitudes are supported on the (chiral) `two-mass-easy' boxes, $\mathcal{I}^1_{a\,a+1,b,b+1}$, for which we introduce the shorthand,
\eq{\tme{{\color{hred}a}\,{\color{hblue}b}}\equiv\mathcal{I}^1_{{\color{hred}a},{\color{hred}a+1},{\color{hblue}b},{\color{hblue}b+1}}\equiv-\frac{\x{X}{N_{{\color{hred}a}{\color{hblue}b}}}\x{\ell}{\overline{N_{{\color{hred}a}{\color{hblue}b}}}}}{\x{\ell}{{\color{hred}a}}\x{\ell}{{\color{hred}a\,\pl1}}\x{\ell}{{\color{hblue}b}}\x{\ell}{{\color{hblue}b\,\pl1}}\x{\ell}{X}}\,,\label{chiral_2me_1_defined}}
with the dual points appearing in the numerator corresponding to those related to lines defined in momentum-twistor space (see \mbox{Appendix \ref{appendix:conventions_and_definitions}}) as,
\eq{N_{ab}\equiv(ab),\qquad\overline{N_{ab}}\equiv\overline{(ab)}\equiv (a\,\mi1\,a\,a\,\pl1)\tcap(b\,\mi1\,b\,b\,\pl1)\,.\label{2me_numerators}}
In terms of these chiral loop integrands, the (finite parts of the) one- and two-loop MHV amplitudes are given by,
\eq{\begin{split}\mathcal{A}_{6,\mathrm{fin}}^{(0),1}&=\big(\tme{1\,3}+\tme{2\,4}+\tme{3\,5}+\tme{4\,6}+\tme{5\,1}+\tme{6\,2}\big)+\big(\tme{1\,4}+\tme{2\,5}+\tme{3\,6}\big)\,,\\
&=\Big[\tme{1\,3}+\frac{1}{2}\tme{1\,4}+\text{cyclic}_6\Big]\,;\\
\mathcal{A}_{6,\mathrm{fin}}^{(0),2}&=\tme{1\,3}\merge\tme{4\,6}+\tme{2\,4}\merge\tme{5\,1}+\tme{6\,2}\merge\tme{3\,5}\,,\\
&=\frac{1}{2}\Big[\tme{1\,3}\merge\tme{4\,6}+\text{cyclic}_6\Big]\,.
\end{split}\label{finite_mhv_parts}}
Notice that using the explicit form of $\tme{a\,b}$ in (\ref{chiral_2me_1_defined}) and the definition of the merge operation (\ref{merge_operation_detail}) the integrand `$\tme{1\,3}\merge\tme{4\,6}$' appearing in (\ref{finite_mhv_parts}) corresponds {\it precisely} to that of $\Omega^{(2)}(\u,\v,\w)$ defined in equation (\ref{omega2_and_omega2tilde_integrands}). 

For the parts in the two-loop ratio function (\ref{six_point_two_loop_ratio_v0}) explicitly related to NMHV ($k\!=\!1$) amplitudes, more building blocks are required. In particular, we will need to disentangle loop-integrand contributions from their loop-independent pre-factors---which in this case, are always simply the familiar `$R$-invariants' \cite{Drummond:2011ic}, for which we define the short-hand `$\r{1}$' for $R[2,3,4,5,6]$, etc.; in momentum-twistor variables, this corresponds to the superfunction
\eq{\r{1}\!\equiv\!R[2,3,4,5,6]\!\equiv\!\frac{\delta^{1\times4}\big(\ab{3456}\eta_2\pl\ab{4562}\eta_3\pl\ab{5623}\eta_4\pl\ab{6234}\eta_5\pl\ab{2345}\eta_6\big)}{\ab{3456}\ab{4562}\ab{5623}\ab{6234}\ab{2345}}\,.\label{r_invariant_defined}}
In terms of these, the NMHV tree amplitude, $\mathcal{A}_{6}^{(1),0}$ would be given by \cite{Drummond:2011ic}
\eq{\begin{split}\mathcal{A}_{6}^{(1),0}&=\r{1}+\r{3}+\r{5}=\r{2}+\r{4}+\r{6}\,\\&=\frac{1}{2}\Big[\r{1}+\text{cyclic}_6\Big]\,.\end{split}\label{six_pt_tree}}
(Functions such as these have been implemented efficiently in open source packages including those of \cite{Bourjaily:2010wh,Dixon:2010ik}.)

To describe the (finite-parts of the) one- and two-loop NMHV amplitudes, $\mathcal{A}_{6,\text{fin}}^{(1),L}$, we require two more examples of chiral one-loop integrands in addition to the two-mass-easy box defined above in (\ref{chiral_2me_1_defined}). Specifically, we also need
\eq{\tmec{{\color{hred}a}\,{\color{hblue}b}}\equiv\mathcal{I}^2_{{\color{hred}a},{\color{hred}a+1},{\color{hblue}b},{\color{hblue}b+1}}\equiv-\frac{\x{X}{\overline{N_{{\color{hred}a}{\color{hblue}b}}}}\x{\ell}{N_{{\color{hred}a}{\color{hblue}b}}}}{\x{\ell}{{\color{hred}a}}\x{\ell}{{\color{hred}a\,\pl1}}\x{\ell}{{\color{hblue}b}}\x{\ell}{{\color{hblue}b\,\pl1}}\x{\ell}{X}}\,\label{chiral_2me_2_defined}}
with numerators given in (\ref{2me_numerators}), and also (chiral versions of) the so-called `two-mass-hard' boxes---e.g.,
\vspace{-0pt}\eq{\mathcal{I}^1_{2346}\equiv\frac{\x{X}{Y(\ell)}}{\x{\ell}{2}\x{\ell}{3}\x{\ell}{4}\x{\ell}{6}\x{\ell}{X}}\,\vspace{-0pt}\label{chiral_2mh_box_defined}}
where the numerator above is given (somewhat indirectly) in twistor-space by \cite{ArkaniHamed:2010gh},
\eq{Y(\ell)\equiv\frac{1}{2}\Big((256)\tncap\big(34(12)\tncap(\ell3)\big)-(234)\tncap\big(56(12)\tncap(\ell3)\big)\Big)\,.\label{numerator_in_twistor_space}}
(This numerator is not immediately in the form required by the definition of the merge operation (\ref{merge_operation_detail}); we refer the reader to \mbox{ref.\ \cite{Bourjaily:2015jna}} for more explicit details.) The parity-conjugate of the integrand (\ref{chiral_2mh_box_defined}) is (a rotation of) $\mathcal{I}^{2}_{5613}$, which is also needed in the representation of the one-loop NMHV integrand. However, it turns out that these two, which differ by the parity of their chiral numerators, are more simply related by a diagrammatic reflection. Let us define the operator `$p$' to be a reflection centered on the external particle labelled by $a\!=\!1$; in terms of this,
\eq{\mathcal{I}^2_{5613}\equiv p\Big[\mathcal{I}^1_{2346}\Big]\,.}%
Concretely, in $x$-coordinates or momentum-twistor variables, $p$ acts according to:
\eq{\begin{split}&p\!:\!\!\{x_1,\!x_2,\!x_3,\!x_4,\!x_5,\!x_6\}\!\mapsto\!\{x_2,\!x_1,\!x_6,\!x_5,\!x_4,\!x_3\}\,,\\
&p\!:\!\!\{z_1,z_2,z_3,\!z_4,z_5,z_6\}\!\mapsto\!\{z_1,z_6,z_5,\!z_4,z_3,z_2\}\,.\end{split}\label{action_of_p_on_xs_and_zs}}
(Here, we have given the action of $p$ on the twistor variables, $z_a$, both because these appear in expressions such as (\ref{numerator_in_twistor_space}) and also because these indices coincide with external momentum labels.)
 For future reference, it is worth noting that $p$ acts on the six-point cross-ratios $u_a$ defined in (\ref{definition_of_uvw_body}) according to \mbox{$p\!\!:\!\!\{{\color{hred}\u},{\color{black}\v},{\color{hblue}\w}\}\!\mapsto\!\{{\color{hblue}\w},\v,{\color{hred}\u}\}$}.

Given these building-blocks, the pieces of (\ref{six_point_two_loop_ratio_v0}) involving NMHV amplitude integrands are given by \cite{Bourjaily:2015jna}
\eq{\begin{split}\mathcal{A}_{6,\mathrm{fin}}^{(1),1}&=\Big[\r{1}\Big(\tme{6\,2}+\tmec{3\,5}+\big(1\pl p\big)\mathcal{I}_{2346}^1\Big)+\mathrm{cyclic}_6\Big]\,;\\\mathcal{A}_{6,\mathrm{fin}}^{(1),2}&=\Big[\r{1}\,\tme{6\,2}\merge\tmec{3\,5}+\mathrm{cyclic}_6\Big]\,.\end{split}\label{nmhv_parts}}

Before combining terms, notice that $\tme{6\,2}\merge\tmec{3\,5}$ appearing in $\mathcal{A}_{6,\text{fin}}^{(1),2}$ corresponds to the integrand of \mbox{$\tilde{\Omega}^{(2)}(\w,\u,\v)\!\equiv\!r^{-1}\big[\tilde{\Omega}^{(2)}\big]$} defined in (\ref{omega2_and_omega2tilde_integrands}); moreover, we may observe that this term {\it also} arises as a contribution in $\mathcal{A}_{6,\text{fin}}^{(1),1}\merge\mathcal{A}_{6,\text{fin}}^{(0),1}$, canceling out from the representation of the ratio function in (\ref{six_point_two_loop_ratio_v0}). 

Combining all the pieces discussed above, we find that the two-loop six-point NMHV ratio function may be represented at the integrand-level according to
\begin{align}\mathcal{R}_6^{(1),2}&\equiv\mathcal{A}_{6,\mathrm{fin}}^{(1),2}-\mathcal{A}_{6,\mathrm{fin}}^{(1),1}\merge\mathcal{A}_{6,\mathrm{fin}}^{(0),1}-\mathcal{A}_{6}^{(k),0}\mathcal{L}_6\,\nonumber\\
&=\Bigg[\r{1}\Bigg(\tme{6\,2}\merge\tmec{3\,5}-\big[\tme{6\,2}+\tmec{3\,5}+\big(1\pl p\big)\mathcal{I}_{2346}^1\big]\merge\mathcal{A}_{6,\text{fin}}^{(0),1}-\frac{1}{2}\mathcal{L}_6\Bigg)+\text{cyclic}_6\Bigg]\,\nonumber\\
&\equiv-\Bigg[\r{1}\Bigg(\frac{1}{2}\mathcal{L}_6+\sum_{i=1}^{15}\mathcal{I}_i\Bigg)+\text{cyclic}_6\Bigg]\,,\label{six_point_two_loop_ratio_detail}\end{align}
where the individual integrands $\mathcal{I}_i$ appearing in (\ref{six_point_two_loop_ratio_detail}) have been defined as follows:
\eq{\hspace{-0pt}\begin{array}{l@{}c@{}l@{$\;\;$}r@{}c@{}l@{$\;\;$}l@{}c@{}l@{$\;\;$}l@{}c@{}l}\mathcal{I}_1&\equiv&\tme{62}\merge\tme{35}&\mathcal{I}_5&\equiv&(1\pl p)\mathcal{I}^1_{2346}\merge\tme{46}&\mathcal{I}_9&\equiv&(1\pl p)\mathcal{I}^1_{2346}\merge\tme{36}&\mathcal{I}_{13}&\equiv&\tme{35}\merge\tmec{35}\\\mathcal{I}_2&\equiv&(1\pl p)\mathcal{I}^1_{2346}\merge\tme{24}&\mathcal{I}_6&\equiv&(1\pl p)\mathcal{I}^1_{2346}\merge\tme{51}&\mathcal{I}_{10}&\equiv&(1\pl p)\tme{13}\merge\tmec{35}&\mathcal{I}_{14}&\equiv&\tme{62}\merge\tme{14}\\
\mathcal{I}_3&\equiv&(1\pl p)\mathcal{I}^1_{2346}\merge\tme{62}&\mathcal{I}_7&\equiv&(1\pl p)\mathcal{I}^1_{2346}\merge\tme{14}&\mathcal{I}_{11}&\equiv&(1\pl p)\tme{25}\merge\tmec{35}&\mathcal{I}_{15}&\equiv&(1\pl r)\tme{62}\merge\tme{51}\\\mathcal{I}_4&\equiv&(1\pl p)\mathcal{I}^1_{2346}\merge\tme{25}&\mathcal{I}_8&\equiv&(1\pl p)\mathcal{I}^1_{2346}\merge\tme{35}&\mathcal{I}_{12}&\equiv&(1\pl p)\mathcal{I}^1_{2346}\merge\tme{13}&\mathcal{I}_{16}&\equiv&\tme{14}\merge\tme{36}\end{array}\hspace{-10pt}\label{individual_integrands_defined}}
Notice that the summation in (\ref{six_point_two_loop_ratio_detail}) involves {\it only} integrands $\{\mathcal{I}_{1},\ldots,\mathcal{I}_{15}\}$---the integral $\mathcal{I}_{16}$ defined in (\ref{individual_integrands_defined}) is relevant only for $\mathcal{L}_6$. Finally, we should clarify that for those integrands in (\ref{six_point_two_loop_ratio_detail}) defined in combination with a reflection `$(1\pl p)$', $\{\mathcal{I}_{2},\ldots,\mathcal{I}_{12}\}$, the reflection acts on the entire integrand {\it after} merging the two parts.

In terms of these building blocks, $\mathcal{L}_6$ would be given by
\eq{\begin{split}\mathcal{L}_6&\equiv \mathcal{A}_{6,\mathrm{fin}}^{(0),2}-\mathcal{A}_{6,\mathrm{fin}}^{(0),1}\merge\mathcal{A}_{6,\mathrm{fin}}^{(0),1}\,\\
&=-\Big[\frac{1}{2}\mathcal{I}_{1}+2\,\mathcal{I}_{14}+\mathcal{I}_{15}+\mathcal{I}_{16}+\text{cyclic}_6\Big]\,.\end{split}\label{calL_in_terms_of_ints}}
As discussed above, $\mathcal{L}_6$ is related to the (two-loop-)logarithm of the MHV amplitude,
\eq{\mathcal{L}_6=8\log\!\Big(\!\mathcal{A}_6^{(0)}\!\Big)-\mathcal{A}_{6,\mathrm{fin}}^{(0),2}\,,}
from which we see that
\eq{4\log\!\Big(\!\mathcal{A}_6^{(0)}\!\Big)=-\Big(\mathcal{I}_{14}+\frac{1}{2}\mathcal{I}_{15}+\frac{1}{2}\mathcal{I}_{16}+\text{cyclic}_6\Big)\,.\label{two_loop_logarithm_integrand_terms}}

Traditionally, the (integrated) six-point ratio function is expressed in terms of two functions $V^{(L)}$ and $\tilde{V}^{(L)}$, differing according to whether or not they change sign under a three-fold rotation:
\eq{R_6^{(1),L}\equiv\frac{1}{2}\left[\Big(\!\r{1}+\r{4}\!\Big)V^{(L)}+\text{cyclic}_3\right]+\frac{1}{2}\left[\Big(\!\r{1}-\r{4}\!\Big)\tilde{V}^{(L)}+\text{cyclic}_3\right]\,.\label{v2_defined}}
The reason for this is that the (unique) identity among six-particle $R$-invariants,
\eq{\r{1}-\r{2}+\r{3}-\r{4}+\r{5}-\r{6}=0\,,}
means that coefficients of $\big(\r{1}\!\pm\!\r{4}\big)$ are necessarily ambiguous---being unchanged by the addition of an odd contribution to the coefficient of $\big(\r{1}\pl\r{4}\big)$, or an even contribution to the coefficient of $\big(\r{1}\mi\r{4}\big)$. This ambiguity is eliminated by the requirement that $V$ ($\tilde{V}$) be defined as {\it exclusively} even (odd) under a three-fold rotation of its arguments. 

To extract integrand-level representations for $V^{(2)}$ and $\tilde{V}^{(2)}$ in (\ref{v2_defined}) from the integrand of the ratio function in (\ref{six_point_two_loop_ratio_detail}), we may decompose each piece into their even/odd parts according to, 
\eq{\mathcal{I}\equiv \mathcal{I}^{\even}+\mathcal{I}^{\odd},\quad\text{where}\quad \mathcal{I}^{\even}\equiv\frac{1}{2}(1\pl r^3)\mathcal{I},\quad \mathcal{I}^{\odd}\equiv\frac{1}{2}(1\mi r^3)\mathcal{I}\,.\label{odd_even_defined_second_time}}
From this, and recognizing that $\mathcal{L}^{\even}_6\!=\!\mathcal{L}_6$, we find the integrand-level representations of the functions $V^{(2)}$ and $\tilde{V}^{(2)}$ appearing in \eqref{v2_defined} would be given by,
\eq{\mathcal{V}^{(2)}=-\frac{1}{2}\left(\frac{1}{2}\mathcal{L}_6+\sum_{i=1}^{15}\mathcal{I}^{\even}_{i}\right)\quad\text{and}\quad \tilde{\mathcal{V}}^{(2)}=-\frac{1}{2}\left(\sum_{i=1}^{15}\mathcal{I}^{\odd}_{i}\right)\,.\label{integrands_for_v2}}
After regularizing and integrating all the terms that appear in these expressions, we find that \eqref{integrands_for_v2} does exactly match the expressions given in \cite{Dixon:2011nj}---but obtained here in terms of (manifestly) dual-conformally invariant, local integral ingredients.\footnote{In terms of the functions $V(y_u,y_v,y_w)$ and $\tilde{V}(y_u,y_v,y_w)$ given in \cite{Dixon:2011nj}, we find that $V^{(2)}\!=\!V(1/\yu,1/\yv,1/\yw)=V(\yu,\yv,\yw)$ and $\tilde{V}^{(2)}\!=\!\tilde{V}(1/\yu,1/\yv,1/\yw)\!=\!-\tilde{V}(\yu,\yv,\yw)$}

In the following subsections, we describe in broad terms how the individual integrands that contribute to the ratio function appear analytically. In particular, we will discuss how the term-wise divergences of integrands encapsulate the cusp anomalous dimension appearing in the two-loop-logarithm of the MHV amplitude and how these divergences cancel in the ratio function. The dual-conformal regulator for the (two-loop-)logarithm of the MHV amplitude represents an interesting and novel regularization scheme; and we find that the (scheme-dependent) collinear anomalous dimension will be different from that of other regularization schemes. 

Another interesting novelty about the dual-conformal regulator is that it introduces both `impurities'---iterated integrals with kinematic-dependent, rational pre-factors---and also terms with lower transcendental weight. Thus, while the ratio function is known to be a pure integral with maximal transcendental weight through at least six loops \cite{sixLoops}, these facts are obscured through the use of the dual-conformal regulator. We will show how these term-wise impurities and lower-weight pollutions cancel in combination, but it would be interesting to know whether such obfuscations may be avoided through a different regularization scheme consistent with dual-conformal invariance. 

\vspace{-2pt}\subsection{Illustrations of Individual Contributions to the Ratio Function}\label{subsec:examples}\vspace{-6pt}

Except for $\mathcal{I}_1\!\equiv\!\tme{62}\merge\tme{35}$ (which is a rotated version of $\Omega^{(2)}$) every term contributing to the two-loop ratio function in (\ref{six_point_two_loop_ratio_detail}) is the mis-aligned merger of a pair of otherwise ordinary one-loop integrands. By this we mean that while the sets of $\ell_1$ and $\ell_2$ propagators in the integrands being merged correspond to local, one-loop Feynman integrands (albeit with unusual numerators), the set of propagators {\it in combination} cannot correspond to any Feynman graph. Nevertheless, these integrands are completely well-defined in dual-momentum space, and can be regulated and evaluated exactly the same way as described in \mbox{section \ref{sec:regularization_and_integration}}. 

Most of the integrals correspond to double-pentagons, which can be regulated and integrated as special instances of the general case discussed in \mbox{section \ref{subsubsec:feynman_parameter_ints_at_two_loop}}. However, three of the integrals contributing to the ratio function in (\ref{six_point_two_loop_ratio_detail}) involve fewer propagators. Specifically, this is the case for $\{\mathcal{I}_{2},\mathcal{I}_{3},\mathcal{I}_4\}$ defined in (\ref{individual_integrands_defined}). For these three, merging results in a two-loop integrand with a numerator directly proportional to one of the propagators, resulting in a (mis-aligned) `penta-box'. Using momentum-twistor variables for the numerators, these integrands would be given by\\[-8pt]
\eq{\begin{split}
\hspace{-16pt}\mathcal{I}_2\!=\!(1\pl p){\color{hred}\mathcal{I}_{{\color{hred}2346}}^1}\merge\tme{{\color{hblue}24}}&\!\equiv\!(1\pl p)\frac{\frac{1}{2}\ab{\ell_2{\color{hblue}(123)\tncap(345)}}\ab{2456}\ab{1234}}{\x{\ell_1}{{\color{hred}2}}\x{\ell_1}{{\color{hred}4}}\x{\ell_1}{{\color{hred}6}}\x{\ell_1}{\ell_2}\x{\ell_2}{{\color{hblue}2}}\x{\ell_2}{{\color{hblue}3}}\x{\ell_2}{{\color{hblue}4}}\x{\ell_2}{{\color{hblue}5}}};\hspace{-15pt}\\
\hspace{-16pt}\mathcal{I}_3\!=\!(1\pl p){\color{hred}\mathcal{I}_{{\color{hred}2346}}^1}\merge\tme{{\color{hblue}62}}&\!\equiv\!(1\pl p)\frac{\frac{1}{2}\ab{\ell_2{\color{hblue}(561)\tncap(123)}}\ab{2634}\ab{5612}}{\x{\ell_1}{{\color{hred}2}}\x{\ell_1}{{\color{hred}4}}\x{\ell_1}{{\color{hred}6}}\x{\ell_1}{\ell_2}\x{\ell_2}{{\color{hblue}6}}\x{\ell_2}{{\color{hblue}1}}\x{\ell_2}{{\color{hblue}2}}\x{\ell_2}{{\color{hblue}3}}};\hspace{-15pt}\\
\hspace{-16pt}\mathcal{I}_4\!=\!(1\pl p){\color{hred}\mathcal{I}_{{\color{hred}2346}}^1}\merge\tme{{\color{hblue}25}}&\!\equiv\!(1\pl p)\frac{\frac{1}{2}\ab{\ell_2{\color{hblue}(123)\tncap(456)}}\ab{5234}\ab{5612}}{\x{\ell_1}{{\color{hred}2}}\x{\ell_1}{{\color{hred}4}}\x{\ell_1}{{\color{hred}6}}\x{\ell_1}{\ell_2}\x{\ell_2}{{\color{hblue}2}}\x{\ell_2}{{\color{hblue}3}}\x{\ell_2}{{\color{hblue}5}}\x{\ell_2}{{\color{hblue}6}}}.\hspace{-15pt}
\end{split}\label{explicit_ints_234}}
In each case, the propagator $1/\x{\ell_1}{{\color{hred}3}}$ from $\mathcal{I}_{2{\color{hred}3}46}^1$ is cancelled against the numerator that results from the merger. It turns out that any double-pentagon integrand (involving at least one massless leg on the pentagon-side (which does not require regularization)) can be represented by a four-fold Feynman parameter integral. To understand this from the discussion in \mbox{section \ref{subsubsec:feynman_parameter_ints_at_two_loop}}, one need only observe that any one-loop box automatically results in a two-fold Feynman parameter integral representation, as does any one-loop pentagon involving at least one massless leg.

Following essentially the same analysis as described in \mbox{section \ref{subsubsec:feynman_parameter_ints_at_two_loop}} of introducing Feynman parameters, recognizing some Feynman parameter integrations as total derivatives, and finally rescaling the Feynman parameters that remain, we find very compact representations of each of the integrals in (\ref{explicit_ints_234}). For example, 
\begin{align}\hspace{-0pt}I_2&\equiv\!(1\pl p)\!\int\limits_0^{\infty}\!\!d^4\vec{\alpha}\frac{(\w\mi1)/2}{(\alpha_1(\w\pl\alpha_2)\pl\alpha_2)(1\pl\alpha_1\pl\alpha_3)(\alpha_1\pl\alpha_3\pl\alpha_4)((\w\pl\alpha_2)(\alpha_1\pl\alpha_3)\pl\alpha_2\pl\alpha_4)}\hspace{-0pt}\nonumber\\
\hspace{-0pt}&=(1\pl p)\frac{1}{2}\Big[H_{1,0,1,1}^{\w}+H_{1,0,0,1}^{\w}-\Zeta{2}H_{1,1}^{\w}-3\Zeta{3}H_{1}^{\w}\Big]\,,\hspace{-0pt}\end{align}
where $H_{\vec{w}}^{x}\!\equiv\!H_{\vec{w}}(1\mi x)$ is an ordinary harmonic polylogarithm. (Recall that \mbox{$p\!:\!\w\!\mapsto\!\u$}.) Interestingly, it turns out that $I_3\!=\!-I_2$, so these two cancel in the representation of the ratio function. 

For $I_4$, we find a similarly compact, four-fold representation:
\begin{align}\hspace{-0pt}I_4\!&\equiv\!(1\pl p)\!\!\int\limits_0^{\infty}\!\!d^4\vec{\alpha}\scalebox{0.80}{$\displaystyle\frac{\mi\w/2}{(\alpha_1(1\pl\alpha_2\w)\pl\alpha_2)(\alpha_1\pl\alpha_3\pl\alpha_4)(1\pl(\alpha_1\pl\alpha_3)\w\pl\alpha_4)(\alpha_1(1\pl\alpha_2\w)\pl\alpha_2(1\pl\alpha_3\w\pl\alpha_4)\pl\alpha_3)}$}\hspace{-0pt}\nonumber\\
\hspace{-0pt}&=-(1\pl p)\Big[H_{0,0,1,1}^{\w}+H_{0,0,0,1}^{\w}-\Zeta{2}H_{0,1}^{\w}+\frac{5}{4}\Zeta{4}\Big]\,.\hspace{-0pt}\end{align}

For all but one of the other individual contributions to the ratio function in (\ref{individual_integrands_defined}), the general procedure described in \mbox{section \ref{subsubsec:feynman_parameter_ints_at_two_loop}} results in a five-fold Feynman-parameter representation. This follows from the fact that for all but one of the integrals, there exists at least one massless leg which need not be regulated. Whenever this happens, we may analytically integrate-out two (of four) Feynman parameters of one of the loop integrals in exactly the same way as we did for the parameters $\gamma_{1},\gamma_2$ in our discussion of $\Omega^{(2)}$ in \mbox{section \ref{subsubsec:exempli_gratia}}. 

The exceptional case is integral $\mathcal{I}_{13}\!\equiv\!\tme{35}\merge\tmec{35}$. Here, every massless leg supports some infrared divergence, requiring that all massless legs be regulated---preventing us from obtaining a five-fold Feynman parameter representation. Nevertheless, the Feynman-parameterization analysis for $\mathcal{I}_{13}$ is far simpler than for the case of a truly general double-pentagon discussed in \mbox{section \ref{subsubsec:feynman_parameter_ints_at_two_loop}}. Indeed, following the now familiar analysis results the following, manifestly dual-conformally invariant regulated representation of $I_{13}$:
\eq{\begin{split}\hspace{-20pt}I_{13}\!&=\!\int\!\!d^4\ell_1\,d^4\ell_2\,\,\frac{\ab{\ell_1(234)\tncap(456)}\ab{\ell_2\,35}\x{4}{6}\x{3}{5}}{\x{\ell_1}{3}\x{\ell_1}{4}\x{\ell_1}{5}\x{\ell_1}{6}\x{\ell_1}{\ell_2}\x{\ell_2}{3}\x{\ell_2}{4}\x{\ell_2}{5}\x{\ell_2}{6}},\\
&=-\!\int\limits_{0}^{\infty}\!\!d^3\!\vec{\alpha}\,d^3\!\vec{\beta}\,\frac{\phantom{-}1\phantom{-}}{f_1\,f_2^2\,f_3}\,,\\[-12pt]\end{split}\label{int13_feynman_params}}
where,
\eq{\begin{split}f_1\!&\equiv \alpha_1(1\pl\alpha_3)\pl\alpha_2\pl{\color{divblue}\eps\big(\alpha_1\alpha_2\pl(u_2\pl\alpha_2)\alpha_3\big)}\,,\\
f_2\!&\equiv f_1\pl\alpha_1\beta_3\pl\beta_1(1\pl\alpha_3\pl\beta_3)\pl\beta_2\pl{\color{divblue}\eps\big(\alpha_2(\beta_1\pl\beta_3)\pl\beta_2(\alpha_1\pl\beta_1\pl\alpha_3\pl\beta_3)\pl u_2\beta_3\big)}\,,\\
f_3\!&\equiv\alpha_1\pl\beta_1\pl\alpha_2\pl\beta_2\pl{\color{divblue}\eps\big(u_2(\alpha_3\pl\beta_3)\big)}\,.
\end{split}\label{int_13_fs}}
In the limit of $\eps\!\to\!0$, this integral can be readily given in terms of logarithms:
\eq{I_{13}=8\Zeta{4}-{\color{wred}24\Zeta{3}}\mi6\Zeta{3}\log(\v)-{\color{divblue}\logdiv\Big[2\Zeta{2}\log(\v)+18\Zeta{3}\Big]}-{\color{divblue}4\Zeta{2}\logsquarediv}\,.\label{int13_integrated}}
Notice that we have highlighted both the divergences and also the contributions that are less-than-maximal-weight---namely the bare ${\color{wred}24\Zeta{3}}$ that contributes to $I_{13}$.

\vspace{-2pt}\subsubsection{Term-Wise Divergent Contributions to the Ratio Function}\label{subsubsec:divergences}\vspace{-6pt}

Among the contributions to the two-loop ratio function, all odd integrals (those contributing to $\tilde{V}^{(2)}$) are finite; but eight of the even (parts of the) integrals are infrared-divergent. Specifically, the divergent integrals are:
\eq{\text{divergent contributions to $V^{(2)}$:}\;\; \{I_{8}^{\even},I_{9}^{\even},I_{10}^{\even},I_{11}^{\even},I_{12}^{\even},I_{13}^{\even},I_{14}^{\even},I_{15}^{\even}\}\,.\label{list_of_divergent_integrals}}
Among these, only three are $\logsquarediv$-divergent: $\{I_{12}^{\even},I_{13}^{\even},I_{15}^{\even}\}$. For all of the infrared-divergent integrals in (\ref{list_of_divergent_integrals}), the origin of its divergence can be understood as arising from insufficient vanishing of the numerators of the integrands being merged in regions where {\it both} loops approach the same soft-collinear region. To see how this may arise, notice that although each loop's numerator protects {\it it} from having support in soft-collinear regions, the propagator $1/\x{\ell_1}{\ell_2}$ that arises through the merge operation introduces new pathways for one loop to access a soft-collinear region of the other loop (for which the numerator does not sufficiently vanish). 

Nevertheless, it is easy to see that the individual divergences cancel in the combination appearing in the ratio function $V^{(2)}$ in (\ref{integrands_for_v2}). The cancellation of $\logsquarediv$ divergences are easiest to see, as each of $\{I_{12}^{\even},I_{13}^{\even},I_{15}^{\even}\}$ include a term, 
\eq{I_{i}^{\even}\supset{\color{black}\!\mi4\Zeta{2}\logsquarediv}\quad\text{for}\quad i\!\in\!\{12,13,15\}\,.}
Thus, $\frac{1}{2}\mathcal{L}_6$ in (\ref{calL_in_terms_of_ints}) includes a divergence of ${\color{divblue}\pl12\Zeta{2}\logsquarediv}$, while the sum over even integrals $\{1,\ldots,15\}$ includes the divergence ${\color{divblue}\mi12\Zeta{2}\logsquarediv}$.

The cancellation of $\logdiv$-divergences from the terms contributing to $V^{(2)}$ are a bit less trivial. For the sake of reference, let us simply quote the $\logdiv$-divergent parts of each integral in (\ref{list_of_divergent_integrals}):
\eq{\begin{array}{l@{$\;\;\;$}l}I^{\even}_{8\phantom{1}}\!\supset\!\logdiv\!\big[6\Zeta{3}\big],&I^{\even}_{12}\!\supset\!\logdiv\!\big[\mi\divergentFunctionSymbol\mi2\Zeta{2}\log(\u \w)\mi12\Zeta{3}\big],\\
I^{\even}_{9\phantom{1}}\!\supset\!\logdiv\!\big[\divergentFunctionSymbol\mi2\Zeta{2}\log(\u\w)\mi6\Zeta{3}\big],&I^{\even}_{13}\!\supset\!\logdiv\!\big[\mi2\Zeta{2}\log(\v)\mi18\Zeta{3}\big],\\
I^{\even}_{10}\!\supset\!\logdiv\!\big[\divergentFunctionSymbol\big],&I^{\even}_{14}\!\supset\!\logdiv\!\big[\mi2\Zeta{2}\log(\v)\big],\\
I^{\even}_{11}\!\supset\!\logdiv\!\big[\mi\divergentFunctionSymbol\pl6\Zeta{3}\big],&I^{\even}_{15}\!\supset\!\logdiv\!\big[\mi12\Zeta{3}\big],
\end{array}}
where we have made use of the expression (which cancels in the sum),
\eq{\divergentFunctionSymbol\equiv(1\pl p)\Big(3\Li{3}{\u}-2\Li{2}{\u}\log(\u)-\frac{1}{2}\logsquare{\u}\log(1\mi \u)\Big)\,.\label{divergentFunctionSymbolDefined}}

Combining all the individual divergences, we see that they cancel according to:
\eq{\begin{split}\frac{1}{2}\mathcal{L}_6&\supset\!+\logdiv\Big[4\Zeta{2}\log(\u\v\w)+36\Zeta{3}\Big]+12\Zeta{2}\logsquarediv\,;\\
\sum_{i=1}^{15}I^{\even}_i&\supset\!-\logdiv\Big[4\Zeta{2}\log(\u\v\w)+36\Zeta{3}\Big]-12\Zeta{2}\logsquarediv\,.\end{split}}
%

\vspace{-2pt}\subsubsection{Other Novelties: Impurities and Lower-Weight Contributions}\label{subsubsec:impurities_and_other_novelties}\vspace{-6pt}

Although the ratio function $V^{(2)}$ is known to be a {\it pure} polylogarithm of uniform (and maximal) weight four, both of these properties are obscured slightly in the individual terms that contribute in equation \eqref{integrands_for_v2}.

Recall that a {\it pure} polylogarithmic function of weight $w$ which depends on some number of (kinematic) variables is one whose total differential with respect to these variables can be expressed in terms of {\it pure} functions of weight $(w\mi1)$ (with coefficients being algebraic functions with only simple poles). In the context of amplitudes, these are functions that can be expressed as iterated integrals of $d\!\log$ forms without any kinematic-dependent pre-factors. These correspond to loop integrals for which {\it all} maximal-co-dimension residues are independent of the external kinematics. Not all loop integrals may be so-normalized, but all the {\it unregulated} integrands contributing to any NMHV ratio function at two loops have this property.\footnote{We have made a weaker statement here than may have been expected: although all N$^2$MHV amplitudes {\it are} (provably) polylogarithmic at two loops, this fact fails to be preservable term-by-term if amplitudes are represented in a basis of local Feynman integrals.} 

That the dual-conformal regulator can spoil the purity of an integral is easy to understand (and hard to avoid): by ignoring any regulator-dependence in the numerators of loop integrands, it is liable to spoil the uniformity of its residues, introducing some relative kinematic-dependence among them. Indeed, this occurs for precisely four of the integrals that contribute to $V^{(2)}$---namely, $\{I^{\even}_{9},I^{\even}_{10},I^{\even}_{11},I^{\even}_{12}\}$. These integrals all have impure contributions that cancel in the sum:
\eq{\begin{array}{l@{$\;\;\;\;\;\;\;\;$}l}I^{\even}_{9\phantom{1}}\!\supset\!(1\pl p)\displaystyle\Big(\frac{\u}{1\mi\u}\impureFunctionSymbol\Big),&I^{\even}_{11}\!\supset\!\mi(1\pl p)\displaystyle\Big(\frac{\u}{1\mi\u}\impureFunctionSymbol\Big),\\
I^{\even}_{10}\!\supset\!(1\pl p)\displaystyle\Big(\frac{\u}{1\mi\u}\impureFunctionSymbol\Big),&I^{\even}_{12}\!\supset\!\mi(1\pl p)\displaystyle\Big(\frac{\u}{1\mi\u}\impureFunctionSymbol\Big),\end{array}\label{impure_function_contributions}}
where we have defined the recurring function
\eq{\begin{split}\impureFunctionSymbol&\equiv\hInt{1,1,0,1}{\u}-\hInt{1,0,1,1}{\u}+6\Zeta{3}H_1^{\u}\\
&=6\big(\Zeta{4}-\Li{4}{\u}\big)+3\big(\Li{3}{\u}-\Zeta{3}\big)\log(\u)+\frac{1}{2}\big(\Li{2}{\u}-\Zeta{2}\big)\logsquare{\u}\,.
\end{split}\label{impure_function_symbol_defined}}

We furthermore find that the dual-conformal regulator does not preserve uniform (let alone maximal) transcendental-weight. Indeed, exactly six of the integrals contributing to $V^{(2)}$ have finite contributions of weight three:
\eq{\begin{array}{l@{$\;\;\;\;\;\;\;\;$}l@{$\;\;\;\;\;\;\;\;$}l}I^{\even}_{8\phantom{1}}\!\supset\!+12\Zeta{3},&I^{\even}_{11}\!\supset\!+12\Zeta{3},&I^{\even}_{13}\!\supset\!-24\Zeta{3},\\
I^{\even}_{9\phantom{0}}\!\supset\!-12\Zeta{3},&I^{\even}_{12}\!\supset\!-12\Zeta{3},&I^{\even}_{15}\!\supset\!-12\Zeta{3};\end{array}\label{wrong_weight_contributions_table}}
which cancel each other in the sum for the ratio function.

\vspace{-2pt}\subsection[Logarithm of the MHV Amplitude, Remainder Function, and $\gamma_{\text{cusp}}$]{Two-Loop MHV Logarithm, Remainder Function, and $\cusp$}\label{subsec:two_loop_log}\vspace{-6pt}

We have already seen in equation (\ref{two_loop_logarithm_integrand_terms}) how the (two-loop-)logarithm of the MHV amplitude is represented in terms of the merged integrands listed in (\ref{individual_integrands_defined}). Expressed in terms of the standard BDS remainder function $R^{(2)}$ as given in \cite{DelDuca:2009au,Goncharov:2010jf}, we find
\begin{align}\hspace{0pt}\log\!\!\Big(\!A_6^{(0)}\!\Big)^{\!L=2}\hspace{-16pt}\!\!&=\!\mi\frac{1}{4}\Big(I_{14}+\frac{1}{2}I_{15}+\frac{1}{2}I_{16}+\text{cyclic}_6\Big)\nonumber\\
    &=\!R^{(2)}\pl\frac{11}{8}\Zeta{4}\pl\hspace{-1pt}\frac{3}{2}\Zeta{3}\hspace{-1pt}\log(\hspace{-1pt}\u\v\w\hspace{-1pt})\pl{\color{wred}9\Zeta{3}}\pl\frac{1}{2}\Zeta{2}\Big[\hspace{-1.5pt}\Li{2}{1\mi\u\hspace{-1pt}}\pl\!\log(\u\hspace{-1pt})\log(\v\hspace{-1pt})\pl\text{cyclic}_3\Big]\nonumber\\
&\phantom{=}\;\;\pl{\color{divblue}\logdiv\Big[\Zeta{2}\log(\u\v\w)\pl9\Zeta{3}\Big]}\pl{\color{divblue}3\Zeta{2}\logsquarediv}\label{integrated_two_loop_logarithm}\,.
\end{align}
It is worthwhile to see how this relates to expressions found using other regularization schemes. 

To best understand the structure of (\ref{integrated_two_loop_logarithm}), it is useful to first consider the dual-conformally regulated form of the one-loop MHV amplitude. Starting from the standard scalar-box representation of the six-point MHV amplitude at one loop \cite{Bern:2004bt}, and using the dual-conformal regulator described in \mbox{section \ref{subsec:dci_regulator}}, we find the following regulated expression:
\eq{\hspace{-10pt}\mi
    A_6^{(0),1}\!\!=\!3\Zeta{2}\pl\!\frac{1}{2}\Big[\Li{2}{1\mi\u}\pl\!\log(\u\hspace{-1pt})\log(\v\hspace{-1pt})\pl\text{cyclic}_3\Big]\pl{\color{divblue}\logdiv\log(\u\v\w)}\pl{\color{divblue}3\logsquarediv}.\label{six_point_one_loop_dci_regulated}}
In terms of this, the two-loop logarithm (\ref{integrated_two_loop_logarithm}) becomes:
\eq{\log\!\Big(\!A_6^{(0)}\!\Big)^{\!L=2}\hspace{-4pt}\!\!=\!\mi\Zeta{2}A_6^{(0),1}\pl{\color{divblue}9\Zeta{3}\logdiv}\pl{\color{wred}9\Zeta{3}}\pl\frac{3}{2}\Zeta{3}\log(\u\v\w)\mi\frac{49}{8}\Zeta{4}\pl
    R^{(2)}\,.\label{two_loop_log_in_terms_of_one_loop}}
This certainly has the right structure to encode the exponentiation of infrared divergences as dictated by the BDS ansatz \cite{Bern:2005iz}---with a scheme-dependent collinear anomalous dimension similar to that encountered in mass-regularization, \cite{Alday:2009zm,Henn:2010bk,Henn:2010ir,Drummond:2010mb}. Considering that dual-conformal regularization is very similar to mass regularization, it is worthwhile to see how sharp this comparison may be made. 

In the mass regularization scheme, one-loop MHV amplitudes are $\logk{2}{m^2}$-divergent. Using leg-label-dependent masses, the mass-regulated one-loop MHV amplitude has the form (see, e.g., \mbox{ref.\ \cite{Drummond:2010mb}}):
\eq{A_{n}^{(0),1}\Bigr|_{m_a^2\to0}=-\sum_{a=1}^n\log^{\hspace{-0.25pt}2}\hspace{-1pt}\!\left(\frac{m_a^2}{\x{a}{a\,\pl2}}\right)+\text{finite}\,.\label{mass_regularization_one_loop_mhv}}
(This is suggestively reminiscent of dimensional-regularization, for which no $1/\epsilon$ divergences arise at one loop; but (\ref{mass_regularization_one_loop_mhv}) does in fact include $\log(m^2)$-divergences too---proportional to $\log(\x{a}{a\,\pl2})$.)  The dual-conformal regulator is closer in spirit to a mass regulator, but with leg-dependent masses. Indeed, making the replacement
\eq{m_a^2\mapsto\eps\frac{\x{a\mi1}{a\,\pl1}\x{a}{a\,\pl2}}{\x{a\mi1}{a\,\pl2}}\,\label{mass_to_dci_transformation}}
the mass-regularization formula (\ref{mass_regularization_one_loop_mhv}) becomes:
\eq{A_{n}^{(0),1}\Bigr|_{m_a^2\propto\eps\to0}\mapsto-n\logsquarediv-2\logdiv\log\!\left(\prod_{a=1}^n\frac{\x{a}{a\,\pl2}}{\x{a}{a\,\pl3}}\right)+\text{finite}'\,,\label{mass_to_dci_regularization_one_loop}}
which reproduces the one-loop divergences of dual-conformally regulated one-loop amplitudes (see equation (\ref{one_loop_divergences_dci_regulator})) up to an overall normalization. For $n\!=\!6$ particles, the $\logdiv$-term in (\ref{mass_to_dci_regularization_one_loop}) is easily seen to become $\logdiv\log(\u\v\w)$, explaining the divergence structure in (\ref{six_point_one_loop_dci_regulated}). (Interestingly, the finite parts of (\ref{mass_to_dci_regularization_one_loop}) do not match those in the dual-conformal regularization scheme.)

The coefficients of $\logdiv^2$ at one loop in equation (\ref{six_point_one_loop_dci_regulated}) and at two loops in equation \eqref{two_loop_log_in_terms_of_one_loop} are determined by the cusp anomalous dimension $\gamma_{c}(a)\!=\!4 a\mi 4 \Zeta{2} a^2\pl\mathcal{O}(a^3)$. In the mass regularization scheme, the quadratic divergence of the $n$-point amplitude is proportional to $\mi\frac{n}{16}\gamma_c(a)$. The above comparison of the divergences between the mass regularization and DCI regulator scheme motivates the conjecture that we can write the coefficient of the quadratic divergence as $\mi\frac{n}{8}\gamma_c(a)$.

In the mass regularization scheme, the coefficient of the linear divergence is proportional to $\mi\frac{n}{2}\tilde{\mathcal{G}}_0(a)$, where $\tilde{\mathcal{G}}_0(a)$ is the scheme dependent collinear anomalous dimension. It takes the value $\tilde{\mathcal{G}}_0(a)\!=\!\mi\Zeta{3} a^2\pl\mathcal{O}(a^3)$ in the mass regularization scheme. In \mbox{ref. \cite{Dixon:2008gr}} it was shown, that in general, the collinear anomalous dimension can be related to two other anomalous dimensions:
\eq{\tilde{\mathcal{G}}_0(a) = \mathcal{G}_{0,\text{eik}}(a) + 2 B_{\delta}(a)\,.}
Here $\mathcal{G}_{0,\text{eik}}$ is the eikonal anomalous dimension, governing the single poles of a Wilson line with single cusp. The quantity $B_{\delta}$ is sometimes called the \emph{virtual} anomalous dimension and governs the coefficient of $\delta(1\mi x)$ in the twist-two anomalous dimensions or DGLAP kernels. In \mbox{ref. \cite{Kotikov:2004er}} the value of the virtual anomalous dimension was determined to be $B_{\delta}\!=\!3\Zeta{3}a^2\pl\mathcal{O}(a^3)$. This motivates the conjecture that in the dual-conformal regulator scheme, at least through two loops, the eikonal anomalous dimension vanishes and the coefficient of the linear divergence can be written as $\frac{n}{2}B_{\delta}$. This is in particular interesting, as the virtual anomalous dimension is known to all orders from integrability~\cite{Freyhult:2007pz}, while the eikonal anomalous dimension is only known perturbatively through three loops.

\newpage
\vspace{-12pt}\section{Conclusions and Future Directions}\label{sec:conclusions}\vspace{-10pt}

In planar maximally supersymmetric ($\mathcal{N}\!=\!4$) Yang-Mills theory (SYM), it seems always possible to represent loop amplitudes and related functions directly in terms of manifestly dual-conformally invariant integrals over loop momenta. Indeed, closed-form expressions involving only local propagators now exist for all amplitudes through three loops \cite{Bourjaily:2017wjl}. Although all infrared-safe combinations of amplitudes related to observables are expected to be dual-conformal invariant, there have long been important obstacles to making such symmetries manifest. The first problem is that infrared divergences require that loop amplitudes be regulated, and the most widely used regularization schemes severely break this symmetry. The second problem, apparent even for integrals that are infrared-finite, is that na\"{i}ve Feynman parameterization deeply obscures dual-conformal invariance---because it involves a sum of terms with different weights under conformal rescalings. The first obstacle to manifest dual-conformal invariance is addressed by the dual-conformal regularization scheme described in \mbox{ref.\ \cite{Bourjaily:2013mma}}; the second can be cured by Feynman-parameterizing one loop at a time, and strategically rescaling the Feynman parameters. 

In this work, we have shown that dual-conformal invariance can be made (and maintained) manifest throughout regularization and loop integration. More specifically, using the dual-conformal regulator, any (ultraviolet-finite) planar loop integral will take the form of a polynomial in $\logdiv$, where each coefficient can be expressed in terms of finite (Feynman-)parametric integrals whose denominators exclusively on `parity-even' dual-conformal cross-ratios---those rationally expressed in terms of Mandelstam invariants. We have proven this by direct construction through two loops for any multiplicity, and we expect it to hold more generally. 

\vspace{-12pt}\section*{Acknowledgements}\vspace{-10pt}
This work began in collaboration with Lance Dixon and we are indebted to him for helping with many of the explicit results for the six particle amplitude. We are grateful to JJ Carrasco for helpful comments on early drafts of this work; to Andrew McLeod, Matt von Hippel, and Matthias Wilhelm for help with the integration of $\Omega^{(L)}$ and its comparison with the results obtained by the hexagon bootstrap program; and to Nima Arkani-Hamed, Marcus Spradlin and Jaroslav Trnka for useful discussions. This work was supported in part by the Danish National Research Foundation (DNRF91), a grant from the Villum Fonden, and a Starting Grant (No.\ 757978) from the European Research Council, and a grant from the Simons Foundation (341344, LA) (JLB), as well as by the U.~S.~Department of Energy (DOE) under contract DE-AC02-76SF00515 (FD). Finally, the authors are grateful for the hospitality and support from the Institute for Advanced Study in Princeton, the Munich Institute for Astro- and Particle Physics (MIAPP), the Mainz Institute for Theoretical Physics (MITP), and the Galileo Galilei Institute in Florence.

\newpage
\appendix

\vspace{-6pt}\section{Conventions and Definitions for Hexagon Functions}\label{appendix:conventions_and_definitions}\vspace{-6pt}

As described above, we use dual-momentum coordinates for which $p_a\!\equiv\!(x_{a+1}\mi x_{a})$, with $\x{a}{b}\!=\!\x{b}{a}\!\equiv\!(x_b\mi x_a)^2\!=\!(p_a\pl\!\ldots\pl p_{b-1})^2$. Each of the dual points $x_{a}$, often denoted simply by `$(a)$', is associated with a line, $\mathrm{span}\{z_{a-1},z_{a}\}$, in momentum-twistor space \cite{Hodges:2009hk} ($\mathbb{P}^3$ represented by homogenous coordinates $z_a\!\in\!\mathbb{C}^4$ for each particle). Momentum-twistor space may be motivated by making the masslessness of particles, \mbox{$p_a^2\!=\!\x{a}{a\,\pl1}$}, manifest---as $\x{a}{b}\!\propto\!\ab{a\mi1\,a\,\,b\mi1\,b}\!\equiv\!\det\{z_{a-1},z_a,z_{b-1},z_b\}$. 

Any `line' in momentum-twistor space defines a (possibly) complex point in dual-momentum space. The `chiral' numerators of $\Omega^{(2)}$ and $\tilde{\Omega}^{(2)}$ were expressed in these terms---see equation (\ref{omega_numerators_defined}). A line in momentum-twistor space can be represented as a bi-twistor $(ab)\!\equiv\!\mathrm{span}\{z_a,z_b\}$, or more indirectly as the intersection of `planes':
\eq{\begin{split}
(a\,b\,c)\tncap(d\,e\,f)&\equiv\mathrm{span}\{z_a,z_b,z_c\}\,\tcap\,\mathrm{span}\{z_d,z_e,z_f\}\,\\
&\equiv(ab)\ab{cdef}+(bc)\ab{adef}+(ca)\ab{bdef}\,.\end{split}}
We include these definitions here only for the sake of completeness---as some of the manipulations involved in recognizing the numerators in equation (\ref{omega2_nums_post_rescaling}) required knowing how these numerators were expressible in terms of momentum-twistor cross-ratios. Interested readers should consult \mbox{ref.\ \cite{ArkaniHamed:2010gh}} for more details. 

For six particles, there are three cyclically-related `parity-even' cross ratios (those expressible rationally in dual-momentum coordinates):
\vspace{-2pt}\eq{\u\equiv\frac{\x{1}{3}\x{4}{6}}{\x{1}{4}\x{3}{6}},\quad \v\equiv\frac{\x{2}{4}\x{1}{5}}{\x{2}{5}\x{1}{4}},\quad \w\equiv\frac{\x{3}{5}\x{2}{6}}{\x{3}{6}\x{2}{5}}\,.\label{definition_of_uvw}\vspace{-2pt}}
When expressed in terms of momentum twistors these become
\vspace{-1pt}\eq{\u\equiv\frac{\ab{61\,23}\ab{34\,56}}{\ab{61\,34}\ab{23\,56}},\quad \v\equiv\frac{\ab{12\,34}\ab{45\,61}}{\ab{12\,45}\ab{34\,61}},\quad \w\equiv\frac{\ab{23\,45}\ab{56\,12}}{\ab{23\,56}\ab{45\,12}}.\quad \vspace{-1pt}}

Recall from section~\ref{subsubsec:exempli_gratia} the appearance of the square root `$\Delta_6$', defined in equation \eqref{delta6_defined} up to a sign. This ambiguity is resolved by re-expressing the cross-ratios $u_a$ in terms of parameters that rationalize $\Delta_6$. This is for example achieved by the variables $y_a$ introduced in \cite{Dixon:2011pw}: In terms of momentum-twistors, we set:\footnote{Concretely, our variables are $(\u,\v,\w)\!=\!(w,u,v)$ and $(\yu,\yv,\yw)\!=\!(1/y_w,1/y_u,1/y_v)$ in terms of the definitions in \cite{Dixon:2011pw}. The translation to the conventions in \cite{Dixon:2011nj} is $(\u,\v,\w)\!=\! (u,v,w)$ and $(\yu,\yv,\yw)\!=\!(1/y_u,1/y_v,1/y_w)$.}
\eq{\hspace{-10pt}\yu\!\equiv\!\frac{\ab{4612}\ab{5123}\ab{3456}}{\ab{3451}\ab{4562}\ab{6123}},\; \yv\!\equiv\!\frac{\ab{2456}\ab{3561}\ab{1234}}{\ab{1235}\ab{2346}\ab{4561}},\; \yw\!\equiv\!\frac{\ab{2346}\ab{3451}\ab{5612}}{\ab{3561}\ab{4612}\ab{2345}}.\hspace{-10pt}}
The parity-even cross-ratios $u_a$ are related to the $y_a$ variables according to:
\eq{\hspace{-60pt}\u\!=\!\frac{y_1(1\mi y_2)(1\mi y_3)}{(1\mi y_1\,y_2)(1\mi y_1\,y_3)},\;\; \v\!=\!\frac{y_2(1\mi y_3)(1\mi y_1)}{(1\mi y_2\,y_3)(1\mi y_2\,y_1)},\;\; \w\!=\!\frac{y_3(1\mi y_1)(1\mi y_2)}{(1\mi y_3\,y_1)(1\mi y_3\,y_2)}.\hspace{-40pt}}
In terms of these, the argument of $\Delta_6$ becomes a perfect square, allowing us to clearly disambiguate (and in fact {\it define}) its sign:
\eq{\Delta_6=\sqrt{(1\mi \u\mi\v\mi\w)^2-4\u\v\w}\equiv\frac{(1\mi\yu)(1\mi\yv)(1\mi\yw)(1\mi\yu\yv\yw)}{(1\mi\yu\yv)(1\mi\yv\yw)(1\mi\yw\yu)}\,.\label{explicit_delta_6_defined}}

In Appendix~\ref{appendix:explicit_integrals} and in the ancillary files attached to this work's submission to the {\tt arXiv}, there is one further set of rationalizing variables used in the explicit representation of iterated integrals, denoted $\{t,x,y\}$, which may be defined by\footnote{A related set of variables was used in \cite{Caron-Huot:2018dsv}: The definitions of $(x,y,z)$ in appendix~A {\it of that paper} translate into our variables as $\frac{x(y+ t)}{t(x-1)}$, $\frac{y(x+t)}{t(y-1)}$, and $\frac{(x+t)(y+t)}{xy(1-x)(1-y)}$, respectively. However, these do \emph{not} rationalize the symbol alphabet.}
\eq{x\equiv\frac{1\mi y_1 y_2 y_3}{y_1y_3(1\mi y_2)},\;\; y\equiv\frac{1\mi y_1y_2y_3}{1\mi y_2},\;\;t\equiv\frac{(1\mi y_3)(1\mi y_1y_2y_3)}{y_3(1\mi y_1)(1\mi y_2)}.\label{txy_definition}}
These parameters have the advantage that they identify the 9-letter symbol alphabet 
\begin{equation*}
	\{\u,\v,\w,1\mi\u,1\mi\v,1\mi\w,\yu,\yv,\yw\}
\end{equation*}
of hexagon functions with the cross-ratios $\frac{(q_i-q_k)(q_j-q_l)}{(q_i-q_l)(q_j-q_k)}$ on the moduli space $\mathfrak{M}_{0,6}$ parametrized as $(q_1\!=\!1,q_2\!=\!0,q_3\!=\!\mi x/t,q_4\!=\!\mi xy/t,q_5\!=\!\infty,q_6\!=\!x)$. The logarithmic differentials of the cross-ratios span the cohomology $H^1(\mathfrak{M}_{0,6})$ which has dimension $9$. In other words, the parametrization \eqref{txy_definition} identifies functions with symbols in the hexagon alphabet precisely with the multiple polylogarithms on $\mathfrak{M}_{0,6}$, which are well-understood \cite{Brown:MZVPeriodsModuliSpaces,Bogner:2014mha}. In particular, we can use algorithms to write functions in a basis.

Finally, in the Feynman-parameter representations of integral contributions to the six point ratio function included among the ancillary files to this work, one final set of (parity-odd) variables are used:
\eq{\hspace{-60pt}v_1\!\equiv\!\frac{(1\mi y_1)(1\mi y_1 y_2 y_3)}{y_1(1\mi y_2\,y_3)},\;\; v_2\!\equiv\!\frac{(1\mi y_2)(1\mi y_1 y_2 y_3)}{y_2(1\mi y_3\,y_1)},\;\; v_3\!\equiv\!\frac{(1\mi y_3)(1\mi y_1 y_2 y_3)}{y_3(1\mi y_1\,y_2)}.\label{vs_defined_by_ys}\hspace{-40pt}}
These odd-variables are simply $v_a\!\equiv\!\Delta_6/u_a$---where the sign of $\Delta_6$ is understood as being defined in terms of the $y_a$'s according to (\ref{explicit_delta_6_defined}).  

Finally, we have made use of two symmetry generators. The first of these is a rotation operator `$r$', which acts on the various hexagon variables according to:
\eq{r(p_{a}, x_{a}, z_{a}, u_{a}, v_a, y_a) = (p_{a+1}, x_{a+1}, z_{a+1}, u_{a+1}, \mi v_{a+1},1/y_{a+1}).\label{rotation_operators_action_defined}}
The minus sign in the action of $r$ on $v_a$ arises because $r\Delta_6=\mi\Delta_6$. 

The final operator used in our work for the two-loop ratio function terms is a reflection `$p$' centered on particle labelled by $a\!=\!1$ which acts on the momentum-twistors and $x$-space coordinates according to equation \eqref{action_of_p_on_xs_and_zs}. 
On the $u_a$ cross-ratios, $p$ acts by $p(\u,\v,\w)=(\w,\v,\u)$; and on the $y_a$ and $v_a$ variables, by:
\eq{p(v_1,v_2,v_3)=(\mi v_3,\mi v_2,\mi v_1),\qquad p(\yu,\yv,\yw)=(\yw^{-1}\!\!,\yv^{-1}\!\!,\yu^{-1})\,.\label{reflection_operator_action_defined}}

\newpage
\vspace{-6pt}\section{Explicit Contributions to the Six-Point Ratio Function}\label{appendix:explicit_integrals}\vspace{-6pt}

In this Appendix, we outline the analytic structure of each of the contributions to the six-point two-loop ratio function, expressed in terms of the integrals tabulated in \eqref{individual_integrands_defined}. We first give the general form of these contributions, then describe the general methodology used to obtain these expression in Appendix~\ref{appendix:technical_aspects_of_integration} and the organization of the ancillary files in Appendix~\ref{appendix:organization_of_ancillary_files}.

We use the shorthand $H_{\vec{w}}^{x}\!\equiv\!H_{\vec{w}}(1\mi x)$ for harmonic polylogarithms \cite{RemiddiVermaseren:HarmonicPolylogarithms} with
\begin{equation}
	H_{0,\vec{w}}(z)
	= \int_0^z \frac{\td z'}{z'} H_{\vec{w}}(z')
	,\quad
	H_{1,\vec{w}}(z)
	= \int_0^z \frac{\td z'}{1-z'} H_{\vec{w}}(z')
	\quad\text{and}\quad
	H_{\emptyset}(z) = 1.
	\label{eq:HPL-def}%
\end{equation}
Let us start with the even parts of these integrals, $I^{\even}_i$, which we can express in terms of harmonic (and ordinary) polylogarithms except for just three of them, $I^{\even}_{5}$, $I^{\even}_{6}$ and $I^{\even}_{16}$, together with the standard hexagon function $\Omega^{(2)}(\u,\v,\w)$ given for example in \cite{Dixon:2011nj}. Concretely, the even integrals are given by:
\eq{\fwboxR{0pt}{I^{\even}_{1\phantom{0}}\!=\,}\fwboxL{350pt}{\Omega^{(2)}(\w,\u,\v)\equiv r^2\!\left[\Omega^{(2)}(\u,\v,\w)\right]\,; }}
\eq{\fwboxR{0pt}{I^{\even}_{2\phantom{0}}\!=\,}\fwboxL{350pt}{\frac{1}{2} \Bigl[H_{1,0,1,1}^{\u}\pl H_{1,0,0,1}^{\u}\pl H_{1,0,1,1}^{\w} \pl H_{1,0,0,1}^{\w}\mi\Zeta{2}\Bigl(H_{1,1}^{\u}\pl H_{1,1}^{\w}\Bigr)\mi 3\Zeta{3}\Bigl(H_{1}^{\u}\pl H_{1}^{\w}\Bigr)\Bigr]\,;}}
\eq{\fwboxR{0pt}{I^{\even}_{3\phantom{0}}\!=\,}\fwboxL{350pt}{\mi I^{\even}_{2}\,;}}
\eq{\fwboxR{0pt}{I^{\even}_{4\phantom{0}}\!=\,}\fwboxL{350pt}{\mi H_{0,0,0,1}^{\u}\mi H_{0,0,1,1}^{\u}\mi H_{0,0,0,1}^{\w}\mi H_{0,0,1,1}^{\w} \pl \Zeta{2}\Bigl[H_{0,1}^{\u}\pl H_{0,1}^{\w}\Bigr]\mi \frac{5}{2}\Zeta{4}\,;} }
\eq{\begin{split}\fwboxR{0pt}{I^{\even}_{7\phantom{0}}\!=\,}&\fwboxL{350pt}{\phantom{\pl}X_3\mi3 I^{\even}_{5}\pl I^{\even}_{6}
            \mi \frac{5}{2}(r+r^2)I^{\even}_{1\phantom{0}}\mi \frac{3}{2} I^{\even}_{1\phantom{0}}\mi\frac{63}{2}\Zeta{4}}\\
&\fwboxL{350pt}{\mi6\Zeta{3}H_{1}^{\v}\pl\Zeta{2}\Bigl[3\Big(H_{1}^{\u}\pl H_{1}^{\w}\Big)H_{1}^{\v}\mi5H_{1}^{\u}H_{1}^{\w}\pl 9H_{0,1}^{\v}\pl 4H_{1,1}^{\v}\pl H_{0,1}^{\u}\pl H_{0,1}^{\w}\Bigr]\,;}\end{split}}
\eq{\fwboxR{0pt}{I^{\even}_{8\phantom{0}}\!=\,}\fwboxL{350pt}{I^{\even}_{2} \mi I^{\even}_{4}\mi\frac{19}{2}\Zeta{4}\pl{\color{wred}12\Zeta{3}}\pl{\color{divblue}6\Zeta{3}\logdiv}\,;}}
\eq{\begin{split}\fwboxR{0pt}{I^{\even}_{9\phantom{0}}\!=\,}&\fwboxL{350pt}{\phantom{\pl}\frac{\u}{1-\u}
            \Bigl[H_{1,1,0,1}^{\u}\mi H_{1,0,1,1}^{\u}{\pl}6\Zeta{3}H_1^{\u}\Bigr]\pl\frac{\w}{1-\w}
            \Bigl[H_{1,1,0,1}^{\w}\mi H_{1,0,1,1}^{\w}{\pl}6\Zeta{3}H_1^{\w}\Bigr]}\\
        &\fwboxL{350pt}{\mi\log(\u)\Bigl[ H_{0,1,1}^{\w}\mi H_{1,0,1}^{\w} \pl3\Zeta{3}\Bigr]\mi H_{0,0,0,1}^{\u}\pl H_{0,0,1,1}^{\u}\mi H_{0,1,0,1}^{\u}\mi2 H_{1,0,0,1}^{\u}\mi H_{1,1,0,1}^{\u}}\\
        &\fwboxL{350pt}{\mi\log(\w)\Bigl[ H_{0,1,1}^{\u}\mi H_{1,0,1}^{\u} \pl3\Zeta{3}\Bigr]\mi H_{0,0,0,1}^{\w}\pl H_{0,0,1,1}^{\w}\mi H_{0,1,0,1}^{\w}\mi2 H_{1,0,0,1}^{\w}\mi H_{1,1,0,1}^{\w}}\\
        &\fwboxL{350pt}{\pl\Zeta{2}\Bigl[H_{0,1}^{\u}\pl H_{1,1}^{\u}\pl H_{0,1}^{\w}\pl H_{1,1}^{\w}\mi 2H_{1}^{\u}H_{1}^{\w}\Bigr]\mi{\color{wred}12\Zeta{3}}\mi\frac{5}{2}\Zeta{4}}\\
&\fwboxL{350pt}{\pl{\color{divblue}\logdiv\Bigl[H_{1,0,1}^{\u}\mi H_{0,1,1}^{\u}\pl H_{1,0,1}^{\w}\mi H_{0,1,1}^{\w}\mi\Zeta{2}\log(\u\w)\Big]}\,;}\end{split}}
\vspace{-12pt}\eq{\begin{split}~\\[-2pt]\fwboxR{0pt}{I^{\even}_{10}\!=\,}&\fwboxL{350pt}{\phantom{\pl}X_{2}\pl
            I_1^{\even}\pl\frac{3}{2}(r\pl r^2)I_1^{\even}\pl I_5^{\even}\mi I_6^{\even}\mi
            I_{11}^{\even}\mi\Zeta{2}\Big[H_{0,1}^{\u}\pl
            H_{1,1}^{\u}\pl4H_{0,1}^{\v}\pl2H_{1,1}^{\v}}\\[-4pt]
            &\fwboxL{350pt}{\pl H_{0,1}^{\w}\pl H_{1,1}^{\w} \pl
                H_{1}^{\u}(H_{1}^{\v}\mi2H_{1}^{\w})\pl
                H_{1}^{\v}H_{1}^{\w}\Big]\pl{\color{wred}12\Zeta{3}}\pl5\Zeta{4}\pl{\color{divblue}6\Zeta{3}\logdiv}\,;}\end{split}}
\eq{\begin{split}\fwboxR{0pt}{I^{\even}_{11}\!=\,}&\fwboxL{350pt}{\mi\frac{\u}{1-\u}
            \Bigl[H_{1,1,0,1}^{\u}\mi
            H_{1,0,1,1}^{\u}{\pl}6\Zeta{3}H_1^{\u}\Bigr]\mi\frac{\w}{1-\w}
            \Bigl[H_{1,1,0,1}^{\w}\mi H_{1,0,1,1}^{\w}{\pl}6\Zeta{3}H_1^{\w}\Bigr]}\\
&\fwboxL{350pt}{\pl\log(\v)\Big[ H_{0,1,1}^{\u}\mi H_{1,0,1}^{\u}\pl H_{0,1,1}^{\w}\mi H_{1,0,1}^{\w}\Big]\pl H_{1,0,0,1}^{\u}\pl H_{1,0,0,1}^{\w}\mi H_{1,1,0,1}^{\u}\mi H_{1,1,0,1}^{\w}}\\
&\fwboxL{350pt}{\mi 2\Big(H_{0,0,1,1}^{\u}\pl H_{0,0,1,1}^{\w}\Big)\pl\Zeta{2}\Big[H_{1,1}^{\u}\pl H_{1,1}^{\w}\mi H_{1}^{\u}H_{1}^{\v}\mi H_{1}^{\v}H_{1}^{\w}\Big]\pl3\Zeta{3}\log(\u\w)}\\
&\fwboxL{350pt}{\pl{\color{wred}12\Zeta{3}}\pl{\color{divblue}\logdiv\Big[H_{0,1,1}^{\u}\mi H_{1,0,1}^{\u}\pl H_{0,1,1}^{\w}\mi H_{1,0,1}^{\w}\mi\Zeta{2}\log(\u\w)\Big]}\,;}
\end{split}}
\eq{\begin{split}\fwboxR{0pt}{I^{\even}_{12}\!=\,}&\fwboxL{350pt}{\mi I_2^{\even}\mi I_9^{\even}\mi I_{15}^{\even}\mi 2 H_{1,1,0,1}^{\u} \mi H_{0,1,0,1}^{\u} \mi H_{1,0,0,1}^{\u}\mi2 H_{1,1,0,1}^{\w} \mi H_{0,1,0,1}^{\w} \mi H_{1,0,0,1}^{\w}}\\
        &\fwboxL{350pt}{\pl{2}\Zeta{2}\Big[H_{1,1}^{\u}\pl H_{1,1}^{\w}\mi 2H_{1}^{\u}H_{1}^{\w}\Big]\pl{\color{black}6}\Zeta{3}\Big[H_1^{\u}\pl H_{1}^{\w}\Big]\mi{\color{wred}36\Zeta{3}}\pl\frac{13}{2}\Zeta{4}}\\
&\fwboxL{350pt}{\mi{\color{divblue}\logdiv\Big[4\Zeta{2}\log(\u\w)\pl30\Zeta{3}\Big]}\mi{\color{divblue}8\Zeta{2}\logsquarediv}\,;}
\end{split}}
\eq{\fwboxR{0pt}{I_{13}^{\even}\!=\,}\fwboxL{350pt}{8\Zeta{4}\mi6\Zeta{3}\log(\v)\mi{\color{wred}24\Zeta{3}}\mi{\color{divblue}\logdiv\Big[2\Zeta{2}\log(\v)\pl18\Zeta{3}\Big]}\mi{\color{divblue}4\Zeta{2}\logsquarediv}\,;}}
\eq{\begin{split}\fwboxR{0pt}{I^{\even}_{14}\!=\,}&\fwboxL{350pt}{\phantom{\pl}X_{1}\pl I_5^{\even}\mi I_6^{\even}\pl(r\pl r^2)I_1^{\even}\pl\Zeta{2}\Big[2H_{0,1}^{\u}\mi4H_{0,1}^{\v}\pl2H_{0,1}^{\w}\mi2H_{1,1}^{\v}\mi H_{1}^{\u}H_{1}^{\v}\pl2H_{1}^{\u}H_{1}^{\w}}\\
        &\fwboxL{350pt}{\mi H_{1}^{\v}H_{1}^{\w}\Big]{\pl}6\Zeta{3}H_{1}^{\v}\pl2\Zeta{4}\mi{\color{divblue}2\Zeta{2}\log(\v)\logdiv}\,;}\\
\end{split}}
\eq{\begin{split}\fwboxR{0pt}{I^{\even}_{15}\!=\,}&\fwboxL{350pt}{\phantom{\pl} H_{0,0,0,1}^{\u}\pl
            H_{0,1,0,1}^{\u}\pl H_{0,0,0,1}^{\w}\pl H_{0,1,0,1}^{\w}\mi
            2\Zeta{2}\Bigl(H_{0,1}^{\u}\pl H_{0,1}^{\w}\Bigr)\mi{\color{wred}12\Zeta{3}}\pl{5\Zeta{4}}}\\
&\fwboxL{350pt}{\mi{\color{divblue}12\Zeta{3}\logdiv}\mi{\color{divblue}4\Zeta{2}\logsquarediv}\,;}\\
\end{split}}
where we have made use of the following expressions, 
\eq{\begin{split}\fwboxR{0pt}{X_{1}\!\equiv\,}&\fwboxL{350pt}{\mi\log(\v/\w) \Big(\!H_{1,0,1}^{\u}\pl H_{0,0,1}^{\u}\!\Big)\mi\log(\v/\u)\Big(\!H_{1,0,1}^{\w}\pl H_{0,0,1}^{\w}\!\Big)\mi\frac{1}{2} \logsquare{\u/\w}\Big(\!H_{0,1}^{\v}\pl H_{1,1}^{\v}\!\Big)}\\[-4pt]
&\fwboxL{350pt}{\pl H_{1,0,1,1}^{\u}\pl H_{0,0,1,1}^{\u}\mi H_{0,1,0,1}^{\u}\pl H_{1,0,0,1}^{\u}\pl H_{1,0,1,1}^{\w}\pl H_{0,0,1,1}^{\w}\mi H_{0,1,0,1}^{\w}\pl H_{1,0,0,1}^{\w}}\\
&\fwboxL{350pt}{\pl H_{1,0,1,1}^{\v}\pl 2 H_{0,0,1,1}^{\v}\pl H_{1,1,0,1}^{\v}\pl H_{0,1,0,1}^{\v}\pl H_{1,0,0,1}^{\v}\pl 2 H_{0,0,0,1}^{\v}\,.}\\[4pt]
\end{split}}
\eq{\begin{split}\fwboxR{0pt}{X_{2}\!\equiv\,}&\fwboxL{350pt}{\phantom{\pl}X_{1}\pl\frac{1}{2}\Big[\!\log(\v/\w) \Big(\!H_{0,0,1}^{\u}\pl H_{0,1,1}^{\u}\!\Big)\pl\log(\v/\u)\Big(\!H_{0,0,1}^{\w}\pl H_{0,1,1}^{\w}\!\Big)}\\
&\fwboxL{350pt}{\mi\frac{1}{2}\logsquare{\v/\w} \Big(\!H_{0,1}^{\u}\pl H_{1,1}^{\u}\!\Big)\mi\frac{1}{2} \logsquare{\v/\u} \Big(\!H_{0,1}^{\w}\pl H_{1,1}^{\w}\!\Big)\pl H_{0,0,1,1}^{\u}\pl 2 H_{1,0,1,1}^{\u}}\\
&\fwboxL{350pt}{\pl H_{1,1,0,1}^{\u}\pl 2 H_{1,0,0,1}^{\u}\pl 3 H_{0,1,0,1}^{\u}\pl3 H_{0,0,0,1}^{\u}\pl H_{0,0,1,1}^{\w}\pl 2 H_{1,0,1,1}^{\w}\pl H_{1,1,0,1}^{\w}\pl 2 H_{1,0,0,1}^{\w}}\\
&\fwboxL{350pt}{\pl3 H_{0,1,0,1}^{\w}\pl3 H_{0,0,0,1}^{\w}\pl2 H_{1,0,1,1}^{\v}\pl2 H_{0,0,1,1}^{\v}\pl2 H_{1,0,0,1}^{\v}\pl2 H_{0,0,0,1}^{\v}\Big]\,;}
\end{split}}
\eq{\begin{split}\fwboxR{0pt}{X_{3\phantom{1}}\!\equiv\,}&\fwboxL{350pt}{\phantom{\pl}\frac{1}{2}X_{1}\mi3X_{2}\pl\frac{3}{2}\Big[\mi\frac{1}{4}\Big(\!\logsquare{\u}\logsquare{\w}\pl\logsquare{\u}\logsquare{\v}\pl\logsquare{\v}\logsquare{\w}\Big)}\\
&\fwboxL{350pt}{\pl\frac{1}{2} \log(\u) \log(\v) \log(\w)\log(\u\w)\pl\log(\v/\w) H_{0,1,1}^{\u} \pl\log(\v/\u) H_{0,1,1}^{\w}}\\
&\fwboxL{350pt}{\pl H_{1,0,1,1}^{\u}\pl H_{1,1,0,1}^{\u}\pl H_{0,1,0,1}^{\u}\pl H_{1,0,0,1}^{\u}\pl H_{0,0,0,1}^{\u}\pl H_{1,0,1,1}^{\w}\pl H_{1,1,0,1}^{\w}\pl H_{0,1,0,1}^{\w}}\\
&\fwboxL{350pt}{\pl H_{1,0,0,1}^{\w}\pl H_{0,0,0,1}^{\w}\pl H_{1,0,1,1}^{\v}\Big]\mi\log(\u)\log(\w) \Big(\frac{1}{2} H_{1,1}^{\v}\mi H_{0,1}^{\v} \Big)}\\
&\fwboxL{350pt}{\pl\log(\u\w)\Big(H_{0,1,1}^{\v}\pl\frac{1}{2} H_{1,0,1}^{\v} \Big)\pl H_{0,0,1,1}^{\v} \mi\frac{1}{2} H_{1,1,0,1}^{\v} \mi H_{0,1,0,1}^{\v}\pl \frac{1}{2} H_{1,0,0,1}^{\v}\pl H_{0,0,0,1}^{\v}\,;}
\end{split}}
Notice that we have highlighted all divergences and all contributions with less-than-maximal-weight.

The three integrals not listed above, $I^{\even}_{5}$, $I^{\even}_{6}$ and $I^{\even}_{16}$, contain two non-hexagon letters,
\eq{1\mi t\mi x\mi y = \frac{1\mi\v\mi\w}{\u \w}, \quad \textrm{and}\quad t\mi xy\mi tx\mi ty=\frac{(1\mi\w)(1\mi\u\mi\v)}{\u \w^2}\,,\label{non_hexagon_letters}}
where we have defined the variables $\{t,x,y\}$ in \mbox{Appendix \ref{appendix:conventions_and_definitions}} (see \mbox{equation (\ref{txy_definition})}). We give explicit, function-level expressions for these integrals in terms of multiple polylogarithms defined in terms of the $\{t,x,y\}$ variables in the ancillary files of this work's submission to the {\tt arXiv}. While we leave the full details to the ancillary files, some of their novelties may be illustrated through the $2\,\tens\,2$ components of their symbols.

The first component of their co-products is expressible in terms of non-Steinmann hexagon functions. The non-hexagon letters in (\ref{non_hexagon_letters}) appear only in the third entry of the symbol. Integrability forbids them from appearing in the first two entries; however, there seems to be no obvious rule forbidding them from appearing in the final entry. The symbol for each is given below, indexed according to the first two entries:
\eq{\begin{split}\fwboxR{0pt}{I^{\even}_{5}\Bigr|_{\u\otimes \v + \v \otimes \u}\!=}&\fwboxL{265pt}{\phantom{\pl}\cop{1\mi\u}{\frac{\u\w}{\v}}\pl\frac{1}{4}\cop{\frac{\w(1\mi\u)}{1\mi\v}}{\frac{\u(1\mi\w)}{\w (1\mi\u)}}\pl\cop{\frac{\w}{\v}}{\w} }\\
&\fwboxL{265pt}{\pl \frac{1}{2}\cop{\frac{1\mi\u\mi\v}{\w}}{\frac{(1\mi\v) \v}{(1\mi\u) \u}}\mi\frac{1}{4} \cop{\yw}{\yu^3 \yv^2 \yw^3};}\\
\end{split}}
\eq{\begin{split}\fwboxR{0pt}{I^{\even}_{5}\Bigr|_{\u\otimes \w + \w\otimes \u}\!=}&\fwboxL{265pt}{\phantom{\pl}\frac{1}{4} \cop{\frac{1\mi\u}{1\mi\w}}{\frac{\w(1\mi\u)}{\u(1\mi\w)}}\mi\cop{\frac{\v}{(1\mi\u) (1\mi\w)}}{\frac{\v}{\u\w}}}\\
	&\fwboxL{265pt}{\pl\frac{1}{4}\cop{\v}{\frac{(1\mi\u)(1\mi\w)\v^2}{\u\w(1\mi\v)^2}}\mi\frac{1}{4} \cop{\yv}{\yu^3 \yv^2 \yw^3}\,;}\\
\end{split}}
\eq{\begin{split}\fwboxR{0pt}{\fwboxL{72pt}{I^{\even}_{5}\Bigr|_{\v\otimes \w + \w \otimes \v}}\!=}&\fwboxL{265pt}{\,\,p\!\left[I^{\even}_{5}\Bigr|_{\u\otimes \v + \v\otimes \u}\right]\,;}
\end{split}}
\eq{\begin{split}\fwboxR{0pt}{\fwboxL{72pt}{I^{\even}_{5}\Bigr|_{\u\otimes(1-\u)}}\!=}&\fwboxL{265pt}{\phantom{\pl}\frac{1}{2}\cop{1\mi\u}{\frac{\u \w}{\v}}\pl\frac{1}{2}\cop{\u}{\frac{\u}{\v\w}}\pl\frac{1}{4} \cop{\yu \yv \yw}{\yu^3 \yv^2 \yw^3}}\\
&\fwboxL{265pt}{\pl\frac{1}{2}\cop{\frac{1\mi\u\mi\v}{\u\v}}{\frac{(1\mi\u) \u}{(1\mi\v)\v}}\pl\frac{1}{4} \cop{\frac{\v}{\u\w}}{\frac{(1\mi\w)\u^5\w^3}{(1\mi\u)\v^4}}\,;}
\end{split}}
\eq{\begin{split}\fwboxR{0pt}{\fwboxL{72pt}{I^{\even}_{5}\Bigr|_{\v\otimes(1-\v)}}\!=}&\fwboxL{265pt}{\phantom{\pl}\frac{1}{2}\cop{\frac{\u\v}{\w}}{\w}\pl\frac{1}{4} \cop{\frac{(1\mi\u\mi\v)^2}{\u\v\w}}{\frac{(1\mi\u) \u}{(1\mi\v)\v}}}\\
&\fwboxL{265pt}{\pl\frac{1}{2} \cop{\frac{\v\w}{\u}}{\u}\pl\frac{1}{4} \cop{\frac{(1\mi\v\mi\w)^2}{\u\v\w}}{\frac{(1\mi\w) \w}{(1\mi\v)\v}}}\\
&\fwboxL{265pt}{\pl\frac{1}{4} \cop{\yu \yv \yw}{\yu^3 \yv^2 \yw^3}\,;}
\end{split}}
\eq{\begin{split}\fwboxR{0pt}{\fwboxL{72pt}{I^{\even}_{5}\Bigr|_{\w\otimes(1-\w)}}\!=}&\fwboxL{265pt}{\,\,p\!\left[I^{\even}_{5}\Bigr|_{\u\otimes(1-\u)}\right]\,;}
\end{split}}
\eq{\begin{split}\fwboxR{0pt}{\fwboxL{72pt}{I^{\even}_{5}\Bigr|_{\v\otimes \v}}\!=}&\fwboxL{265pt}{\,\mi\cop{\u}{\w} \mi \cop{\w}{\u}\,.}
\end{split}}

\vspace{10pt}\noindent While for $I^{\even}_6$, we have,
\eq{\begin{split}\fwboxR{0pt}{\fwboxL{72pt}{I^{\even}_{6}\Bigr|_{\u\otimes \v + \v\otimes \u}}\!=}&\fwboxL{265pt}{\phantom{\pl}\cop{\frac{1\mi\v}{\v}}{1\mi\u}\pl\cop{\frac{\w(1\mi\u)}{1\mi\v}}{1\mi\w}}\\
&\fwboxL{265pt}{\pl\cop{\frac{1\mi\u\mi\v}{(1\mi\u) \w}}{\frac{\v}{1\mi\u}} \pl \cop{\yw}{\yv}\,;}
\end{split}}
\eq{\begin{split}\fwboxR{0pt}{\fwboxL{72pt}{I^{\even}_{6}\Bigr|_{\v\otimes \w + \w \otimes \v}}\!=}&\fwboxL{265pt}{\,\,p\!\left[I^{\even}_{6}\Bigr|_{\u\otimes \v + \v\otimes \u}\right]\,;}
\end{split}}
\eq{\begin{split}\fwboxR{0pt}{\fwboxL{72pt}{I^{\even}_{6}\Bigr|_{\u\otimes \w + \w\otimes \u}}\!=}&\fwboxL{265pt}{\phantom{\pl}\cop{\yv}{\yv}\mi\cop{\frac{\v}{(1\mi\u) (1\mi\w)}}{\frac{\v}{(1\mi\u) (1\mi\w)}}\,;}
\end{split}}
\eq{\begin{split}\fwboxR{0pt}{\fwboxL{72pt}{I^{\even}_{6}\Bigr|_{\u\otimes(1-\u)}}\!=}&\fwboxL{265pt}{\phantom{\pl}\cop{\frac{1\mi\u}{\u}}{1\mi\u}\pl\cop{\frac{\v}{\u \w}}{1\mi\w}}\\
&\fwboxL{265pt}{\pl \cop{\frac{\u \w}{1\mi\u\mi\v}}{\frac{\v}{1\mi\u}} \mi \cop{\yu \yv \yw}{\yv}\,;}
\end{split}}
\eq{\begin{split}\fwboxR{0pt}{\fwboxL{72pt}{I^{\even}_{6}\Bigr|_{\v\otimes(1-\v)}}\!=}&\fwboxL{265pt}{\phantom{\pl}\cop{\frac{1\mi\v\mi\w}{\u}}{\frac{1\mi\w}{\v}}\pl \cop{\frac{1\mi\u\mi\v}{\w}}{\frac{1\mi\u}{\v}}}\\
&\fwboxL{265pt}{\pl\cop{\frac{\v}{1\mi\v}}{\v} \mi\cop{\yu \yv \yw}{\yv}\,;}
\end{split}}
\eq{\begin{split}\fwboxR{0pt}{\fwboxL{72pt}{I^{\even}_{6}\Bigr|_{\w\otimes(1-\w)}}\!=}&\fwboxL{265pt}{\,\,p\!\left[I^{\even}_{6}\Bigr|_{\u\otimes(1-\u)} \right]\,;}
\end{split}}
\eq{\begin{split}\fwboxR{0pt}{\fwboxL{72pt}{I^{\even}_{6}\Bigr|_{\v\otimes \v}}\!=}&\fwboxL{265pt}{\,\mi\cop{\u}{1\mi\u}\pl \cop{1\mi\v}{\v}\mi \cop{\w}{1\mi\w}\,.}
\end{split}}

\vspace{10pt}\noindent Finally, for $I^{\even}_{16}$, we have,
\eq{\begin{split}\fwboxR{0pt}{\fwboxL{72pt}{2I^{\even}_{16}\Bigr|_{\u\otimes \v + \v\otimes \u}}\!=}&\fwboxL{265pt}{\phantom{\pl}\cop{1\mi\u}{\frac{(1\mi\u) (1\mi\v)^3}{\u\v^3}}\pl\cop{1\mi\v}{\frac{(1\mi\u)^3(1\mi\v)}{\u^3\v}}}\\
&\fwboxL{265pt}{\pl\cop{\frac{(1\mi\u\mi\v)^4}{\w}}{\frac{\u\v}{(1\mi\u)(1\mi\v)}}\pl\cop{\yw}{\yu\yv}\,;}
\end{split}}
\eq{\begin{split}\fwboxR{0pt}{\fwboxL{72pt}{2I^{\even}_{16}\Bigr|_{\u\otimes \w + \w\otimes \u}}\!=}&\fwboxL{265pt}{\phantom{\pl}\cop{\frac{1\mi\u}{\v}}{\frac{\u\v}{(1\mi\u)(1\mi\v)}}\pl\cop{\yv}{\yu\yv}}\\
&\fwboxL{265pt}{\pl\cop{1-\w}{\frac{\u(1\mi\v)}{(1\mi\u)\v}}\,;}
\end{split}}
\eq{\begin{split}\fwboxR{0pt}{\fwboxL{72pt}{2I^{\even}_{16}\Bigr|_{\v\otimes \w + \w \otimes \v}}\!=}&\fwboxL{265pt}{\,\,r^2p\!\left[2I^{\even}_{6}\Bigr|_{\u\otimes \w + \w\otimes \u}\right]\,;}
\end{split}}
\eq{\begin{split}\fwboxR{0pt}{\fwboxL{72pt}{2I^{\even}_{16}\Bigr|_{\u\otimes (1-\u)}}\!=}&\fwboxL{265pt}{\phantom{\pl}4\cop{\frac{1\mi\u\mi\v}{\u}}{\frac{1\mi\v}{\v}}
\pl4\cop{\frac{(1\mi\u\mi\v)(1\mi\u)}{\u\v}}{\frac{1-\u}{\u}}}\\
&\fwboxL{265pt}{\pl2I^{\even}_{16}\Bigr|_{\w\otimes (1-\w)}\,;}\\
\end{split}}
\eq{\begin{split}\fwboxR{0pt}{\fwboxL{72pt}{2I^{\even}_{16}\Bigr|_{\v\otimes (1-\v)}}\!=}&\fwboxL{265pt}{\mi4\cop{\frac{\u}{1\mi\u}}{\frac{\u}{1\mi\u}}
\pl4\cop{\frac{\v}{1\mi\v}}{\frac{\v}{1\mi\v}}\pl2I^{\even}_{16}\Bigr|_{\u\otimes (1-\u)}\,;}\\
\end{split}}
\eq{\begin{split}\fwboxR{0pt}{\fwboxL{72pt}{2I^{\even}_{16}\Bigr|_{\w\otimes (1-\w)}}\!=}&\fwboxL{265pt}{\cop{\frac{\u}{\v}}{\frac{\u(1\mi\v)}{(1\mi\u)\v}}
\pl\cop{\w}{\frac{\u\v}{(1\mi\u)(1\mi\v)}}\mi\cop{\yu\yv\yw}{\yu\yv}.}\\
\end{split}}

\subsection*{Parity-Odd Contributions}
Regarding the parity-odd parts $I^{\odd}_{i}$ of the integrals, all but four of them vanish:
\eq{I^{\odd}_{1}=I^{\odd}_{2}=I^{\odd}_{3}=I^{\odd}_{4}=I^{\odd}_{6}=I^{\odd}_{8}=I^{\odd}_{9}=I^{\odd}_{11}=I^{\odd}_{12}=I^{\odd}_{13}=I^{\odd}_{15}=I^{\odd}_{16}=0\,.}
The only non-zero odd parts are $\{I^{\odd}_{5},I^{\odd}_{7},I^{\odd}_{10},I^{\odd}_{14}\}$. These are very simply related to each other and to the odd part of the ratio function
$\tilde{V}^{(2)} = -\tilde{V}(y_1,y_2,y_3)= \tilde{V}(y_u,y_v,y_w)$ in terms of the function $\tilde{V}(y_u,y_v,y_w)$ given in \cite{Dixon:2011nj}:
\eq{I^{\odd}_{5}=-I^{\odd}_{7}=-I^{\odd}_{14}=I^{\odd}_{10}+2\tilde{V}^{(2)}.}

Explicit expressions for these integrals (given in terms of hyperlogarithms) are a bit too unwieldy to warrant being presented here---but complete expressions will be provided in the ancillary files as part of this work's submission to the {\tt arXiv}. However, some of their structure can be understood through the symbol entries, which we provide below for the function $I^{\odd}_{10}$. As for the even integrals, the non-hexagon letters appear only in the third slot of the symbol. Therefore, we give the $2\,\tens\,2$ part of the co-product, where the first entry can be written in terms of non-Steinmann hexagon functions. 
\vspace{-24pt}\eq{\begin{split}~\\[12pt]\fwboxR{0pt}{\fwboxL{72pt}{I^{\odd}_{10}\Bigr|_{\u\otimes \v
                    + \v\otimes \u}}\!=}&\fwboxL{265pt}{\phantom{\pl}\frac{1}{2} \cop{1\mi\u}{\yu
                \yv \yw}\pl\cop{1\mi\v}{\yu \yv \yw}\mi \frac{1}{2} \cop{\u}{\yw}}\\
        &\fwboxL{265pt}{\mi\frac{1}{2} \cop{\w}{\frac{\yu}{\yw}}\!\mi\frac{1}{2} \cop{1\mi\u\mi\v}{\yu \yv \yw^2}\!;}
\end{split}}
\eq{\begin{split}\fwboxR{0pt}{\fwboxL{72pt}{I^{\odd}_{10}\Bigr|_{\u\otimes \w + \w\otimes \u}}\!=}&\fwboxL{265pt}{\phantom{\pl}\frac{1}{2} \cop{\frac{1\mi\w}{1\mi\u}}{\yu \yv \yw} \pl \frac{1}{2} \cop{\v}{\frac{\yw}{\yu}}\pl\frac{1}{2} \cop{\frac{\w}{\u}}{\yv}\,;}
\end{split}}
\eq{\begin{split}\fwboxR{0pt}{\fwboxL{72pt}{I^{\odd}_{10}\Bigr|_{\v\otimes \w + \w \otimes \v}}\!=}&\fwboxL{265pt}{\,\,p\!\left[I^{\odd}_{10}\Bigr|_{\u\otimes \v + \v \otimes \u}\right]\,;}
\end{split}}
\eq{\begin{split}\fwboxR{0pt}{\fwboxL{72pt}{I^{\odd}_{10}\Bigr|_{\u\otimes(1-\u)}}\!=}&\fwboxL{265pt}{\phantom{\pl}\frac{1}{2} \cop{\frac{1\mi\u\mi\v}{\u\v}}{\yw}\pl\frac{1}{2} \cop{\frac{1\mi\u\mi\v}{\w}}{\yu \yv \yw}\,;}
\end{split}}
\eq{\begin{split}\fwboxR{0pt}{\fwboxL{72pt}{I^{\odd}_{10}\Bigr|_{\v\otimes(1-\v)}}\!=}&\fwboxL{265pt}{\phantom{\pl}\frac{1}{2} \cop{\u}{\yu \yv}\pl \frac{1}{2} \cop{\v}{\frac{\yu}{\yw}}\mi\frac{1}{2}\cop{\w}{\yv\yw}}\\
&\fwboxL{265pt}{\pl\frac{1}{2} \cop{1\mi\u\mi\v}{\yu \yv \yw^2}\mi\frac{1}{2} \cop{1\mi\v\mi\w}{\yu^2 \yv \yw}\,;}
\end{split}}
\eq{\begin{split}\fwboxR{0pt}{\fwboxL{72pt}{I^{\odd}_{10}\Bigr|_{\w\otimes(1-\w)}}\!=}&\fwboxL{265pt}{\,\,p\!\left[I^{\odd}_{10}\Bigr|_{\u\otimes(1-\u)}\right]\,;}
\end{split}}
\eq{\begin{split}\fwboxR{0pt}{\fwboxL{72pt}{I^{\odd}_{10}\Bigr|_{\u\otimes \u}}\!=}&\fwboxL{265pt}{\phantom{\pl}\frac{1}{2} \cop{1\mi\u}{\yu \yv \yw}\mi\frac{1}{2} \cop{\v}{\yw}\mi\frac{1}{2} \cop{\w}{\yv} \,;}
\end{split}}
\eq{\begin{split}\fwboxR{0pt}{\fwboxL{72pt}{I^{\odd}_{10}\Bigr|_{\w\otimes \w}}\!=}&\fwboxL{265pt}{\,\,p\!\left[I^{\odd}_{10}\Bigr|_{\u\otimes \u}\right]\,.}
\end{split}}
%

\vspace{-0pt}\subsection{Technical Details of Integration Methods}\label{appendix:technical_aspects_of_integration}\vspace{-6pt}
In the ancillary files to this paper---whose organizations will be discussed below---we provide everything necessary to compute all of the integrals defined in \eqref{individual_integrands_defined}. All of them are linearly reducible, but their dependence of the resulting polylogarithms on the regulator $\eps$  has varying degrees of complexity.

The integrals $\{I_1, I_2, I_3, I_4, I_5, I_6, I_7, I_8, I_{11}, I_{15}, I_{16}\}$ are either finite or hyperlogarithms in $\eps$, so they can be written as hyperlogarithms in $\eps$ and the expansion at $\eps\!=\!0$ is straightforward. Example code which performs these integrations and takes the small $\eps$ expansion analytically is included in the ancillary files. 

For integrals $\{I_9, I_{10},I_{14}\}$, the arguments of the polylogarithms obtained by direct integration are not rational in $\eps$, and they are indeed not hyperlogarithms `in $\eps$'. However, we can parametrize $\eps\!=\!\eps(\rho)\!=\!\rho\pl\mathcal{O}(\rho^2)$ such that the arguments become rational in $\rho$. Hence we can write the integrals as hyperlogarithms in $\rho$ and the expansion at $\rho\!=\!0$ is straightforward again. This transformation is demonstrated in the ancillary files. 

For $I_{13}$, we found it most convenient to re-parametrize $\v\!=\!\v(\tilde{\v},\eps)\!=\!\tilde{\v}\pl \mathcal{O}(\eps)$ in terms of a suitable coordinate $\tilde{\v}$, to achieve the rationalization in $\eps$. After expanding the hyperlogarithms at $\eps\!=\!0$, we have polylogarithms of $\tilde{\v}|_{\eps=0}\!=\!\v$.

Finally, for $I_{12}$, we have decided to take a much more general approach---illustrating that all the transformations just discussed are not truly necessary, but merely convenient shortcuts to obtain the small-$\eps$ expansion from hyperlogarithms depending on $\eps$. Recall from the discussion in \mbox{section \ref{subsec:analytic_integration_methods}} that these integrals may be represented generally in the form
\begin{equation*}
	\Li{n_1,\ldots,n_d}{f_1(\eps),\ldots,f_d(\eps)}
\end{equation*}
with arguments $f_i(\eps)$ that are algebraic functions of $\eps$ (and the cross-ratios). We are merely interested in the first few terms in the expansion at $\eps\!=\!0$, and not in the full dependence on $\eps$. The expansion at $\eps\!=\!0$ can be calculated from the differential equation of polylogarithms, which is known in terms of the differentials $\td\log (f_i(\eps)\mi f_j(\eps))$. We can simply expand these in $\eps\!=\!0$ and thereby recursively compute the expansion of the polylogarithms as a series in $\eps^j \logkdiv{k}$ term-by-term. This procedure is illustrated for the case of $I_{12}$ in the ancillary files.

\vspace{-6pt}\subsection{Organization of Ancillary Files}\label{appendix:organization_of_ancillary_files}\vspace{-6pt}

Included with this article's submission to the {\tt arXiv} is a collection of `ancillary' files---available by following the link on the article's abstract page. These files are also available under DOI \href{http://doi.org/10.5287/bodleian:BRyawJrRN}{10.5287/bodleian:BRyawJrRN}. In particular, we have provided both the Feynman-parametric and final hyperlogarithmic representations of each integral discussed above, together with example code that demonstrates how the latter can be obtained from the former using \HyperInt.

The examples discussed in this paper are provided and documented in human-readable plain-text files. The file `{\tt integral\_data.txt}' can be read directly by {\Mathematica} and `{\tt IntDefs.mpl}' contains the same definitions in {\Maple} syntax. In particular, these files defines all coordinates explicitly in terms of $\{\yu,\yv,\yw\}$ as discussed in Appendix~\ref{appendix:conventions_and_definitions}, and clarify the definitions of each integral $I_{i}$ of \eqref{individual_integrands_defined} in terms of syntax used by the packages associated with \mbox{refs. \cite{Bourjaily:2010wh,Bourjaily:2012gy,Bourjaily:2013mma,Bourjaily:2015jna}}.

After this, each Feynman-parametric integrand, {\tt feynInt[i\_]} is defined in two parts: $\{I^{\even}_i,I^{\odd}_i\}$. Each of these is expressed in terms of rational, parametric expressions such as the examples discussed in section~\ref{sec:six_point_example}. In particular, the variables $\alpha_i,\beta_i$---denoted as {\tt a1},\ldots, {\tt b1},\ldots---should be understood as homogeneous coordinates on $\mathbb{P}^n$, and---after de-projectivization according to (\ref{projective_measure_defined})---should be integrated over the entire positive, real domain. 

The file `{\tt integral\_data.txt}' also contains hyperlogarithm representations of the odd and even parts obtained after parametric integration, expressed in terms described above in the beginning of this Appendix. At the very end of this file, the combinations of $I_i$ required for the representation of the logarithm of the MHV amplitude, the function $\mathcal{L}_6$ defined in \eqref{calL_in_terms_of_ints}, and $V^{(2)}$ and $\widetilde{V}^{(2)}$ are given. These are also given in {\Maple} format in the files `{\tt results.mpl}' and `{\tt comparisons.mpl}'.

In addition to these source files, we include {\Maple} programs that take as input the parametric integrands and then use {\HyperInt} to explicitly reproduce the hyperlogarithm expressions for the integrals.

%

\providecommand{\href}[2]{#2}\begingroup\raggedright\endgroup

\end{document}